\documentclass[onecolumn,unpublished,a4paper]{quantumarticle}
\usepackage[utf8]{inputenc}
\usepackage[backend=biber,style=alphabetic,maxbibnames=999,backref=true]{biblatex}
\usepackage{amsmath}
\usepackage{amssymb}
\usepackage{amsthm}
\usepackage{amsfonts}
\usepackage{changepage}
\usepackage[caption=false]{subfig}
\usepackage[colorlinks]{hyperref}
\usepackage[all]{hypcap}
\usepackage{tikz}
\usepackage{relsize}
\usepackage{color,soul}
\usepackage{capt-of}
\usepackage{mathtools}
\usepackage{float}
\usepackage[section]{placeins}
\usepackage{listings}
\usepackage[T1]{fontenc}  % Without this, underscores lost when copy-pasting output!
\usetikzlibrary{decorations.pathreplacing}
\usepackage{braket}

\DefineBibliographyStrings{english}{
  backrefpage  = {Cited on page},
  backrefpages = {Cited on pages}
}
\renewcommand\thesection{\arabic{section}}

\theoremstyle{definition}
\newtheorem{definition}{Definition}[section]

\newtheorem{assumption}[definition]{Assumption}
\newcommand{\eq}[1]{\hyperref[eq:#1]{Equation~\ref*{eq:#1}}}
\renewcommand{\sec}[1]{\hyperref[sec:#1]{Section~\ref*{sec:#1}}}
\DeclareRobustCommand{\app}[1]{\hyperref[app:#1]{Appendix~\ref*{app:#1}}}
\newcommand{\ass}[1]{\hyperref[ass:#1]{Assumption~\ref*{ass:#1}}}
\newcommand{\fig}[1]{\hyperref[fig:#1]{Figure~\ref*{fig:#1}}}
\newcommand{\tbl}[1]{\hyperref[tbl:#1]{Table~\ref*{tbl:#1}}}
\newcommand{\theoremref}[1]{\hyperref[theorem:#1]{Theorem~\ref*{theorem:#1}}}
\newcommand{\definitionref}[1]{\hyperref[definition:#1]{Definition~\ref*{definition:#1}}}

\begin{document}
\title{Magic state cultivation: growing T states as cheap as CNOT gates}

\date{\today}
\author{Craig Gidney}
\email{craig.gidney@gmail.com}
\affiliation{Google Quantum AI, California, USA}

\author{Noah Shutty}
\affiliation{Google Quantum AI, California, USA}

\author{Cody Jones}
\affiliation{Google Quantum AI, California, USA}

\begin{abstract}
We refine ideas from \cite{knill1996threshold,jones2016colorcode,chamberland2020colorinjection,bombin2024gap,gidney2023colorcode,gidney2024yoked,hirano2024zeroleveldistill} to efficiently prepare good $|T\rangle$ states.
We call our construction ``magic state cultivation'' because it gradually grows the size and reliability of one state.
Cultivation fits inside a surface code patch and uses roughly the same number of physical gates as a lattice surgery CNOT gate of equivalent reliability.
We estimate the infidelity of cultivation (from injection to idling at distance 15) using a mix of state vector simulation, stabilizer simulation, error enumeration, and Monte Carlo sampling.
Compared to prior work, cultivation uses an order of magnitude fewer qubit·rounds to reach logical error rates as low as $2 \cdot 10^{-9}$ when subjected to $10^{-3}$ uniform depolarizing circuit noise.
Halving the circuit noise to $5 \cdot 10^{-4}$ improves the achievable logical error rate to $4 \cdot 10^{-11}$.
Cultivation's efficiency and strong response to improvements in physical noise suggest that further magic state distillation may never be needed in practice.

\textbf{Data availability}: \emph{Code, circuits, and stats are \href{https://doi.org/10.5281/zenodo.13777072}{available on Zenodo}~\cite{gidneyy2024cultivationdata}.}
\end{abstract}

{
  \hypersetup{linkcolor=blue}
  \tableofcontents
}

\begin{figure}
    \begin{adjustwidth}{-1cm}{-1cm}
        \centering
        \resizebox{\linewidth}{!}{
            \includegraphics{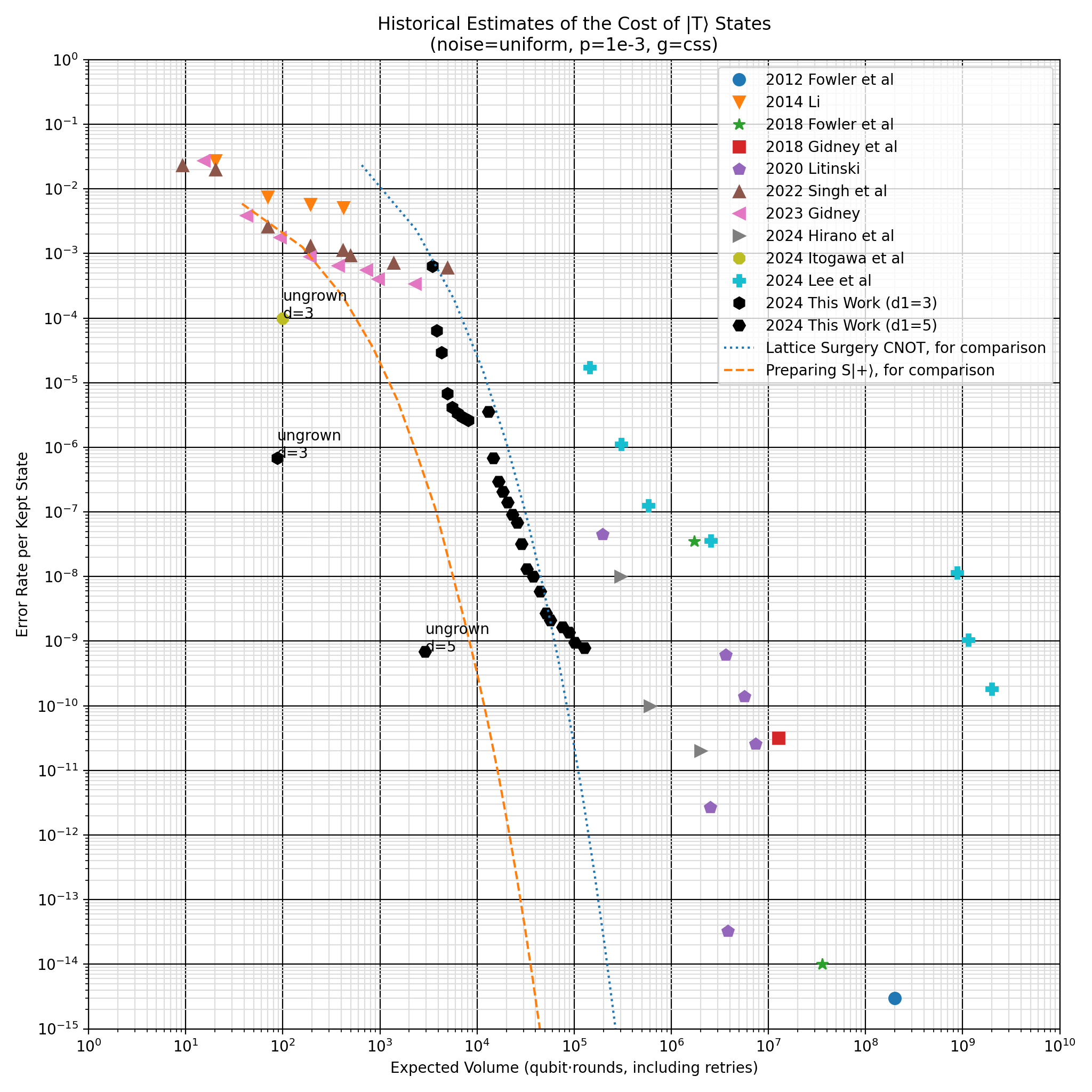}
        }
    \end{adjustwidth}
    \caption{
        \textbf{Scatter plot of historical estimates of $T|+\rangle$ cost trade-offs} from \cite{fowler2012surfacecodereview,li2015,fowler2018latticesurgery,gidney2019autoccz,litinski2019notascostly,singh2022,gidney2023hook,hirano2024zeroleveldistill,itogawa2024zeroleveldistilldistill,lee2024colordistillation,gidney2024ybasis}, under circuit noise with a noise strength of $10^{-3}$.
        Points marked with ``ungrown'' omit the escape stage; they don't account for the cost of growing the state to a large code distance.
        They represent the part of the process that's experimentally accessible today.
    }
    \label{fig:historical_progression}
\end{figure}
\begin{figure}
    \centering
    \resizebox{\linewidth}{!}{
    \begin{tabular}{|r|c|c||c|c||c|c|}
         \hline
         Paper      
            &Noise
            &Fault
            &Error Rate
            &Discard Rate
            &Error Rate
            &Discard Rate
        \\
            &Strength
            &Distance
            &(Ungrown)
            &(Ungrown)
            &(End to End)
            &(End to End)\\
         \hline
         Li 2015~\cite{li2015}
            &$10^{-3}$
            &7
            &N/A
            &N/A

            % https://arxiv.org/pdf/1410.7808
            % by eye from Figure 4
            &$5 \cdot 10^{-3}$

            % https://arxiv.org/pdf/1410.7808 page 4 right hand side
            % "the first phase succeeds with a probability in the range
            % 0.38 ∼ 0.59 depending on the rate of single-qubit errors"
            &50\%               

            \\
         Chamberland et al 2020~\cite{chamberland2020colorinjection} 
            &${5 \cdot 10^{-4}}$
            &7

            % https://www.nature.com/articles/s41534-020-00319-5/tables/1
            % last row, third column
            &$3 \cdot 10^{-7}$

            % derived from being 50% at 1e-4 and the ratio of the expected
            % cost at 1e-4 vs 5e-4 in https://www.nature.com/articles/s41534-020-00319-5/tables/1
            &95\%

            &N/A
            &N/A
            \\
         Chamberland et al 2020~\cite{chamberland2020colorinjection} 
            &${10^{-4}}$
            &7

            % https://www.nature.com/articles/s41534-020-00319-5/tables/1
            % third row, third column
            &$5 \cdot 10^{-10}$

            % https://arxiv.org/pdf/2003.03049
            % page 13, bottom left
            % "repeat the protocol until it is accepted (with an acceptance probability"
            &50\%

            &N/A
            &N/A
            \\
         
         Gidney 2023~\cite{gidney2023hook}
            &$10^{-3}$
            &7
            &N/A
            &N/A
            
            % https://arxiv.org/pdf/2302.12292
            % by eye from figure 6, page 10
            &$6 \cdot 10^{-4}$
            % https://arxiv.org/pdf/2302.12292
            % by eye from figure 6, page 10
            &75\%

            \\
         Itogawa et al 2024~\cite{itogawa2024zeroleveldistilldistill}   
            &$10^{-3}$
            &3
            
            % https://arxiv.org/pdf/2403.03991
            % by eye from figure 10 page 7
            &$1 \cdot 10^{-4}$

            % https://arxiv.org/pdf/2403.03991
            % by eye from figure 11 page 8
            &30\%

            &N/A
            &N/A
            \\
         \hline
         (This Paper) 
            &$10^{-3}$
            &3
            &$6 \cdot 10^{-7}$
            &35\%
            &$3 \cdot 10^{-6}$
            &80\%
            \\
         (This Paper) 
            &$10^{-3}$
            &5
            &$6 \cdot 10^{-10}$
            &85\%
            &$2 \cdot 10^{-9}$
            &99\%
            \\
         (This Paper) 
            &${5 \cdot 10^{-4}}$
            &5
            &$2 \cdot 10^{-11}$
            &65\%
            &$4 \cdot 10^{-11}$
            &90\%
            \\
         (This Paper) 
            &${10^{-4}}$
            &5
            &$6 \cdot 10^{-15}$
            &20\%
            &N/A
            &N/A
            \\
         \hline
    \end{tabular}
    }
    \caption{
        \textbf{Selected historical estimates of $T|+\rangle$ cost trade-offs}, without logical distillation.
    }
    \label{fig:historical-comparison}
\end{figure}
\clearpage

\begin{figure}
    \centering
    \resizebox{\linewidth}{!}{
        \includegraphics{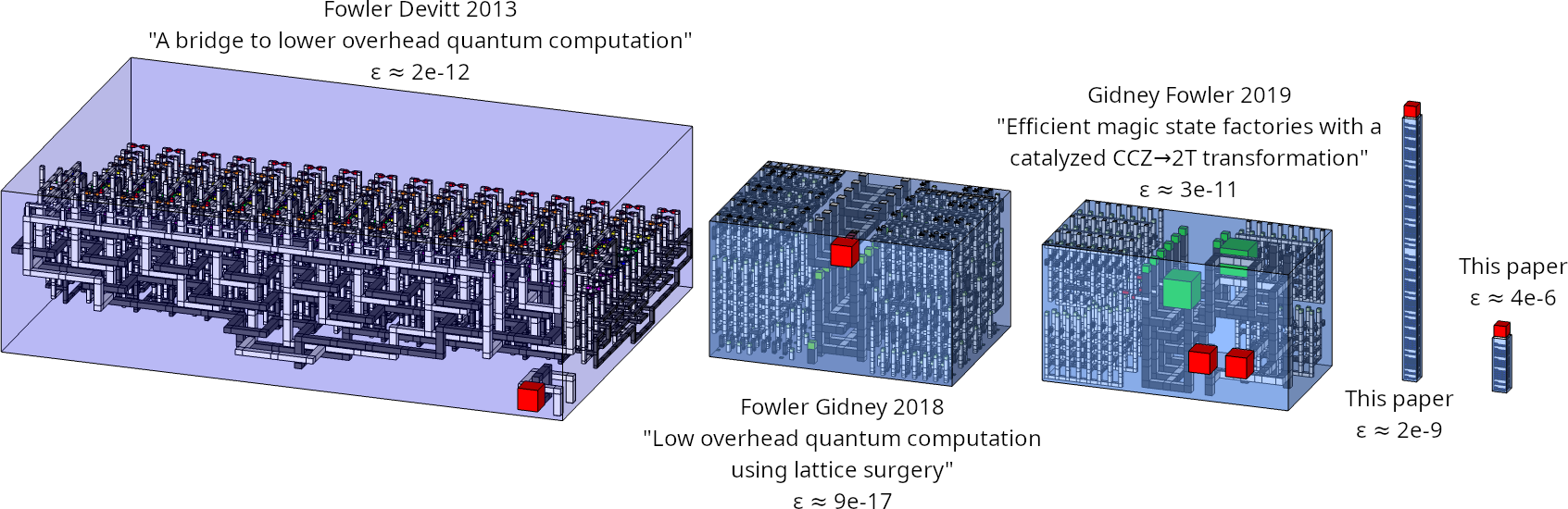}
    }
    \caption{
        \textbf{To-scale spacetime defect diagrams of historical constructions} for producing $|T\rangle$ states, showing improvement over time.
        (Limited to papers that included 3d models.)
        Assumes $10^{-3}$ uniform depolarizing circuit noise.
        $\epsilon$ annotations indicate logical error rates.
        Red boxes indicate output locations.
        Left: the braided factory from \cite{fowler2012bridge}.
        Middle left: the lattice surgery factory from \cite{fowler2018latticesurgery}.
        Middle right: the double-output catalyzed factory from \cite{gidney2019catalyzeddistillation}.
        Right: our construction retrying-until-success to cultivate a magic state, with a target fault distance of 5 and a target fault distance of 3.
    }
    \label{fig:historical-comparison-3d}
\end{figure}

\section{Introduction}
\label{sec:introduction}

The surface code is a quantum error correcting code with a high threshold and simple planar connectivity requirements~\cite{fowler2012surfacecodereview}.
These properties have made it a leading contender in the design of fault tolerant quantum computers (though recently the competition has been growing~\cite{hastings2021dynamically,bravyi2024ldpcibm,gidney2023colorcode}).
One downside of the surface code is the high cost of performing non-Clifford gates, like Ts and Toffolis.
For two decades, the cheapest known way to perform these gates has been by magic state distillation and gate teleportation~\cite{bravyi2005distillation,gidney2019catalyzeddistillation,litinski2019notascostly,hirano2024zeroleveldistill}.

The main inefficiency of magic state distillation is the fact that it's performed with logical operations~\cite{fowler2012surfacecodereview,gidney2019catalyzeddistillation,litinski2019notascostly}.
The large size of these operations, relative to physical operations, has resulted in magic state distillation being a leading contributor to the expected cost of fault tolerant quantum computations.
Many papers have attempted to optimize the cost of magic state distillation or to replace it with other techniques~\cite{fowler2012surfacecodereview,fowler2012bridge,li2015,fowler2018latticesurgery,gidney2019autoccz,litinski2019notascostly,singh2022,gidney2023hook,hirano2024zeroleveldistill,itogawa2024zeroleveldistilldistill,lee2024colordistillation,brown2019transversalccz,beverland2021codeswitch,vasmer2019threedsurfacecode,bombin2018threedcolorcode,chamberland2020colorinjection,heussen2024}.
Despite these efforts, the cost has remained substantial.

The enduring cost of preparing magic states has shaped proposals for quantum computer architectures.
For example, \cite{litinski2018} is a commonly cited architecture that pays large routing overheads to ensure magic state factories can run continuously.
As another example, high-depth ripple carry adders were expected to run faster than low-depth carry-lookahead adders because ripple carry adders use fewer T gates~\cite{gidney2017factoring,gidney2020blockadder}.
More generally, the cost of quantum algorithms has often been approximated and optimized by focusing mainly on gates that require magic states~\cite{lee2021hypercontraction,campbell2017shortgate,maslov2016,heyfron2018,kissinger2020}.

In \cite{chamberland2020colorinjection}, Chamberland and Noh proposed a way to improve magic states using physical operations within a code, instead of logical operations between codes.
The color code has transversal Clifford gates, including the $H_{XY} = (X + Y) / \sqrt{2}$ gate which has $|T\rangle = T|+\rangle = |0\rangle + \sqrt{i}|1\rangle$ as its +1 eigenstate and $Z|T\rangle$ as its -1 eigenstate.
\cite{chamberland2020colorinjection} noted that it was possible to measure in the eigenbasis of $H_{XY}$ by transversally controlling $H_{XY}$ gates with a GHZ state $|00..0\rangle + |11..1\rangle$.
Phase kickback from these controlled operations would transform the GHZ state into the state $|00..0\rangle + s|11..1\rangle$, where $s = \pm 1$ is the projective measurement outcome in the eigenbasis of $H_{XY}$ and can be extracted using local measurements and classical communication.
Therefore, after encoding a $|T\rangle$ state into a color code in some noisy fashion, the state can be cross-checked by performing this GHZ-controlled transversal $H_{XY}$ gate.
These cross-checks increase the fault distance of the state.
The difficulty of preparing large GHZ states prevents this from being viable at arbitrarily large code distances, but it works well at small code distances.

\cite{chamberland2020colorinjection} assumed a complex planar connectivity, and required an ambitious physical noise strength of between $10^{-4}$ and $5 \cdot 10^{-4}$.
Itogawa et al~\cite{itogawa2024zeroleveldistilldistill} improved on this by showing that the idea still works with a simple square grid connectivity and $10^{-3}$ uniform depolarizing circuit noise.
These improvements are what brought the construction to our attention.
In this paper we further refine the construction, enormously improving its performance.
We make five main improvements.

First, we don't attempt to correct \emph{any} errors when at small code distances.
Instead, we postselect using error detection.
Carefully managing the amount of error detection, and the transition into error correction, is key to achieving good efficiency.
Detecting too much ruins the retry rate.
Detecting too little ruins the error rate.
We think prior work has been detecting too little.

Second, we don't keep the size of the code fixed.
We iteratively improve the fault distance of the state while gradually increasing the size of the code hosting it.
This incremental growth in the reliability of the state and the size of the code is why we call our construction ``cultivation''.
Incremental growth reduces the amount of postselection required to reach a target fault distance, making it tractable to target a fault distance of 5 despite assuming uniform depolarizing circuit noise with a noise strength of $10^{-3}$.

Third, we reduce the cost of checking the logical state.
Conceptually we're still performing a transversal $H_{XY}$ controlled by a GHZ state, but the form of the circuit is streamlined.
In particular, we perform checks in pairs with one check being the time reverse of the other.
This allows qubits to be more freely rearranged during a check, because its time-reversed partner will undo these rearrangements.
Also, the time-reversed check restores helper qubits to their original states, creating many accompanying flag checks.

Fourth, we measure color code stabilizers using the superdense cycle from \cite{gidney2023colorcode}.
This measures the stabilizers using fewer circuit layers than other known circuits, under square grid connectivity, while preserving code distance by having the different stabilizer measurements flag each other.
The superdense cycle is difficult to decode, but this downside isn't an issue in this paper because we can simply postselect away any complicated cases.

Fifth, we introduce a technique (``grafting'') for rapidly growing the size of the code hosting the prepared state.
We include this step in our simulations and in our estimates.
Prior work has tended to ignore the need to grow the prepared state to large distances.
However, we find that this step is surprisingly difficult and costly.
Accounting for it is crucial to getting accurate numbers.

Magic state cultivation is a construction whose goals are practical, not theoretical.
Due to exponential growth in postselection costs versus target fault distance, cultivation isn't relevant to performance in the asymptotic limit.
Regardless, cultivation is an efficient way to produce T states with error rates relevant to practical quantum computations.
In particular, our simulations will show that cultivation makes it substantially cheaper to reach intermediate logical error rates like $10^{-6}$ or $10^{-9}$ (see \fig{historical_progression} and \fig{historical-comparison}).
These error rates are nearly sufficient to run well known quantum algorithms like \cite{gidney2017factoring,haner2020ellipticcurve,babbush2018}.
Our simulations will further show that cultivation's costs and error rates improve dramatically as physical noise strength is reduced.
Even modest ongoing improvements to physical qubits could easily result in cultivation meeting the practical requirements of quantum computers.
So we suspect cultivation will eventually obsolete magic state distillation, in practice if not in theory.

The paper is organized as follows.
In \sec{construction} we describe our construction along with some alternatives that we considered.
In \sec{results} we describe our simulations of the construction, quantifying its performance.
We conclude in \sec{conclusion}.
\app{noise-model} describes our noise model.
\app{chunk} summarizes the process by which we created the circuits we simulated.
\app{additional-figures} includes bonus figures that didn't fit well in the main text.

\section{Construction}
\label{sec:construction}

The code we wrote to produce our construction is \href{https://doi.org/10.5281/zenodo.13777072}{available on Zenodo}~\cite{gidneyy2024cultivationdata}.

\subsection{Stages}

We split our construction into three stages: the injection stage, the cultivation stage, and the escape stage.

\begin{itemize}
    \item
    Injection.
    The injection stage creates the initial encoded $|T\rangle$ state.
    The state should end up stored in a distance 3 code (or better), even though the state itself will have a fault distance of 1.
    This stage is the easiest stage; there are many ways to do it that perform similarly.
    \item
    Cultivation.
    This stage gradually increases the fault distance of the stored $|T\rangle$ state by performing cross-checks on it.
    The main challenge of this stage is that retry costs grow exponentially with the number of gates, harshly punishing any inefficiencies.
    \item
    Escape!
    Cultivation ends with a $|T\rangle$ state too good for the code it's trapped inside.
    Staying for even one unpostselected round would substantially damage the state, because the logical error rate without postselection is much higher.
    The goal of the escape stage is to fix this, by rapidly boosting the size of the code hosting the state.
    The challenge of the escape stage is that its circuit is far too large to postselect every possible detection event.
    Fine grained measures of whether or not to discard a shot are needed.
\end{itemize}

\begin{figure}
    \centering
    \resizebox{\linewidth}{!}{
    \includegraphics{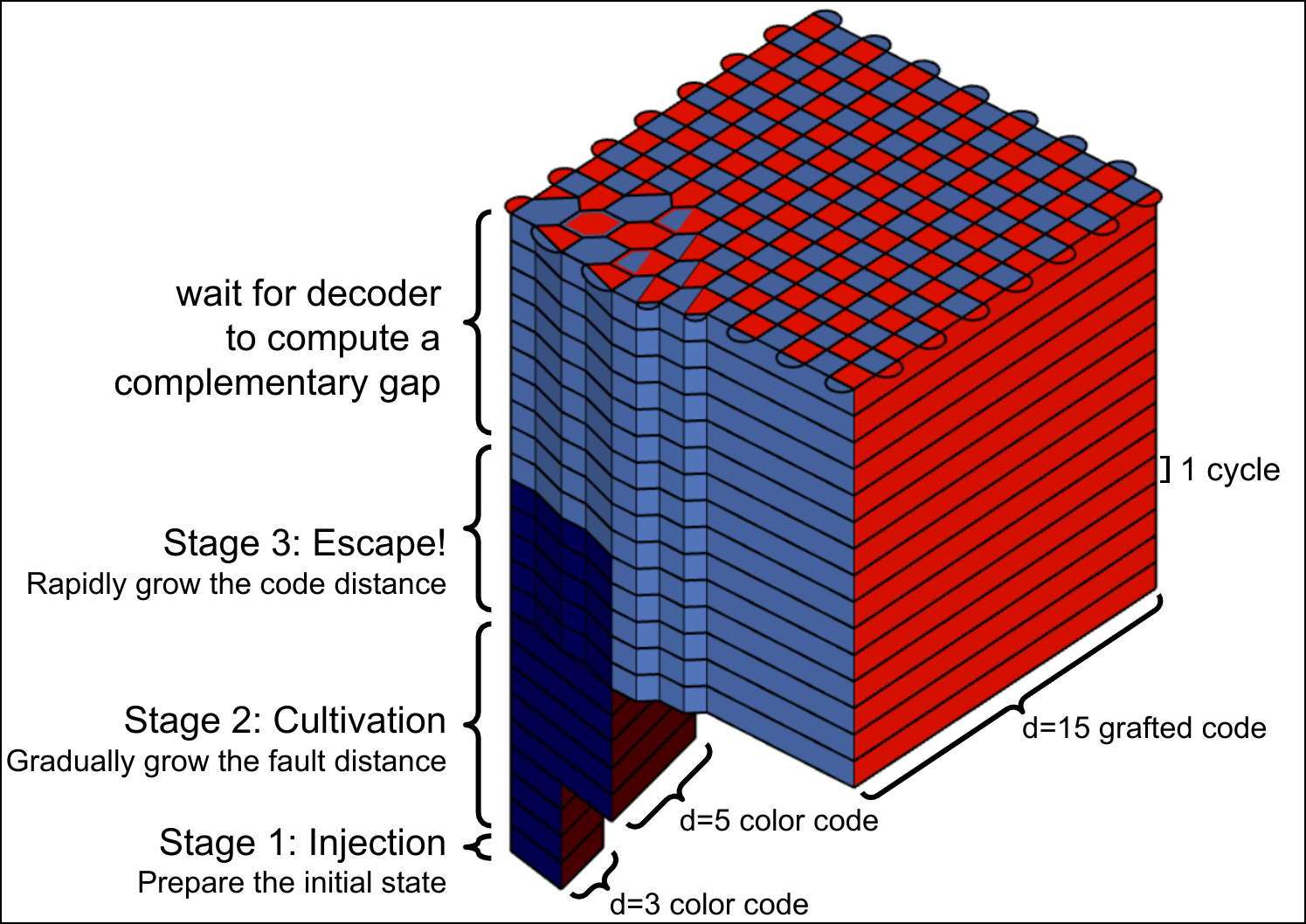}
    }
    \caption{
        \textbf{Overview of a magic state cultivation.}
        Time moves upward.
        Darker colors are color code boundaries.
        Lighter colors are X/Z boundaries, and X/Z stabilizers.
        In practice cultivation must be attempted several times before it succeeds, and multiple attempts at stage 1 and stage 2 would be run in parallel.
    }
    \label{fig:construction-overview}
\end{figure}

These stages are visually summarized in \fig{construction-overview}.

\subsection{Injection Stage}

\begin{figure}
    \begin{adjustwidth}{-2.5cm}{-2.5cm}
        \centering
        \resizebox{\linewidth}{!}{
            \includegraphics{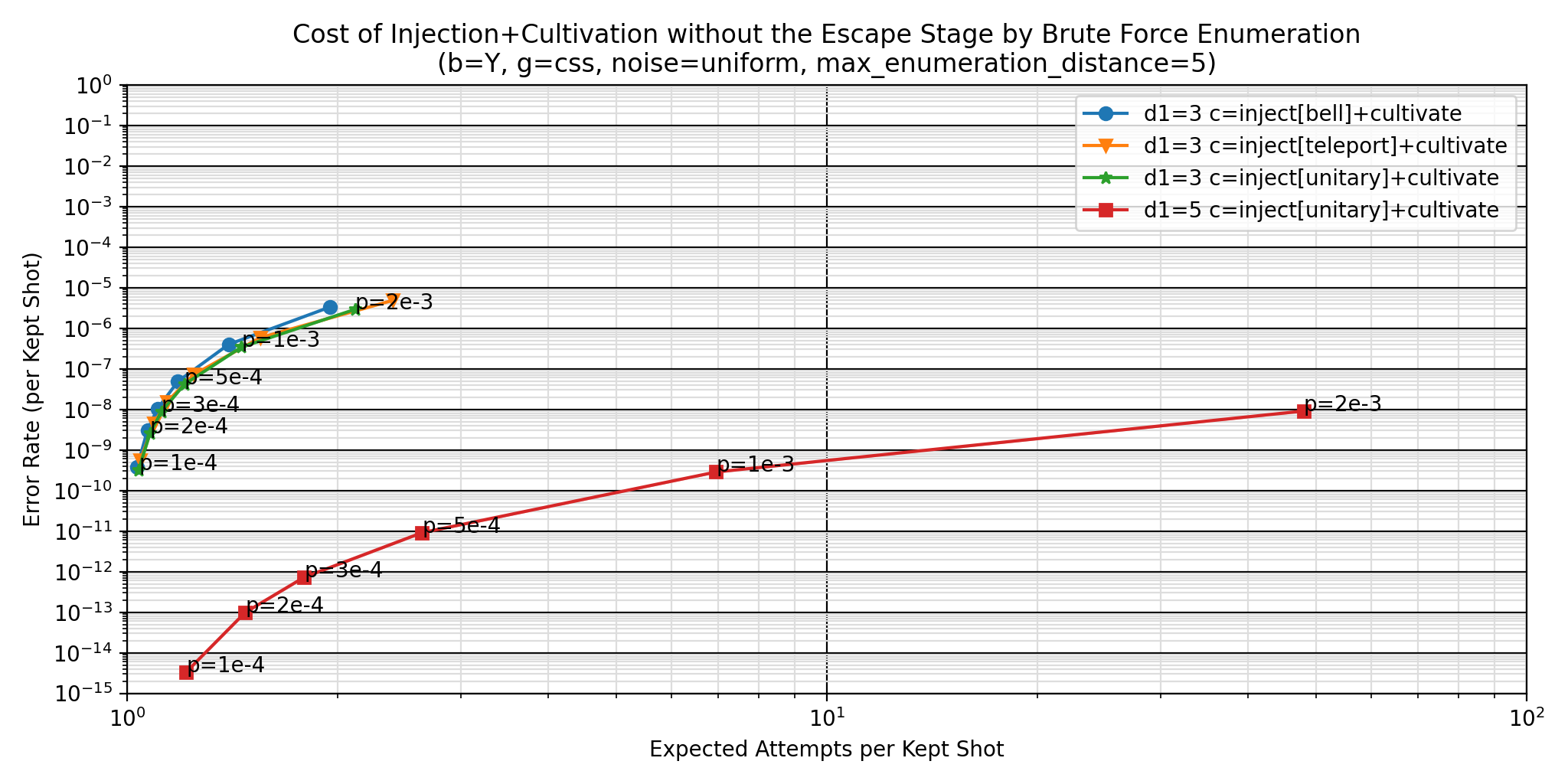}
        }
    \end{adjustwidth}
    \caption{
        \textbf{Cost of cultivation without an escape stage}.
        Shows the similar performance of different injection stages.
        Estimated by enumerating all possible logical errors up to distance 5.
    }
    \label{fig:inject-only-enumerated}
\end{figure}

\begin{figure}[p]
    \centering
    \begin{adjustwidth}{-1.0cm}{-1.0cm}
        \centering
        \resizebox{\linewidth}{!}{
            \includegraphics{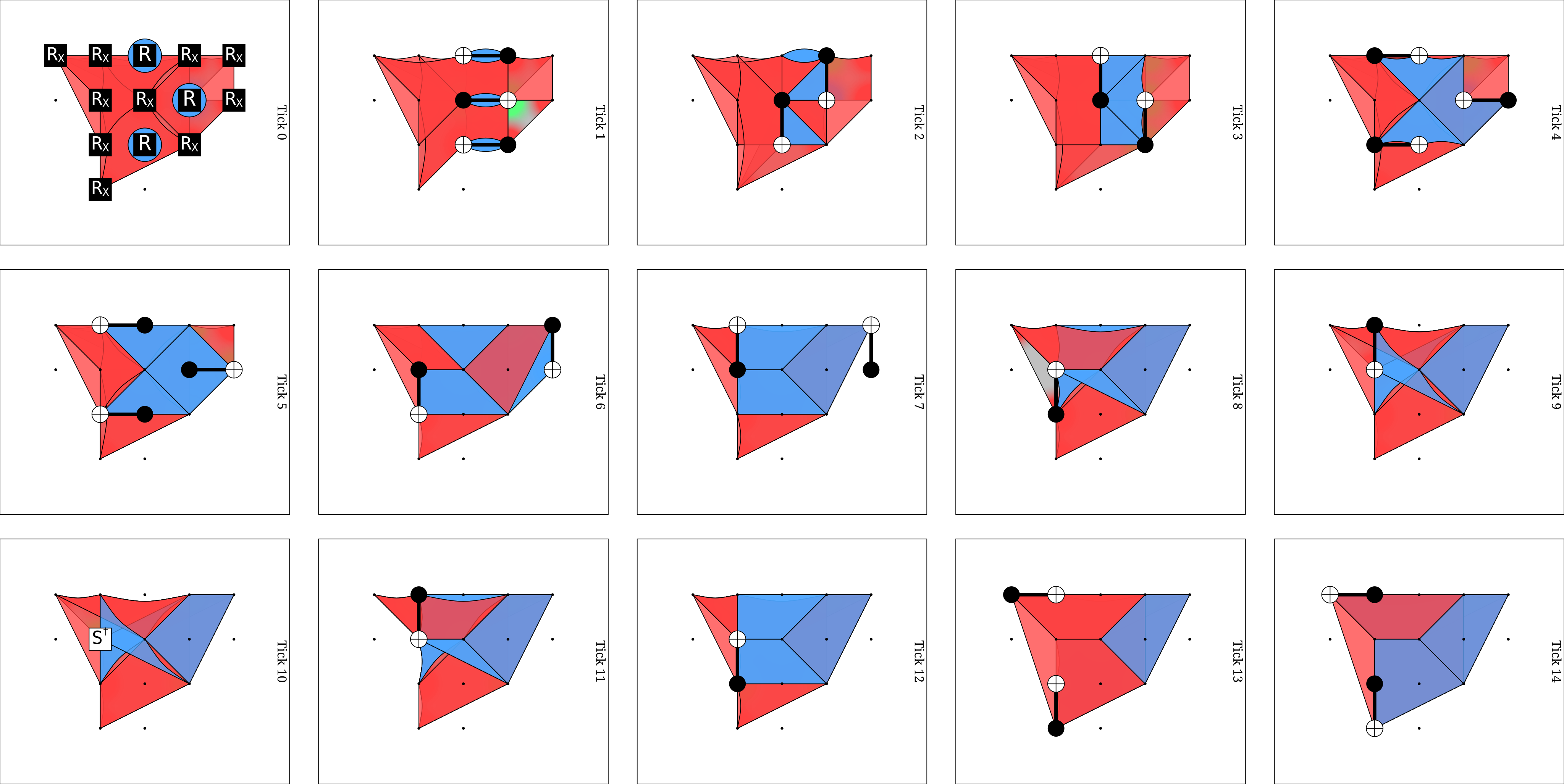}
        }
    \end{adjustwidth}
    \caption{
        \textbf{Detector slice diagram of unitary injection}.
        Replace the S gate with a T gate to prepare $T|+\rangle$ instead of $S|+\rangle$.
        Ends with the color code patch slightly deformed, requiring the following superdense cycle to be adjusted.
        \href{
            https://algassert.com/crumble\#circuit=Q(0,0)0;Q(1,0)1;Q(1,1)2;Q(1,2)3;Q(1,3)4;Q(2,0)5;Q(2,1)6;Q(2,2)7;Q(2,3)8;Q(3,0)9;Q(3,1)10;Q(3,2)11;Q(4,0)12;Q(4,1)13;POLYGON(0,0,1,0.25)9_13_11_6;POLYGON(0,1,0,0.25)2_6_11_3;POLYGON(1,0,0,0.25)1_9_6_2;TICK;R_10_7_5;RX_6_9_11_2_1_3_13_0_4_12;TICK;CX_6_10_9_5_11_7;TICK;CX_6_7_9_10;TICK;CX_6_5_11_10;TICK;CX_1_5_3_7_13_10;TICK;CX_5_1_7_3_10_13;TICK;CX_2_3_12_13;TICK;CX_2_1_13_12;TICK;CX_3_2;TICK;CX_1_2;TICK;S_DAG_2;TICK;CX_1_2;TICK;CX_3_2;TICK;CX_0_1_4_3;TICK;CX_1_0_3_4;TICK;POLYGON(0,0,1,0.25)12_9_6_11;POLYGON(0,1,0,0.25)8_11_6_2;POLYGON(0,1,0,0.25)2_6_11_4;POLYGON(1,0,0,0.25)2_6_9_0;TICK;R_5_13_7_8;RX_1_10_3;TICK;CX_1_5_3_7_4_8_10_13;TICK;CX_1_0_5_9_7_11_8_4_10_6;TICK;CX_1_2_5_6_7_8_10_11;TICK;CX_3_2_7_6_10_9_13_12;TICK;CX_2_3_6_7_9_10_12_13;TICK;CX_0_1_6_10_9_5_11_7;TICK;CX_2_1_6_5_8_7_11_10;TICK;CX_1_5_3_7_10_13;TICK;M_5_13_7_4;MX_1_10_3;DT(2,0,0)rec[-7];DT(4,1,0)rec[-6];DT(2,2,0)rec[-5];DT(1,3,0)rec[-4];DT(1,0,0)rec[-3];DT(3,1,0)rec[-2];DT(1,2,0)rec[-1];TICK;POLYGON(0,0,1,0.25)12_9_6_11;POLYGON(0,1,0,0.25)8_11_6_2;POLYGON(1,0,0,0.25)2_6_9_0;TICK;MPP_X0*X2*X6*X9;DT(1,2,1)rec[-1]_rec[-2]_rec[-4];TICK;MPP_Z0*Z2*Z6*Z9;DT(2,0,2)rec[-1]_rec[-9];TICK;MPP_X2*X6*X8*X11;DT(1,0,3)rec[-1]_rec[-6];TICK;MPP_Z2*Z6*Z8*Z11;DT(2,2,4)rec[-1]_rec[-9];TICK;MPP_X6*X9*X11*X12;DT(2,1,5)rec[-1];TICK;MPP_Z6*Z9*Z11*Z12;DT(4,1,6)rec[-1]_rec[-12];TICK;MPP_Y0*Y2*Y6*Y8*Y9*Y11*Y12;OI(0)rec[-1]_rec[-8]_rec[-9]
        }{Click here to open this circuit (plus the modified superdense cycle) in Crumble.}
    }
    \label{fig:d3-init-unitary}
\end{figure}

In \cite{itogawa2024zeroleveldistilldistill}, the injection stage is performed by a unitary encoding circuit (\href{
    https://arxiv.org/pdf/2403.03991v1#page=2
}{the left half of figure 2 in their paper}).
In \cite{chamberland2020colorinjection}, the injection stage is performed by lattice surgery between a data qubit storing $|T\rangle$ and a degenerate color code (\href{
    https://www.nature.com/articles/s41534-020-00319-5/figures/7
}{figure 7a of their paper}).
In this work we considered three different designs for the injection stage: ``teleport injection'' (see \fig{d3-init-teleport}), ``Bell injection'' (see \fig{d3-init-bell}), and ``unitary injection'' (see \fig{d3-init-unitary}).
We used brute force enumeration of error mechanisms to compare these methods (see \fig{inject-only-enumerated}).
Of the three approaches, we found that unitary injection had the best performance, though all three were surprisingly close despite their conceptual and physical differences.

Our unitary injection circuit performs a series of Clifford gates that deterministically prepare the stabilizers and observables of a distance 3 color code storing a $|T\rangle$ state.
Note that the circuit \emph{isn't} an encoding circuit; it doesn't first prepare a physical $|T\rangle$ state and then expand it into a color code.
Instead, similar to \cite{gidney2023hook}, we arrange the circuit so that the T gate appears in the middle.
This allows the T gate to occur on a qubit protected by a Z stabilizer, catching X or Y errors on the T gate.

\subsection{Cultivation Stage}

In \cite{chamberland2020colorinjection}, cultivation is done by alternating between a flagged GHZ-controlled transversal $H$ check and a flagged stabilizer check.
At code distance 3, ignoring single qubit Clifford gate layers, their T state check spans 10 layers and 16 qubits while their stabilizer check spans 9 layers and 49 qubits (see figures
\href{
    https://www.nature.com/articles/s41534-020-00319-5/figures/1
}{1},
\href{
    https://www.nature.com/articles/s41534-020-00319-5/figures/2
}{2},
and
\href{
    https://www.nature.com/articles/s41534-020-00319-5/figures/3
}{3}
of \cite{chamberland2020colorinjection}).
To reach fault distance $d$, they initialize a distance $2d+1$ color code and run $d$ alternations.

In \cite{itogawa2024zeroleveldistilldistill}, cultivation is restricted to growing the fault distance from 1 to 2, which is done by the flagged creation of a GHZ state followed by using the GHZ state to control a transversal $H_{XY}$ gate.
This process spans 15 layers and 25 qubits.

Our cultivation stage works on a check-grow-stabilize cycle.
The ``check'' step increases the fault distance of the T state by checking on the logical value of the code.
The ``grow'' step enables further increases in the fault distance by increasing the size of the color code.
The ``stabilize'' step measures the stabilizers of the code a few times, to ensure new stabilizers can detect errors and to prevent correlated errors between T state checks.

\begin{figure}[p]
    \centering
    \begin{adjustwidth}{-1.0cm}{-1.0cm}
        \centering
        \resizebox{\linewidth}{!}{
            \includegraphics{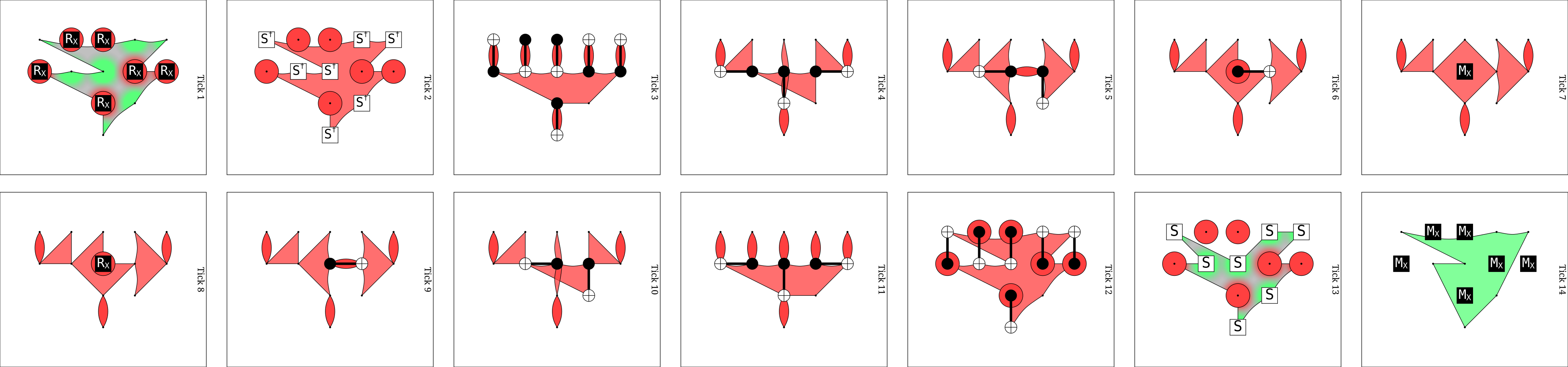}
        }
    \end{adjustwidth}
    \caption{
        \textbf{Detector slice diagram of double-checking a $d=3$ color code's logical value}.
        Replace $S$ gates with $T$ gates (and $S^\dagger$ with $T^\dagger$) to double-check $T|+\rangle$ instead of $S|+\rangle$.
        Large green shapes are slices of the logical value being measured and re-prepared.
        The stabilizers of the color code (not shown) are preserved by this circuit, up to deferrable Pauli Z feedback.
        \href{
            https://algassert.com/crumble\#circuit=Q(0,0)0;Q(0,1)1;Q(1,0)2;Q(1,1)3;Q(2,0)4;Q(2,1)5;Q(2,2)6;Q(2,3)7;Q(3,0)8;Q(3,1)9;Q(3,2)10;Q(4,0)11;Q(4,1)12;POLYGON(0,0,1,0.25)8_11_10_5;POLYGON(0,1,0,0.25)7_10_5_3;POLYGON(1,0,0,0.25)3_5_8_0;TICK;MPP_Y0*Y3*Y5*Y7*Y8*Y10*Y11;TICK;MPP_X0*X3*X5*X8;TICK;MPP_Z0*Z3*Z5*Z8;TICK;MPP_X3*X5*X7*X10;TICK;MPP_Z3*Z5*Z7*Z10;TICK;MPP_X5*X8*X10*X11;TICK;MPP_Z5*Z8*Z10*Z11;TICK;RX_12_9_4_2_6_1;TICK;S_DAG_0_3_5_7_8_10_11;TICK;CX_1_0_9_8_6_7_4_5_2_3_12_11;TICK;CX_3_1_5_6_9_12;TICK;CX_5_3_9_10;TICK;CX_5_9;TICK;MX_5;DT(0,0,0)rec[-1]_rec[-8];TICK;RX_5;TICK;CX_5_9;TICK;CX_5_3_9_10;TICK;CX_3_1_5_6_9_12;TICK;CX_1_0_9_8_6_7_12_11_4_5_2_3;TICK;S_5_10_11_8_7_3_0;TICK;MX_12_9_4_2_6_1;DT(4,1,1)rec[-6];DT(3,1,1)rec[-5];DT(2,1,1)rec[-4]_rec[-7];DT(1,0,1)rec[-3];DT(2,2,1)rec[-2];DT(0,1,1)rec[-1];TICK;MPP_Z5*Z8*Z10*Z11;DT(2,1,2)rec[-1]_rec[-9];TICK;MPP_X5*X8*X10*X11;DT(2,1,3)rec[-1]_rec[-9]_rec[-11];TICK;MPP_Z3*Z5*Z7*Z10;DT(1,1,4)rec[-1]_rec[-13];TICK;MPP_X3*X5*X7*X10;DT(1,1,5)rec[-1]_rec[-11]_rec[-15];TICK;MPP_Z0*Z3*Z5*Z8;DT(0,0,6)rec[-1]_rec[-17];TICK;MPP_X0*X3*X5*X8;DT(0,0,7)rec[-1]_rec[-13]_rec[-19];TICK;MPP_Y0*Y3*Y5*Y7*Y8*Y10*Y11;OI(0)rec[-1]_rec[-8]_rec[-9]_rec[-12]_rec[-13]
        }{Click here to open this circuit in crumble.}
    }
    \label{fig:cat-check-d3}
\end{figure}

\begin{figure}[p]
    \centering
    \begin{adjustwidth}{-1.0cm}{-1.0cm}
        \centering
        \resizebox{\linewidth}{!}{
            \includegraphics{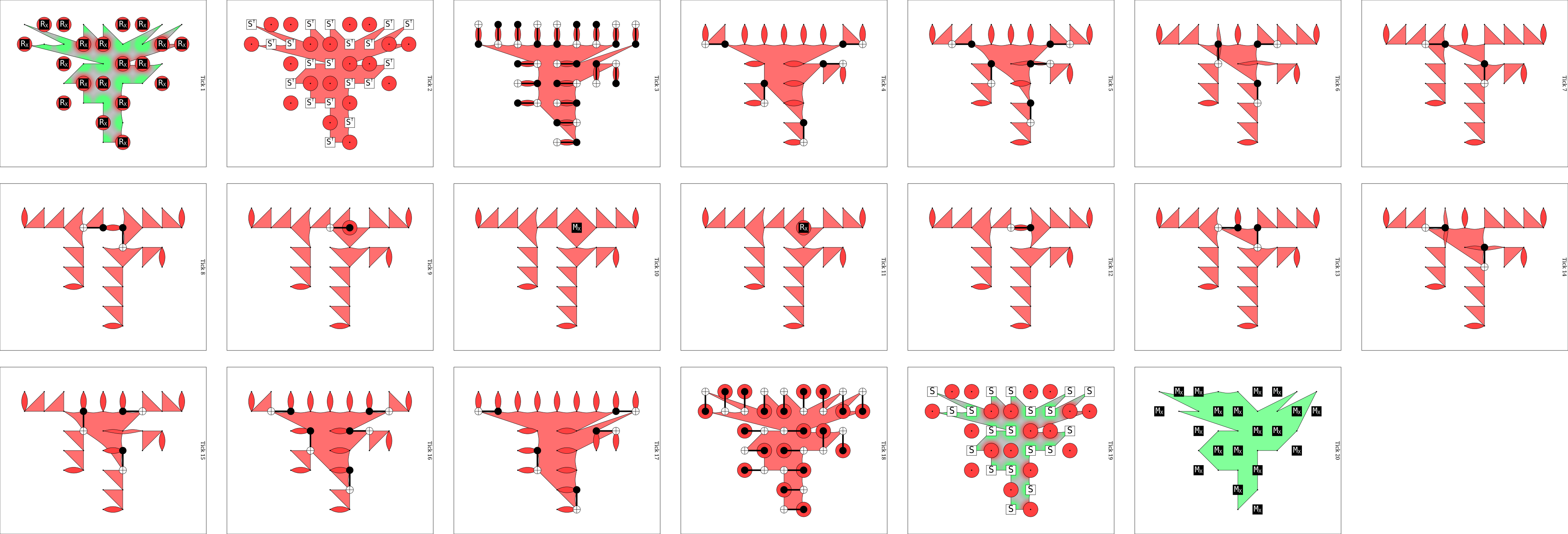}
        }
    \end{adjustwidth}
    \caption{
        \textbf{Detector slice diagram of double-checking a $d=5$ color code's logical value}.
        Distance 5 variant of \fig{cat-check-d3}.
        \href{
            https://algassert.com/crumble\#circuit=Q(0,0)0;Q(0,1)1;Q(1,0)2;Q(1,1)3;Q(2,0)4;Q(2,1)5;Q(2,2)6;Q(2,3)7;Q(2,4)8;Q(3,0)9;Q(3,1)10;Q(3,2)11;Q(3,3)12;Q(3,4)13;Q(4,0)14;Q(4,1)15;Q(4,2)16;Q(4,3)17;Q(4,4)18;Q(4,5)19;Q(4,6)20;Q(5,0)21;Q(5,1)22;Q(5,2)23;Q(5,3)24;Q(5,4)25;Q(5,5)26;Q(5,6)27;Q(6,0)28;Q(6,1)29;Q(6,2)30;Q(6,3)31;Q(7,0)32;Q(7,1)33;Q(7,2)34;Q(7,3)35;Q(8,0)36;Q(8,1)37;POLYGON(0,0,1,0.25)9_14_22_16_11_5;POLYGON(0,0,1,0.25)29_32_36_34;POLYGON(0,0,1,0.25)24_31_26_18;POLYGON(0,1,0,0.25)7_11_5_3;POLYGON(0,1,0,0.25)22_29_34_31_24_16;POLYGON(0,1,0,0.25)13_18_26_20;POLYGON(1,0,0,0.25)3_5_9_0;POLYGON(1,0,0,0.25)14_32_29_22;POLYGON(1,0,0,0.25)11_16_24_18_13_7;TICK;MPP_Y0*Y3*Y5*Y7*Y9*Y11*Y13*Y14*Y16*Y18*Y20*Y22*Y24*Y26*Y29*Y31*Y32*Y34*Y36;TICK;MPP_X0*X3*X5*X9_X7*X11*X13*X16*X18*X24_X14*X22*X29*X32;TICK;MPP_Z0*Z3*Z5*Z9_Z7*Z11*Z13*Z16*Z18*Z24_Z14*Z22*Z29*Z32;TICK;MPP_X3*X5*X7*X11_X13*X18*X20*X26_X16*X22*X24*X29*X31*X34;TICK;MPP_Z3*Z5*Z7*Z11_Z13*Z18*Z20*Z26_Z16*Z22*Z24*Z29*Z31*Z34;TICK;MPP_X5*X9*X11*X14*X16*X22_X18*X24*X26*X31_X29*X32*X34*X36;TICK;MPP_Z5*Z9*Z11*Z14*Z16*Z22_Z18*Z24*Z26*Z31_Z29*Z32*Z34*Z36;TICK;RX_21_28_2_4_6_8_17_19_35_1_10_15_33_30_12_23_25_27_37;TICK;S_DAG_9_11_13_14_16_18_20_22_24_26_29_31_34_36_32_3_5_7_0;TICK;CX_1_0_2_3_4_5_10_9_15_14_21_22_28_29_33_32_30_31_35_34_23_16_17_24_25_18_19_26_27_20_6_11_12_7_8_13_37_36;TICK;CX_12_13_3_1_26_27_33_37_30_34;TICK;CX_25_26_11_12_5_3_29_33_23_30;TICK;CX_24_25_10_11_22_29;TICK;CX_10_5_23_24;TICK;CX_15_10_22_23;TICK;CX_22_15;TICK;MX_22;DT(0,0,0)rec[-1]_rec[-20];TICK;RX_22;TICK;CX_22_15;TICK;CX_15_10_22_23;TICK;CX_10_5_23_24;TICK;CX_24_25_10_11_22_29;TICK;CX_25_26_11_12_5_3_29_33_23_30;TICK;CX_12_13_3_1_26_27_33_37_30_34;TICK;CX_1_0_2_3_4_5_10_9_15_14_21_22_28_29_33_32_30_31_35_34_23_16_17_24_25_18_19_26_27_20_12_7_8_13_6_11_37_36;TICK;S_9_11_13_14_16_18_20_22_24_26_29_31_34_32_3_5_7_0_36;TICK;MX_17_19_6_8_2_4_21_28_35_1_10_15_33_30_12_23_25_27_37;DT(4,3,1)rec[-19];DT(4,5,1)rec[-18];DT(2,2,1)rec[-17];DT(2,4,1)rec[-16];DT(1,0,1)rec[-15];DT(2,0,1)rec[-14];DT(5,1,1)rec[-13]_rec[-20];DT(6,0,1)rec[-12];DT(7,3,1)rec[-11];DT(0,1,1)rec[-10];DT(3,1,1)rec[-9];DT(4,1,1)rec[-8];DT(7,1,1)rec[-7];DT(6,2,1)rec[-6];DT(3,3,1)rec[-5];DT(5,2,1)rec[-4];DT(5,4,1)rec[-3];DT(5,6,1)rec[-2];DT(8,1,1)rec[-1];TICK;MPP_X0*X3*X5*X9_X7*X11*X13*X16*X18*X24_X14*X22*X29*X32;DT(0,0,2)rec[-3]_rec[-41];DT(2,3,2)rec[-2]_rec[-40];DT(4,0,2)rec[-1]_rec[-23]_rec[-39];TICK;MPP_Z0*Z3*Z5*Z9_Z7*Z11*Z13*Z16*Z18*Z24_Z14*Z22*Z29*Z32;DT(0,0,3)rec[-3]_rec[-41];DT(2,3,3)rec[-2]_rec[-40];DT(4,0,3)rec[-1]_rec[-39];TICK;MPP_X3*X5*X7*X11_X13*X18*X20*X26_X16*X22*X24*X29*X31*X34;DT(1,1,4)rec[-3]_rec[-41];DT(3,4,4)rec[-2]_rec[-40];DT(4,2,4)rec[-1]_rec[-29]_rec[-39];TICK;MPP_Z3*Z5*Z7*Z11_Z13*Z18*Z20*Z26_Z16*Z22*Z24*Z29*Z31*Z34;DT(1,1,5)rec[-3]_rec[-41];DT(3,4,5)rec[-2]_rec[-40];DT(4,2,5)rec[-1]_rec[-39];TICK;MPP_X5*X9*X11*X14*X16*X22_X18*X24*X26*X31_X29*X32*X34*X36;DT(2,1,6)rec[-3]_rec[-35]_rec[-41];DT(4,4,6)rec[-2]_rec[-40];DT(6,1,6)rec[-1]_rec[-39];TICK;MPP_Z5*Z9*Z11*Z14*Z16*Z22_Z18*Z24*Z26*Z31_Z29*Z32*Z34*Z36;DT(2,1,7)rec[-3]_rec[-41];DT(4,4,7)rec[-2]_rec[-40];DT(6,1,7)rec[-1]_rec[-39];TICK;MPP_Y0*Y3*Y5*Y7*Y9*Y11*Y13*Y14*Y16*Y18*Y20*Y22*Y24*Y26*Y29*Y31*Y32*Y34*Y36;OI(0)rec[-1]_rec[-20]_rec[-21]_rec[-22]_rec[-23]_rec[-24]_rec[-25]_rec[-26]_rec[-27]_rec[-28]_rec[-29]
        }{Click here to open this circuit in crumble.}
    }
    \label{fig:cat-check-d5}
\end{figure}

Our check step is actually a ``double-check''; it performs two checks with the second check being the time reverse of the first.
Example double-check circuits are shown in \fig{cat-check-d3} and \fig{cat-check-d5}.
These circuits start by applying $T^\dagger$ gates to the data qubits, so that the goal becomes checking the $X$ parity of all data qubits (instead of the $H_{XY}$ parity).
Each data qubit is paired with an adjacent ancilla, which we call the data qubit's partner.
The partners are initialized into the $|+\rangle$ state and then each data qubit is targeted by a CNOT gate controlled by its partner qubit.
After these CNOTs are applied, each data/partner pair is in the +1 eigenspace of the swap operation, meaning we can ``swap'' a data qubit and its partner by doing nothing.
We ``swap'' certain pairs, so that the partner qubits now form a contiguous region within the qubit grid.
We measure the X parity of this contiguous region by choosing a spanning tree, and folding the X parity towards the root of the tree using CNOT gates along the tree.
The root is then measured in the X basis, revealing the parity of the first check.
At this point the data qubits are still entangled with many of the partner qubits and need to be restored to their original configuration.
We restore them by performing the same check again, but time-reversed.
We prepare the root of the spanning tree into a $|+\rangle$ state (this is the time-reverse of the X basis measurement of the root), we perform the CNOT gates that folded the parity into the root in the reverse order to unfold the parity back to where it started, we perform the data/partner CNOTs to unentangle the partners from the data qubits, we measure the partners in the X basis (this is the time-reverse of the $|+\rangle$ preparation of the partners), and finally we restore the orientations of the data qubits by applying $T$ gates.
Note that, because the time-reversed check unscrambles everything done by the time-forward check, and because the partner qubits started in the $|+\rangle$ state, every partner qubit should end in the $|+\rangle$ state.
This means the X basis measurement of every partner should be +1.
The partners aren't just workspace; they double as flag checks.
The time-reversed check of the observable is embedded amongst these flag checks.
This approach to checking the T state spans 15 qubits and 6 layers per check at $d=3$, making it faster and more compact than both \cite{chamberland2020colorinjection} and \cite{hirano2024zeroleveldistill}.

\begin{figure}
    \centering
    \resizebox{\linewidth}{!}{
        \includegraphics{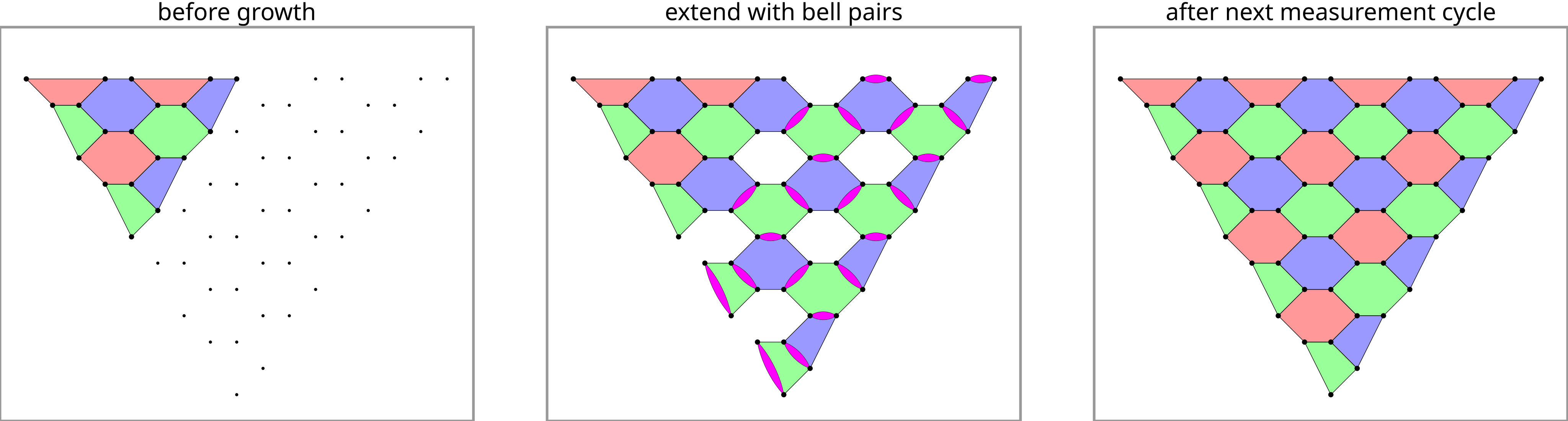}
    }
    \caption{
        \textbf{How to grow a color code} by initializing added data qubits into Bell pairs (shown as wedges).
        Because the blue-green boundary is the one being extended, the Bell pairs must be positioned to imply the values of the blue and green stabilizers.
        The missing red stabilizers are initialized by the next measurement cycle.
    }
    \label{fig:color-code-growth}
\end{figure}

To grow a color code, we prepare the data qubits covered by the new larger color code, but not the old smaller color code, into Bell pairs (see \fig{color-code-growth}).
The Bell pairs are chosen so that they determine the stabilizers whose colors match the boundary that is being moved~\cite{jones2016colorcode}.
Note that, because the preparation of the Bell pairs happens only on new data qubits, the growth step can be overlapped with the preceding check step.

To stabilize the grown color code, we use the superdense color code cycle~\cite{gidney2023colorcode}.
A key detail is how many times to repeat the cycle before moving on to the next T state checks.
We initially used automated fault distance searches to guide this choice.
In a phenomenological noise model, it would be sufficient to measure the stabilizers twice.
But under circuit noise the superdense cycle doesn't reach the desired fault distance when repeated only twice; it contains spacetimelike hook errors that increase the required number of repetitions to four.
However, when estimating logical error rates, we found that the fault distance wasn't predictive of the final performance.
Adding a third superdense cycle noticeably improves the logical error rate, but adding a fourth cycle has negligible effects.
The benefit of the fourth cycle isn't worth the cost of postselecting another cycle, and so we decided that a stabilize step would use three superdense cycles despite this technically not hitting the fault distance target.

\subsection{Escape Stage}

When we initially began working on this paper, we assumed that the escape stage would be trivial compared to the cultivation stage.
This expectation seems to be echoed in previous work.
In \cite{chamberland2020colorinjection} there's recognition of the fact that the patch grows, but the cost of doing so isn't included in estimates.
In \cite{itogawa2024zeroleveldistilldistill}, the need to grow the patch beyond distance 3 isn't considered.
In this paper, we simulated the escape stage as part of producing our estimates and found that it was by far the hardest to get working well.

Initially, we tried simply growing the color code into a larger color code as in \fig{color-code-growth}.
Unfortunately, our current color code decoder (Chromobius~\cite{gidney2023colorcode}) isn't good enough for this strategy to work.
Chromobius' predictions aren't accurate enough, and it doesn't provide a way to report how confident it is in its predictions (which is crucial for postselecting efficiently during the escape).
We tried a few hacks for fixing this, but none of them worked well enough to report.
We believe that a pure color code growth strategy should be viable, but it requires better color code decoders than we have access to today.

With the simple route blocked, we were forced down a more complicated path: escaping into a surface code.
In prior work, turning color codes into surface codes has been done by local unitary transformations~\cite{kubicacolor2surface,itogawa2024zeroleveldistilldistill} and by lattice surgery~\cite{poulsennautrup2017,shutty2022mergedcolorcode,itogawa2024zeroleveldistilldistill}.
Unfortunately, both of these solutions weren't efficient enough for our purposes.
The unitary transformations were unworkable because they failed to preserve the fault distance of the system.
The lattice surgery constructions preserved the fault distance, but prevented the code distance of the system from increasing until \emph{after} the lattice surgery finished (instead of immediately).
These aren't small problems; the error rates of these methods are orders of magnitude higher than the error rate of the states we're trying to preserve.
To achieve sufficient fidelity we had to invent a new solution, which we call ``grafting''.

Our grafting construction grows a large surface code out of a small color code.
It creates a grafted code that is a combination of a surface code and a color code.
Topologically, the grafted code is a ``partially folded surface code'', with the color code region corresponding to the folded region.
This grafted code has a code distance set by the large surface code, not by the small color code.
We stabilize the grafted code by measuring its stabilizers several times, until it becomes possible to discard the color code region and transition into a fully matchable code.
Once we've escaped to a large matchable code the state is safe, because we have excellent decoders for this situation~\cite{higgott2021pymatching,paler2023correlated,gidney2024yoked}.

\begin{figure}
    \centering
    \resizebox{\linewidth}{!}{
        \includegraphics{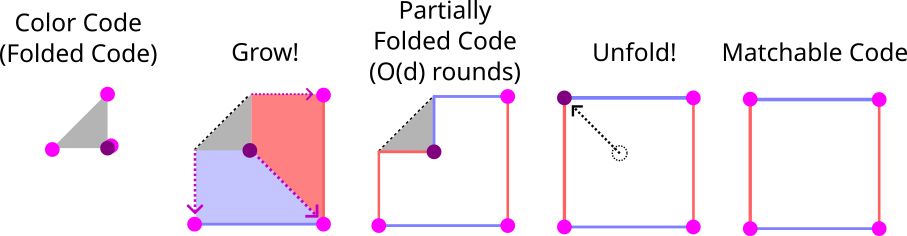}
    }
    \caption{
        \textbf{Topological summary of transforming a small color code into a large matchable code}.
        Left: the initial color code is topologically equivalent to a folded surface code.
        The folded surface code has four twists spread over its three corners.
        The two twists sharing a corner are on ''opposite sides of the fold''.
        Left middle: three of the twists are moved outward, creating a partially folded surface code.
        Middle: the remaining twist stays still for $O(d)$ rounds where $d$ is the distance of the initial color code.
        This avoids twists passing near the worldlines of other twists.
        Right middle: eventually, the remaining twist is moved to the corner.
        Right: the system has fully unfolded into a matchable code.
    }
    \label{fig:topological-growth}
\end{figure}

Grafting is topologically distinct from lattice surgery between a color code and a surface code~\cite{poulsennautrup2017,shutty2022mergedcolorcode}.
They look similar, but can be easily distinguished if you know what to look for.
First, lattice surgery attaches \emph{one} boundary of the color code to the surface code (visually: the color code points \emph{away} from the surface code, as in \href{
    https://journals.aps.org/prapplied/pdf/10.1103/PhysRevApplied.18.014072#page=17
}{figure 6 of} \cite{shutty2022mergedcolorcode}).
Grafting the codes results in \emph{two} boundaries of the color code being attached to the surface code (visually: the color code points \emph{into} the surface code).
Second, in lattice surgery the surface code boundaries that are attached to the corners of the color code have the same type.
For example, when doing a logical $M_{XX}$ measurement, both attached boundaries would be Z type boundaries.
Grafting results in these boundaries having opposite type.
Third, and most importantly, the code distance of intermediate configurations created by lattice surgery cannot be larger than the distance of either code.
Grafting a surface code with a color code produces a stabilizer code whose distance is larger than the distance of the color code.

\begin{figure}
    \begin{adjustwidth}{-2.5cm}{-2.5cm}
        \centering
        \resizebox{\linewidth}{!}{
            \includegraphics{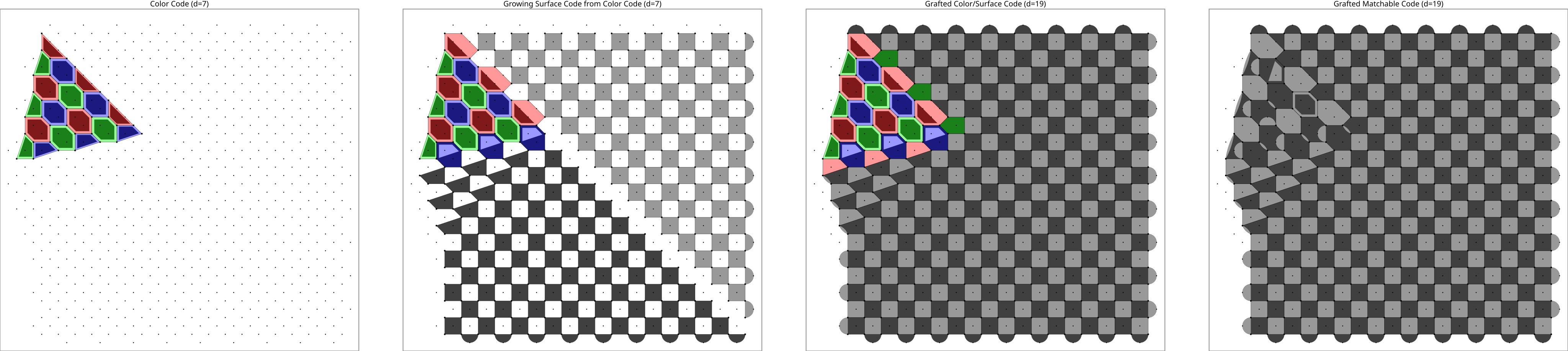}
        }
    \end{adjustwidth}
    \caption{
        \textbf{Stabilizer configurations while escaping to a matchable code.}
        Dark tiles are X basis stabilizers.
        Light tiles are Z basis stabilizers.
        Red, green, and blue tiles are part of the unmatchable color code region, where there are data errors producing three detection events instead of two.
        When two checked stabilizers overlap, one of them is inset to avoid visual ambiguity.
        Left: the initial color code.
        Middle-left: detector slice immediately after initializing new data qubits for the grown patch, implying the initialization bases of the new data qubits.
        Middle-right: stabilizers to measure $d-1$ times before ablating the color code region.
        Right: the final fully matchable configuration that we ultimately transition into.
        \href{
            https://algassert.com/crumble\#circuit=Q(0.5,4.5)0;Q(1,2)1;Q(1,3)2;Q(1,4)3;Q(1,5)4;Q(1.5,0.5)5;Q(1.5,1.5)6;Q(1.5,2.5)7;Q(1.5,3.5)8;Q(1.5,4.5)9;Q(1.5,5.5)10;Q(2,0)11;Q(2,1)12;Q(2,2)13;Q(2,3)14;Q(2,4)15;Q(2,5)16;Q(2,6)17;Q(2.5,0.5)18;Q(2.5,1.5)19;Q(2.5,2.5)20;Q(2.5,3.5)21;Q(2.5,4.5)22;Q(2.5,5.5)23;Q(3,1)24;Q(3,2)25;Q(3,3)26;Q(3,4)27;Q(3,5)28;Q(3.5,0.5)29;Q(3.5,1.5)30;Q(3.5,2.5)31;Q(3.5,3.5)32;Q(3.5,4.5)33;Q(3.5,5.5)34;Q(4,0)35;Q(4,1)36;Q(4,2)37;Q(4,3)38;Q(4,4)39;Q(4,5)40;Q(4,6)41;Q(4.5,0.5)42;Q(4.5,1.5)43;Q(4.5,2.5)44;Q(4.5,3.5)45;Q(4.5,4.5)46;Q(4.5,5.5)47;Q(5,1)48;Q(5,2)49;Q(5,3)50;Q(5,4)51;Q(5,5)52;Q(5.5,0.5)53;Q(5.5,1.5)54;Q(5.5,2.5)55;Q(5.5,3.5)56;Q(5.5,4.5)57;Q(5.5,5.5)58;Q(6,0)59;Q(6,1)60;Q(6,2)61;Q(6,3)62;Q(6,4)63;Q(6,5)64;Q(6,6)65;Q(6.5,0.5)66;Q(6.5,1.5)67;Q(6.5,2.5)68;Q(6.5,3.5)69;Q(6.5,4.5)70;Q(6.5,5.5)71;Q(7,2)72;Q(7,4)73;POLYGON(0.25,0.25,0.25,0.5)18_5;POLYGON(0.25,0.25,0.25,0.5)42_29;POLYGON(0.25,0.25,0.25,0.5)29_30_18;POLYGON(0.25,0.25,0.25,0.5)21_22_9_3;POLYGON(0.25,0.25,0.25,0.5)66_53;POLYGON(0.25,0.25,0.25,0.5)53_54_43_42;POLYGON(0.25,0.25,0.25,0.5)67_68_55_54;POLYGON(0.25,0.25,0.25,0.5)55_56_45_44;POLYGON(0.25,0.25,0.25,0.5)45_46_33_32;POLYGON(0.25,0.25,0.25,0.5)33_34_23_22;POLYGON(0.25,0.25,0.25,0.5)23_10;POLYGON(0.25,0.25,0.25,0.5)69_70_57_56;POLYGON(0.25,0.25,0.25,0.5)57_58_47_46;POLYGON(0.25,0.25,0.25,0.5)47_34;POLYGON(0.25,0.25,0.25,0.5)71_58;POLYGON(0.25,0.25,0.6875,0.5)25_31_32_21_14_13;POLYGON(0.25,0.6875,0.25,0.5)6_13_14_2;POLYGON(0.25,0.6875,0.25,0.5)43_44_31_25_30;POLYGON(0.625,0.625,0.625,0.5)42_43_30_29;POLYGON(0.625,0.625,0.625,0.5)3_9;POLYGON(0.625,0.625,0.625,0.5)66_67_54_53;POLYGON(0.625,0.625,0.625,0.5)54_55_44_43;POLYGON(0.625,0.625,0.625,0.5)44_45_32_31;POLYGON(0.625,0.625,0.625,0.5)32_33_22_21;POLYGON(0.625,0.625,0.625,0.5)22_23_10_9;POLYGON(0.625,0.625,0.625,0.5)68_67;POLYGON(0.625,0.625,0.625,0.5)68_69_56_55;POLYGON(0.625,0.625,0.625,0.5)56_57_46_45;POLYGON(0.625,0.625,0.625,0.5)46_47_34_33;POLYGON(0.625,0.625,0.625,0.5)70_69;POLYGON(0.625,0.625,0.625,0.5)70_71_58_57;POLYGON(0.625,0.625,0.99609375,0.5)25_31_14_13;POLYGON(0.625,0.99609375,0.625,0.5)6_13_14_2;POLYGON(0.6875,0.25,0.25,0.5)25_13_6_5;POLYGON(0.99609375,0.625,0.625,0.5)18_30_25_13_6_5;POLYGON(0.99609375,0.625,0.625,0.5)14_21_3_2;TICK;MPP_Y5*Z18*Y6*Z29*Y2*Z42*X3*Z53*X9*Z66*X10_X13*X25*X14*X31*X21*X32_Z43*Z54*Z44*Z55_X22*X33*X23*X34_Z67*Z68_Z45*Z56*Z46*Z57_Z69*Z70_X58*X71;TICK;MPP_X5*X18_X6*X13*X2*X14_X29*X42_X30*X25*X43*X31*X44_X3*X21*X9*X22_X53*X66_X54*X67*X55*X68_X32*X45*X33*X46_X10*X23_X56*X69*X57*X70_X34*X47;TICK;MPP_X5*X6*X13*X25_X18*X29*X30_Z2*Z14*Z3*Z21_X42*X53*X43*X54_Z31*Z44*Z32*Z45_Z9*Z22*Z10*Z23_Z55*Z68*Z56*Z69_Z33*Z46*Z34*Z47_Z57*Z70*Z58*Z71;TICK;MPP_Z5*Z18*Z6*Z13*Z30*Z25_Z3*Z9_Z53*Z66*Z54*Z67_Z21*Z32*Z22*Z33_X44*X55*X45*X56_X46*X57*X47*X58;TICK;MPP_Z6*Z13*Z2*Z14_Z29*Z42*Z30*Z43;TICK;MPP_Z13*Z25*Z14*Z31;TICK;R_19_7_36_26_15_4_60_49_38_27_16_72_62_51_40_73_64;RX_11_12_1_35_24_20_8_0_59_48_37_61_50_39_28_17_63_52_41_65;TICK;CX_11_5_12_19_1_7_35_29_24_30_20_26_8_15_0_4_59_53_37_44_17_23_68_72_41_47_70_73_65_71;TICK;CX_11_18_12_5_35_42_19_25_7_14_20_13_59_66_26_32_15_22_67_72_17_10_69_73_41_34_65_58;TICK;CX_12_6_24_29_19_13_7_2_30_25_20_14_8_3_37_31_26_21_15_9;TICK;CX_1_6_24_18_7_13_42_36_20_25_48_53_37_43_26_31_15_21_66_60_54_49_44_38_32_27_22_16_61_67_50_55_39_45_28_33_68_62_56_51_46_40_63_69_52_57_70_64;TICK;CX_5_12_13_20_2_8_25_19_14_7_48_42_43_36_37_30_67_60_61_54_55_49_50_44_45_38_39_32_33_27_28_22_23_16_69_62_63_56_57_51_52_46_47_40_71_64;TICK;CX_6_12_29_36_13_19_2_7_30_25_14_20_3_8_53_60_48_54_43_49_31_38_21_27_9_16_61_68_55_62_50_56_45_51_39_46_33_40_28_34_63_70_57_64_52_58;TICK;CX_18_12_6_1_13_7_25_20_14_8_3_0_31_26_21_15_9_4;TICK;CX_12_19_1_7_30_36_20_26_8_15_0_4_48_43_54_60_44_49_32_38_22_27_10_16_61_55_50_45_39_33_28_23_56_62_46_51_34_40_63_57_52_47_58_64;TICK;M_19_7_36_26_15_4_60_49_38_27_16_72_62_51_40_73_64;MX_11_12_1_35_24_20_8_0_59_48_37_61_50_39_28_17_63_52_41_65;DT(1.5,0.5,0)rec[-37]_rec[-46];DT(1.5,1.5,0)rec[-36]_rec[-40];DT(3.5,0.5,0)rec[-35]_rec[-39];DT(2,2,0)rec[-34]_rec[-38];DT(1,3,0)rec[-33]_rec[-53];DT(1,4,0)rec[-32]_rec[-45];DT(5.5,0.5,0)rec[-31]_rec[-44];DT(4.5,1.5,0)rec[-30]_rec[-72];DT(3.5,2.5,0)rec[-29]_rec[-51];DT(2.5,3.5,0)rec[-28]_rec[-43];DT(1.5,4.5,0)rec[-27]_rec[-50];DT(6.5,1.5,0)rec[-26]_rec[-70];DT(5.5,2.5,0)rec[-25]_rec[-49];DT(4.5,3.5,0)rec[-24]_rec[-69];DT(3.5,4.5,0)rec[-23]_rec[-48];DT(6.5,3.5,0)rec[-22]_rec[-68];DT(5.5,4.5,0)rec[-21]_rec[-47];DT(2.5,0.5,0)rec[-20]_rec[-66];DT(3,2,0)rec[-19]_rec[-55];DT(2,3,0)rec[-18]_rec[-65];DT(4.5,0.5,0)rec[-17]_rec[-64];DT(3.5,1.5,0)rec[-16]_rec[-54];DT(3.5,3.5,0)rec[-15]_rec[-73];DT(2.5,4.5,0)rec[-14]_rec[-62];DT(0.5,4.5,0)rec[-13];DT(6.5,0.5,0)rec[-12]_rec[-61];DT(5.5,1.5,0)rec[-11]_rec[-52];DT(4.5,2.5,0)rec[-10]_rec[-63];DT(6.5,2.5,0)rec[-9]_rec[-60];DT(5.5,3.5,0)rec[-8]_rec[-42];DT(4.5,4.5,0)rec[-7]_rec[-59];DT(2.5,5.5,0)rec[-6]_rec[-71];DT(1.5,5.5,0)rec[-5]_rec[-58];DT(6.5,4.5,0)rec[-4]_rec[-57];DT(4.5,5.5,0)rec[-3]_rec[-41];DT(3.5,5.5,0)rec[-2]_rec[-56];DT(5.5,5.5,0)rec[-1]_rec[-67];TICK;MPP_Y5*Z18*Y6*Z29*Y2*Z42*X3*Z53*X9*Z66*X10_X13*X25*X14*X31*X21*X32_Z43*Z54*Z44*Z55_X22*X33*X23*X34_Z67*Z68_Z45*Z56*Z46*Z57_Z69*Z70_X58*X71;DT(1.5,3.5,1)rec[-7]_rec[-22];DT(5,2,1)rec[-6]_rec[-38];DT(3,5,1)rec[-5]_rec[-14];DT(7,2,1)rec[-4]_rec[-34];DT(5,4,1)rec[-3]_rec[-32];DT(7,4,1)rec[-2]_rec[-30];DT(6,6,1)rec[-1]_rec[-9];OI(0)rec[-8]_rec[-26]_rec[-82];TICK;MPP_X5*X18_X6*X13*X2*X14_X29*X42_X30*X25*X43*X31*X44_X3*X21*X9*X22_X53*X66_X54*X67*X55*X68_X32*X45*X33*X46_X10*X23_X56*X69*X57*X70_X34*X47;DT(2,0,2)rec[-11]_rec[-39];DT(2,1,2)rec[-10]_rec[-38];DT(4,0,2)rec[-9]_rec[-36];DT(4,2,2)rec[-8]_rec[-29]_rec[-34];DT(1,4,2)rec[-7];DT(6,0,2)rec[-6]_rec[-31];DT(6,2,2)rec[-5]_rec[-28];DT(4,4,2)rec[-4]_rec[-26];DT(2,6,2)rec[-3]_rec[-24];DT(6,4,2)rec[-2]_rec[-23];DT(4,6,2)rec[-1]_rec[-21];TICK;MPP_X5*X6*X13*X25_X18*X29*X30_Z2*Z14*Z3*Z21_X42*X53*X43*X54_Z31*Z44*Z32*Z45_Z9*Z22*Z10*Z23_Z55*Z68*Z56*Z69_Z33*Z46*Z34*Z47_Z57*Z70*Z58*Z71;DT(1,2,3)rec[-9]_rec[-46]_rec[-47];DT(3,1,3)rec[-8]_rec[-44]_rec[-47];DT(2,4,3)rec[-7]_rec[-61];DT(5,1,3)rec[-6]_rec[-39];DT(4,3,3)rec[-5]_rec[-57];DT(2,5,3)rec[-4]_rec[-55];DT(6,3,3)rec[-3]_rec[-53];DT(4,5,3)rec[-2]_rec[-51];DT(6,5,3)rec[-1]_rec[-49];TICK;MPP_Z5*Z18*Z6*Z13*Z30*Z25_Z3*Z9_Z53*Z66*Z54*Z67_Z21*Z32*Z22*Z33_X44*X55*X45*X56_X46*X57*X47*X58;DT(2.5,1.5,4)rec[-6]_rec[-71];DT(1,5,4)rec[-5]_rec[-66];DT(6,1,4)rec[-4]_rec[-65];DT(3,4,4)rec[-3]_rec[-62];DT(5,3,4)rec[-2]_rec[-42];DT(5,5,4)rec[-1]_rec[-37];TICK;MPP_Z6*Z13*Z2*Z14_Z29*Z42*Z30*Z43;DT(1.5,2.5,5)rec[-2]_rec[-72];DT(4,1,5)rec[-1]_rec[-71];TICK;MPP_Z13*Z25*Z14*Z31;DT(3,3,6)rec[-1]_rec[-71]
        }{Click here to open an example of the middle-right patch's cycle circuit} in Crumble.
        \href{
            https://algassert.com/crumble\#circuit=Q(0.5,2.5)0;Q(0.5,4.5)1;Q(1,2)2;Q(1,3)3;Q(1,4)4;Q(1,5)5;Q(1.5,0.5)6;Q(1.5,1.5)7;Q(1.5,2.5)8;Q(1.5,3.5)9;Q(1.5,4.5)10;Q(1.5,5.5)11;Q(2,0)12;Q(2,1)13;Q(2,2)14;Q(2,3)15;Q(2,4)16;Q(2,5)17;Q(2,6)18;Q(2.5,0.5)19;Q(2.5,1.5)20;Q(2.5,2.5)21;Q(2.5,3.5)22;Q(2.5,4.5)23;Q(2.5,5.5)24;Q(3,1)25;Q(3,2)26;Q(3,3)27;Q(3,4)28;Q(3,5)29;Q(3.5,0.5)30;Q(3.5,1.5)31;Q(3.5,2.5)32;Q(3.5,3.5)33;Q(3.5,4.5)34;Q(3.5,5.5)35;Q(4,0)36;Q(4,1)37;Q(4,2)38;Q(4,3)39;Q(4,4)40;Q(4,5)41;Q(4,6)42;Q(4.5,0.5)43;Q(4.5,1.5)44;Q(4.5,2.5)45;Q(4.5,3.5)46;Q(4.5,4.5)47;Q(4.5,5.5)48;Q(5,1)49;Q(5,2)50;Q(5,3)51;Q(5,4)52;Q(5,5)53;Q(5.5,0.5)54;Q(5.5,1.5)55;Q(5.5,2.5)56;Q(5.5,3.5)57;Q(5.5,4.5)58;Q(5.5,5.5)59;Q(6,0)60;Q(6,1)61;Q(6,2)62;Q(6,3)63;Q(6,4)64;Q(6,5)65;Q(6,6)66;Q(6.5,0.5)67;Q(6.5,1.5)68;Q(6.5,2.5)69;Q(6.5,3.5)70;Q(6.5,4.5)71;Q(6.5,5.5)72;Q(7,2)73;Q(7,4)74;POLYGON(0,0,1,0.25)7_3;POLYGON(0,0,1,0.25)19_31_26_14_7_6;POLYGON(0,0,1,0.25)14_15;POLYGON(0,0,1,0.25)43_44_31_30;POLYGON(0,0,1,0.25)26_32;POLYGON(0,0,1,0.25)15_22_4_3;POLYGON(0,0,1,0.25)4_10;POLYGON(0,0,1,0.25)67_68_55_54;POLYGON(0,0,1,0.25)55_56_45_44;POLYGON(0,0,1,0.25)45_46_33_32;POLYGON(0,0,1,0.25)33_34_23_22;POLYGON(0,0,1,0.25)23_24_11_10;POLYGON(0,0,1,0.25)69_68;POLYGON(0,0,1,0.25)69_70_57_56;POLYGON(0,0,1,0.25)57_58_47_46;POLYGON(0,0,1,0.25)47_48_35_34;POLYGON(0,0,1,0.25)71_70;POLYGON(0,0,1,0.25)71_72_59_58;POLYGON(1,0,0,0.25)19_6;POLYGON(1,0,0,0.25)7_14_15_3;POLYGON(1,0,0,0.25)43_30;POLYGON(1,0,0,0.25)30_31_19;POLYGON(1,0,0,0.25)26_32_33_22_15_14;POLYGON(1,0,0,0.25)22_23_10_4;POLYGON(1,0,0,0.25)67_54;POLYGON(1,0,0,0.25)54_55_44_43;POLYGON(1,0,0,0.25)44_45_32_26_31;POLYGON(1,0,0,0.25)68_69_56_55;POLYGON(1,0,0,0.25)56_57_46_45;POLYGON(1,0,0,0.25)46_47_34_33;POLYGON(1,0,0,0.25)34_35_24_23;POLYGON(1,0,0,0.25)24_11;POLYGON(1,0,0,0.25)70_71_58_57;POLYGON(1,0,0,0.25)58_59_48_47;POLYGON(1,0,0,0.25)48_35;POLYGON(1,0,0,0.25)72_59;TICK;MPP_Y6*Z19*X7*Z30*X3*Z43*X4*Z54*X10*Z67*X11_X14*X26*X15*X32*X22*X33_Z44*Z55*Z45*Z56_X23*X34*X24*X35_Z68*Z69_Z46*Z57*Z47*Z58_Z70*Z71_X59*X72;TICK;MPP_X6*X19_X7*X14*X3*X15_X30*X43_X31*X26*X44*X32*X45_X4*X22*X10*X23_X54*X67_X55*X68*X56*X69_X33*X46*X34*X47_X11*X24_X57*X70*X58*X71_X35*X48;TICK;MPP_Z6*Z19*Z7*Z14*Z31*Z26_Z3*Z15*Z4*Z22_X43*X54*X44*X55_Z32*Z45*Z33*Z46_Z10*Z23*Z11*Z24_Z56*Z69*Z57*Z70_Z34*Z47*Z35*Z48_Z58*Z71*Z59*Z72;TICK;MPP_X19*X30*X31_Z7*Z3_Z14*Z15_Z26*Z32_Z4*Z10_Z54*Z67*Z55*Z68_Z22*Z33*Z23*Z34_X45*X56*X46*X57_X47*X58*X48*X59;TICK;MPP_Z30*Z43*Z31*Z44;TICK;R_0_20_8_37_27_16_5_61_50_39_28_17_73_63_52_41_74_65;RX_12_13_2_36_25_21_9_1_60_49_38_62_51_40_29_18_64_53_42_66;TICK;CX_12_6_13_20_2_8_36_30_25_31_21_27_9_16_1_5_60_54_38_45_18_24_69_73_42_48_71_74_66_72;TICK;CX_12_19_36_43_8_15_21_14_60_67_27_33_16_23_68_73_18_11_70_74_42_35_66_59;TICK;CX_2_0_25_30_8_3_31_26_21_15_9_4_38_32_27_22_16_10;TICK;CX_2_7_25_19_8_14_3_0_43_37_21_26_49_54_38_44_27_32_16_22_67_61_55_50_45_39_33_28_23_17_62_68_51_56_40_46_29_34_69_63_57_52_47_41_64_70_53_58_71_65;TICK;CX_6_13_7_2_3_9_26_20_15_8_49_43_44_37_38_31_68_61_62_55_56_50_51_45_46_39_40_33_34_28_29_23_24_17_70_63_64_57_58_52_53_47_48_41_72_65;TICK;CX_7_13_2_0_30_37_14_20_31_26_4_9_54_61_49_55_44_50_32_39_22_28_10_17_62_69_56_63_51_57_46_52_40_47_34_41_29_35_64_71_58_65_53_59;TICK;CX_19_13_7_2_14_8_26_21_15_9_4_1_32_27_22_16_10_5;TICK;CX_13_20_2_8_31_37_21_27_9_16_1_5_49_44_55_61_45_50_33_39_23_28_11_17_62_56_51_46_40_34_29_24_57_63_47_52_35_41_64_58_53_48_59_65;TICK;M_0_20_8_37_27_16_5_61_50_39_28_17_73_63_52_41_74_65;MX_12_13_2_36_25_21_9_1_60_49_38_62_51_40_29_18_64_53_42_66;DT(1.5,1.5,0)rec[-38]_rec[-47];DT(1.5,0.5,0)rec[-37]_rec[-56];DT(2,2,0)rec[-36]_rec[-46];DT(3.5,0.5,0)rec[-35]_rec[-39];DT(3,2,0)rec[-34]_rec[-45];DT(1,3,0)rec[-33]_rec[-55];DT(1,4,0)rec[-32]_rec[-44];DT(5.5,0.5,0)rec[-31]_rec[-43];DT(4.5,1.5,0)rec[-30]_rec[-73];DT(3.5,2.5,0)rec[-29]_rec[-53];DT(2.5,3.5,0)rec[-28]_rec[-42];DT(1.5,4.5,0)rec[-27]_rec[-52];DT(6.5,1.5,0)rec[-26]_rec[-71];DT(5.5,2.5,0)rec[-25]_rec[-51];DT(4.5,3.5,0)rec[-24]_rec[-70];DT(3.5,4.5,0)rec[-23]_rec[-50];DT(6.5,3.5,0)rec[-22]_rec[-69];DT(5.5,4.5,0)rec[-21]_rec[-49];DT(2.5,0.5,0)rec[-20]_rec[-67];DT(2,1,0)rec[-19];DT(2,3,0)rec[-18]_rec[-66];DT(4.5,0.5,0)rec[-17]_rec[-65];DT(3.5,1.5,0)rec[-16]_rec[-48];DT(3.5,3.5,0)rec[-15]_rec[-74];DT(2.5,4.5,0)rec[-14]_rec[-63];DT(0.5,4.5,0)rec[-13];DT(6.5,0.5,0)rec[-12]_rec[-62];DT(5.5,1.5,0)rec[-11]_rec[-54];DT(4.5,2.5,0)rec[-10]_rec[-64];DT(6.5,2.5,0)rec[-9]_rec[-61];DT(5.5,3.5,0)rec[-8]_rec[-41];DT(4.5,4.5,0)rec[-7]_rec[-60];DT(2.5,5.5,0)rec[-6]_rec[-72];DT(1.5,5.5,0)rec[-5]_rec[-59];DT(6.5,4.5,0)rec[-4]_rec[-58];DT(4.5,5.5,0)rec[-3]_rec[-40];DT(3.5,5.5,0)rec[-2]_rec[-57];DT(5.5,5.5,0)rec[-1]_rec[-68];TICK;MPP_Y6*Z19*X7*Z30*X3*Z43*X4*Z54*X10*Z67*X11_X14*X26*X15*X32*X22*X33_Z44*Z55*Z45*Z56_X23*X34*X24*X35_Z68*Z69_Z46*Z57*Z47*Z58_Z70*Z71_X59*X72;DT(1.5,3.5,1)rec[-7]_rec[-22];DT(5,2,1)rec[-6]_rec[-38];DT(3,5,1)rec[-5]_rec[-14];DT(7,2,1)rec[-4]_rec[-34];DT(5,4,1)rec[-3]_rec[-32];DT(7,4,1)rec[-2]_rec[-30];DT(6,6,1)rec[-1]_rec[-9];OI(0)rec[-8]_rec[-83];TICK;MPP_X6*X19_X7*X14*X3*X15_X30*X43_X31*X26*X44*X32*X45_X4*X22*X10*X23_X54*X67_X55*X68*X56*X69_X33*X46*X34*X47_X11*X24_X57*X70*X58*X71_X35*X48;DT(2,0,2)rec[-11]_rec[-39];DT(1,2,2)rec[-10]_rec[-37];DT(4,0,2)rec[-9]_rec[-36];DT(4,2,2)rec[-8]_rec[-29]_rec[-34];DT(1,4,2)rec[-7];DT(6,0,2)rec[-6]_rec[-31];DT(6,2,2)rec[-5]_rec[-28];DT(4,4,2)rec[-4]_rec[-26];DT(2,6,2)rec[-3]_rec[-24];DT(6,4,2)rec[-2]_rec[-23];DT(4,6,2)rec[-1]_rec[-21];TICK;MPP_Z6*Z19*Z7*Z14*Z31*Z26_Z3*Z15*Z4*Z22_X43*X54*X44*X55_Z32*Z45*Z33*Z46_Z10*Z23*Z11*Z24_Z56*Z69*Z57*Z70_Z34*Z47*Z35*Z48_Z58*Z71*Z59*Z72;DT(2.5,1.5,3)rec[-8]_rec[-64];DT(2,4,3)rec[-7]_rec[-60];DT(5,1,3)rec[-6]_rec[-38];DT(4,3,3)rec[-5]_rec[-56];DT(2,5,3)rec[-4]_rec[-54];DT(6,3,3)rec[-3]_rec[-52];DT(4,5,3)rec[-2]_rec[-50];DT(6,5,3)rec[-1]_rec[-48];TICK;MPP_X19*X30*X31_Z7*Z3_Z14*Z15_Z26*Z32_Z4*Z10_Z54*Z67*Z55*Z68_Z22*Z33*Z23*Z34_X45*X56*X46*X57_X47*X58*X48*X59;DT(3,1,4)rec[-9]_rec[-52];DT(0.5,2.5,4)rec[-8]_rec[-74];DT(1.5,2.5,4)rec[-7]_rec[-72];DT(3,3,4)rec[-6]_rec[-70];DT(1,5,4)rec[-5]_rec[-68];DT(6,1,4)rec[-4]_rec[-67];DT(3,4,4)rec[-3]_rec[-64];DT(5,3,4)rec[-2]_rec[-44];DT(5,5,4)rec[-1]_rec[-39];TICK;MPP_Z30*Z43*Z31*Z44;DT(4,1,5)rec[-1]_rec[-72]
        }{Click here to open an example of the right patch's cycle circuit} in Crumble.
    }
    \label{fig:code-transitions}
\end{figure}

After the grafted code has stabilized, we need to replace the color code region of the grafted code with some sort of matchable code.
We settled on a strategy that drops some of the color code stabilizers, keeps some of the color code stabilizers, and decomposes some of the color code stabilizers into two body stabilizers.
This produces a code with the correct code distance, which can be transitioned into with a fault distance nearly matching that code distance, while not oversubscribing the available measurement qubits.
We show a topological summary of our escape stage in \fig{topological-growth}, and the exact stabilizer code configurations in \fig{code-transitions}.
We considered a few different ways of dropping the color code region, summarized in \fig{code-transition-alternates}.

\subsection{Decoding}
\label{sec:decoding}

During the injection stage and the cultivation stage, we use full postselection.
If any detector from these stages fires, we discard the system and start over.
We did some exploratory experiments where we allowed correcting simple detection event patterns, to attempt to reduce retry rates, but ultimately decided to stick with full postselection of the early stages.

A secondary benefit of the choice to postselect the early stages is it simplifies feedback.
Using error correction during the cultivation stage would require complex just-in-time decoding~\cite{chamberland2020colorinjection}, to prevent X errors from turning $T$ into $T^\dagger$.
Fixing such an X error too late ruins the shot.
When postselecting, a shot with a detected X error was going to be discarded anyways.
The postselected outcome is the same, regardless of the latency.
That said, there are still two kinds of fast feedback needed by cultivation.
First, when growing the color code patch, stabilizers prepared non-deterministically need to be forced into a specific eigenstate before the next layer of T gates.
This is a hard time limit that must be met.
Second, when a detection event is caught, the state needs to be discarded and another cultivation attempt started.
This is a soft time limit, but restarting quickly improves expected cost by increasing attempt rates.
Because these fast feedback operations have simple conditions and simple effects, we assume they can be done with a latency of less than one microsecond.

\begin{figure}
    \centering
    \resizebox{\linewidth}{!}{
        \includegraphics{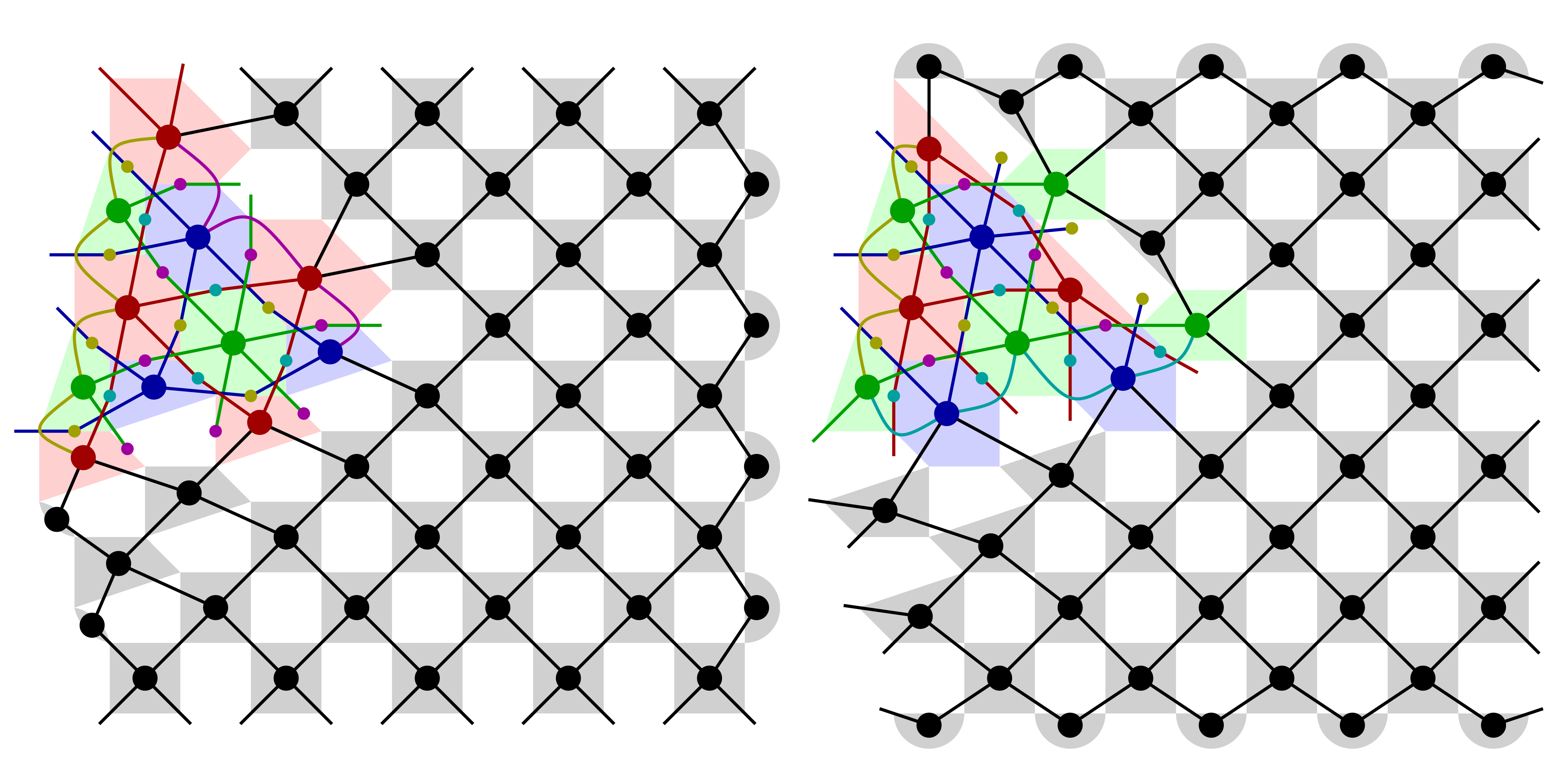}
    }
    \caption{
        \textbf{How the color code region is exposed to matching-based decoders}.
        Left: X basis subgraph.
        Right: Z basis subgraph.
        Small circles are ``doublet'' nodes, added for each adjacent pair of colored nodes.
        Errors that flip three colored nodes are exposed to the matcher as three separate errors, each flipping a colored node and a doublet node corresponding to the other two.
    }
    \label{fig:color-code-for-matcher}
\end{figure}

To decode the escape stage, we transform the decoding problem from a color code problem (which we don't have good decoders for) into a matching problem.
We do this by postselecting any shots that contain detection events on red X basis stabilizers or green Z basis stabilizers in the color code region.
Postselecting these stabilizers ensures the number of unpostselected detection events caused by X basis or Z basis data errors is at most 2, enabling matching to be used.
The stabilizers to postselect were chosen based on the layout of the grafted code.
They are the stabilizers that, if removed, don't reduce its code distance to $O(1)$.
These stabilizers are also the stabilizers that are discarded (or decomposed) when transitioning to a fully matchable code, meaning postselection of these stabilizers has a natural stopping point.

To present the color code region to the matcher in a form it can understand, we add ``doublet'' nodes, which represent a pair of excitations on adjacent detectors.
Example doublet nodes are included in \fig{color-code-for-matcher}.
Unmatchable errors in the color code region are decomposed into matchable errors involving these doublet nodes.
For example, a red + green + blue error mechanism would be exposed to the matcher as three independent error mechanisms: a red + (green-blue-doublet) error, a green + (red-blue-doublet) error, and a blue + (red-green-doublet) error.
This makes it possible for the matcher to understand the cost of travelling across the color code region, allowing a confidence to be estimated via the complementary gap~\cite{gidney2024yoked}.

Beware that exposing the decoding problem to the matcher using doublet nodes won't work well without postselecting one of the colors.
The issue is that matching can't get rid of an odd number of excitations, except by matching at least one of them to a boundary.
So there's no way to locally resolve the excitations of a red + green + blue error.
Postselecting one of the colors prevents this case from occurring.
There are still suboptimal cases, such as two red+green+blue errors overlapping on the postselected color, but the achieved performance was sufficient for our purposes in this paper.

The final step of decoding is to decide whether or not to keep the state, based on the complementary gap of the escape stage.
This requires waiting for the decoder to finish, which we assume takes 10 microseconds.
By varying the cutoff for an acceptable gap, we can achieve a variety of trade-offs between expected cost and fidelity.

\subsection{Errata}
\label{sec:Errata}

Shortly before releasing this paper, we noticed a minor mistake in our circuits. 
The issue is that we're preparing the stabilizers of the color code so that its X and Z stabilizers are in their +1 eigenstate.
For stabilizers with 6 qubits, this results in the Y stabilizer being in the -1 eigenstate (because $X^{\otimes 2} \cdot Z^{\otimes 2} = -Y^{\otimes 2}$).
This means the $H_{XY}$ gate isn't \emph{trivially} transversal.
Applying $H_{XY}$ to each physical qubit doesn't perform a logical $H_{XY}$, it performs one up to Pauli gates.
It also flips the sign of the 6-body X-basis stabilizers.
As a result, a GHZ state controlling $H_{XY}$ gates targeting the physical qubits of a color code will become entangled with the 6-body X-basis stabilizers instead of only experiencing phase kickback.
We missed this mistake because we only performed state vector simulation of the $d=3$ color code, which has no 6-body operators, and because in our stabilizer simulations of the $d=5$ case (with T gates replaced by S gates) Stim was silently correcting for the problem when computing the expected values of detectors and observables.

Note that this sign-entangling problem isn't an issue in \cite{chamberland2020colorinjection}.
They use the $H$ gate, instead of the $H_{XY}$ gate.
$H$ exchanges the X and Z stabilizers, and the state is in their +1 eigenstates, so no sign flip occurs.
For this paper we don't want to use $H$, because consuming the resulting magic state to perform a non-Clifford gate requires a Y basis interaction (which is more costly for planar surface codes~\cite{gidney2024ybasis}).

The sign-entangling problem can be fixed in two ways.
The first option is that we could prepare the color code so that the sign of overlapping stabilizers is always identical across the X, Y, and Z bases.
This requires the state to be in the +1 eigenstate of 4-body stabilizers, because $+X^{\otimes 4} \cdot +Z^{\otimes 4} = +Y^{\otimes 4}$ and $-X^{\otimes 4} \cdot -Z^{\otimes 4} \neq -Y^{\otimes 4}$.
Conversely, it requires the state to be in the -1 eigenstate of 6-body stabilizers, because $+X^{\otimes 6} \cdot +Z^{\otimes 6} \neq +Y^{\otimes 6}$ and $-X^{\otimes 6} \cdot -Z^{\otimes 6} = -Y^{\otimes 6}$.
Implementing this option would mean folding Pauli gates into the injection stage (\fig{d3-init-unitary}) as well as into the Bell pair preparation that occurs when growing the color code (\fig{color-code-growth}).
The second option is to cancel out the problematic sign flips while performing the transversal $H_{XY}$.
This can be done by folding Pauli gates into the single qubit rotations that appear in \fig{cat-check-d5}.
There could also be changes to \fig{cat-check-d3}, though technically none are needed because the $d=3$ color code has no 6-body operators.
See \cite{bombin2013gauge} for more discussion of adjusting signs to ensure gates are transversal.

Neither of these options would impact our stabilizer simulations, because Pauli gates commute with digitized errors.
It's conceivable that the state vector simulations with T gates, from \fig{t-vs-s}, could be affected.
Choosing the second option (fix signs during logical state checks), while leaving \fig{cat-check-d3} alone, would avoid this possibility.

We spent weeks performing Monte Carlo sampling of the current circuits, so updating and resampling them will take some time.
We intend to fix the issue in a future version of the paper.
We don't expect the changes to affect the final results.

\section{Results}
\label{sec:results}

The generated circuits we benchmarked, and the statistics we collected, are \href{https://doi.org/10.5281/zenodo.13777072}{available on Zenodo}~\cite{gidneyy2024cultivationdata}.

\subsection{Conventions}

In plots, and throughout the paper, we use the following conventions and terminology:

\begin{itemize}
    \item The ``code distance'' of a code is the minimum number of single qubit Pauli errors needed to flip a logical observable without flipping a stabilizer.
    \item The ``fault distance'' of a circuit or a state is the minimum number of errors from the noise model needed to cause a logical error without tripping any detectors.
    \item 
$d_1$ is the target fault distance of the magic state.
It's the code distance at the end of the cultivation stage (equivalently: at the beginning of the escape stage).
    \item 
$d_2$ is the code distance at the end of the escape stage.
    \item A ``round'' (also called a ``cycle'') is a series of circuit layers measuring/preparing the stabilizers of a code. A round contains multiple two qubit gate layers and starts/ends with reset/measure layers. We count rounds, instead of circuit layers, due to reset/measure layers lasting substantially longer than unitary layers when executed on a superconducting quantum computer.
    \item 
$r_1$ is the number of rounds spent idling in the grafted code during the escape stage, before transitioning into the final matchable code.
    \item 
$r_2$ is the number of rounds spent idling in the final matchable code, until the simulation is ended. 
    \item 
$c$ is the name of the circuit task being simulated.
    \item 
$q$ is the automatically computed number of qubits present in the simulated circuit.
    \item 
$p$ is the noise strength.
    \item 
$r$ is the automatically computed number of rounds present in the simulated circuit.
    \item 
$b$ is the basis being cultivated, with $b=Y$ corresponding to cultivation of the $Y$ eigenbasis $S|+\rangle$.
    \item 
$g$ is the name of the gateset. $g=\text{css}$ means the gate set is single qubit gates, plus CNOT as the two qubit gate, plus single qubit Pauli basis measurement and reset.
    \item 
``$\text{noise}$'' is the noise model being used, with $\text{noise}=\text{uniform}$ referring to uniform depolarizing circuit noise.
    \item 
``$\text{decoder}$'' is the decoder being used.
``perfectionist'' means full postselection.
``desaturation'' means the decoding described in \sec{decoding}.
``pymatching-gap'' means pymatching was used to compute a complementary gap~\cite{higgott2023sparseblossom,gidney2024yoked}.
\end{itemize}

\subsection{Assumptions}

Simulating our construction was (and is) a major challenge.
Our circuit for preparing a fault distance 5 T state performs more than 50 T gates and escapes into a code with hundreds of qubits.
It's possible to simulate circuits with this many T gates and qubits, for example see \cite{kissinger2022zxtsim}, but the per-shot simulation time tends to be measured in minutes.
We want to use Monte Carlo sampling to validate our constructions, meaning we need billions or even trillions of shots.
For this to be tractable we need hundreds of shots per second, not hundreds of seconds per shot.

In \cite{chamberland2020colorinjection}, large cultivation circuits were simulated by tracking the propagation of Pauli errors through the circuit.
When a Pauli error propagated across a T gate, becoming a non-Pauli error, it was twirled back into a Pauli error.
For example, when an $X$ error crosses a T gate, the $X$ will transform into $X+Y$ and then twirling will replace it with an X gate 50\% of the time and a Y gate the other 50\% of the time.
We attempted to use this technique, but couldn't get it to work.
The basic problem we encountered is that it's common in fault tolerant circuits for there to be benign errors (such as an X error on a $|+\rangle$ state) that unfold into stabilizers of the code under the action of the circuit.
An error equal to a stabilizer is still a benign error, so this kind of unfolding is fine and in fact quite normal.
Unfortunately, the twirling process would transform this benign stabilizer error into a malignant error by acting differently on its individual Pauli terms.
The result was that when we used this simulation technique, it would report logical error rates of $\Theta(p)$ when using a noise strength of $p$, even though we knew the circuits are more reliable than that.
We made a few attempts at fixing this issue, but couldn't find a solution.

In \cite{gidney2023hook} and \cite{zhou2024constantdepth}, a different hack is used for avoiding simulating T gates: $Z$ gates and $S$ gates are used as proxies for the $T$ gate.
Our circuits (similar to \cite{gidney2023hook,zhou2024constantdepth}) have the property that replacing every T gate with an S gate (or Z gate) will prepare an $S|+\rangle$ state (or $Z|+\rangle$ state) instead of a $T|+\rangle$ state.
The idea is that, because the overall structure of the circuit doesn't depend on which of the gates is being used, the error rate of a $T|+\rangle$ preparation should be similar to the error rate of a $Z|+\rangle$ preparation or $S|+\rangle$ preparation.
Unfortunately, when we attempted to verify this idea, we found that the situation is complicated.

\begin{figure}
    \begin{adjustwidth}{-1.5cm}{-1.5cm}
        \centering
        \resizebox{\linewidth}{!}{
            \includegraphics{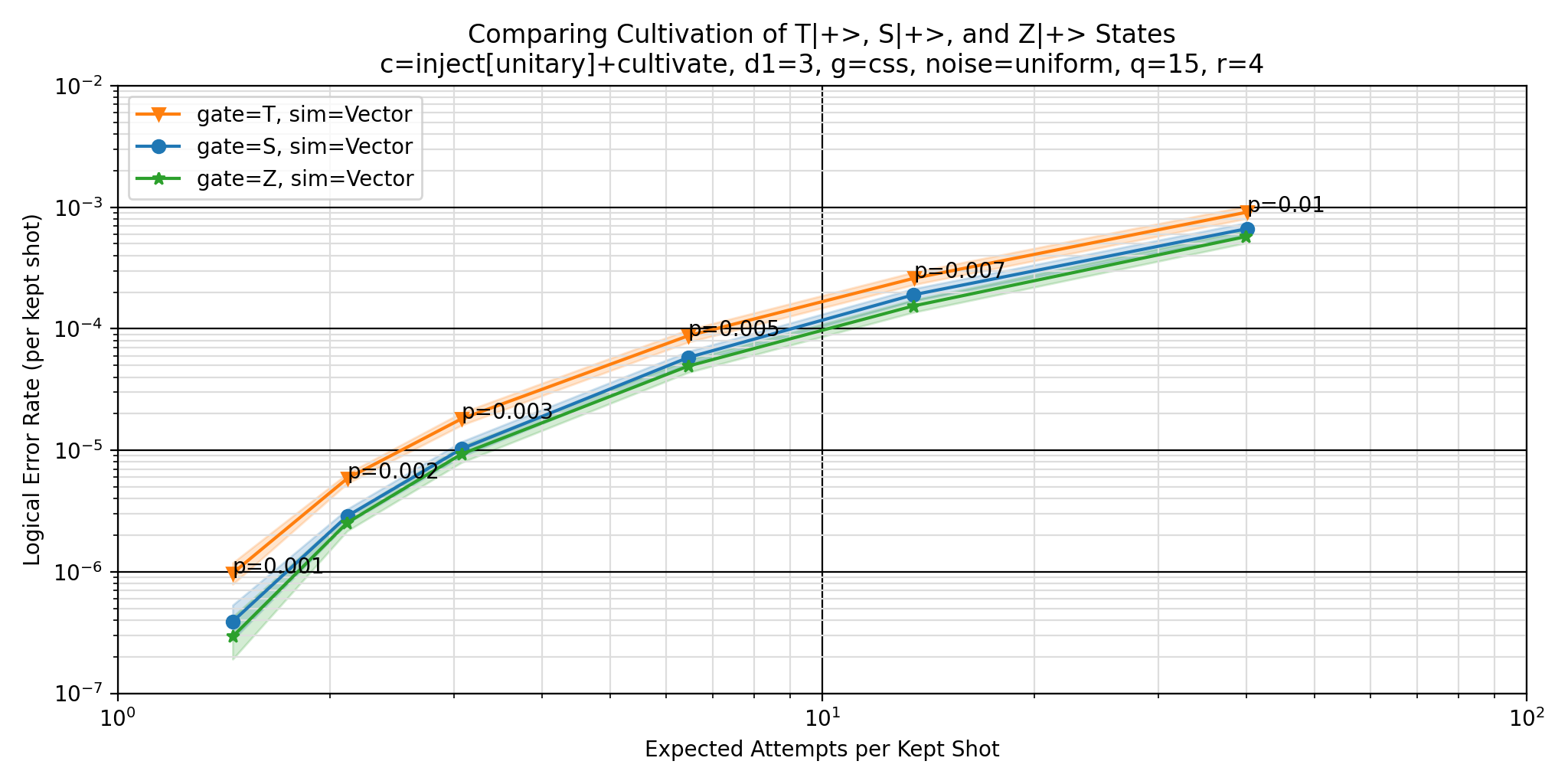}
        }
    \end{adjustwidth}
    \caption{
        \textbf{Small-scale comparison of cultivating T, S, and Z states without an escape stage.}
        Uses the exact same circuit for each case, except each $T$ gate is replaced by $T$, $S$, or $Z$.
        The $T$ curve has an error rate roughly twice as high as the $S$ and $Z$ curves, with the ratio becoming gradually larger as the noise strength lowers.
        Discard rates appear identical.
        Shaded regions indicate error rate hypotheses with a likelihood within a factor of 1000 of the max likelihood hypothesis, given the sampled data.
    }
    \label{fig:t-vs-s}
\end{figure}

We performed full state vector simulations of the injection and cultivation stage, targeting a fault distance of 3, for $T|+\rangle$, $S|+\rangle$, and $Z|+\rangle$ cultivation.
These simulations clearly show that the logical error rate for cultivating $T|+\rangle$ is higher than the logical error rates for cultivating $S|+\rangle$ or $Z|+\rangle$ (see \fig{t-vs-s}).
For the cases we simulated, cultivating $T|+\rangle$ had a logical error rate around twice as high as cultivating $S|+\rangle$.
We conjecture that this is a reasonable approximation for all cases we consider in this paper.
This is the assumption we use for converting the numbers produced by our simulations into the estimates we show in \fig{historical_progression} and \fig{historical-comparison}.

\begin{assumption}
For the cases considered by this paper, the logical error rate of $T|+\rangle$ cultivation can be estimated by doubling the logical error rate of $S|+\rangle$ cultivation (using the same circuit but with $T$ replaced by $S$).
\label{ass:tbys}
\end{assumption}

Beware that it's clear from \fig{t-vs-s} that the gap between $S|+\rangle$ and $T|+\rangle$ cultivation is growing slowly as the physical noise strength is decreased.
\ass{tbys} will clearly break if the noise strength is made substantially lower.
\ass{tbys} could also potentially break as the fault distance of cultivation is increased.
We were able to check that the assumption holds for fault distance 3 cultivation, but we don't have an easy way of checking that it continues to hold for fault distance 5 cultivation.
Regardless, due to a lack of better alternatives, we are forced to rely on \ass{tbys}.

\subsection{Simulations}

\begin{figure}
    \begin{adjustwidth}{-2.5cm}{-2.5cm}
        \centering
        \resizebox{\linewidth}{!}{
            \includegraphics{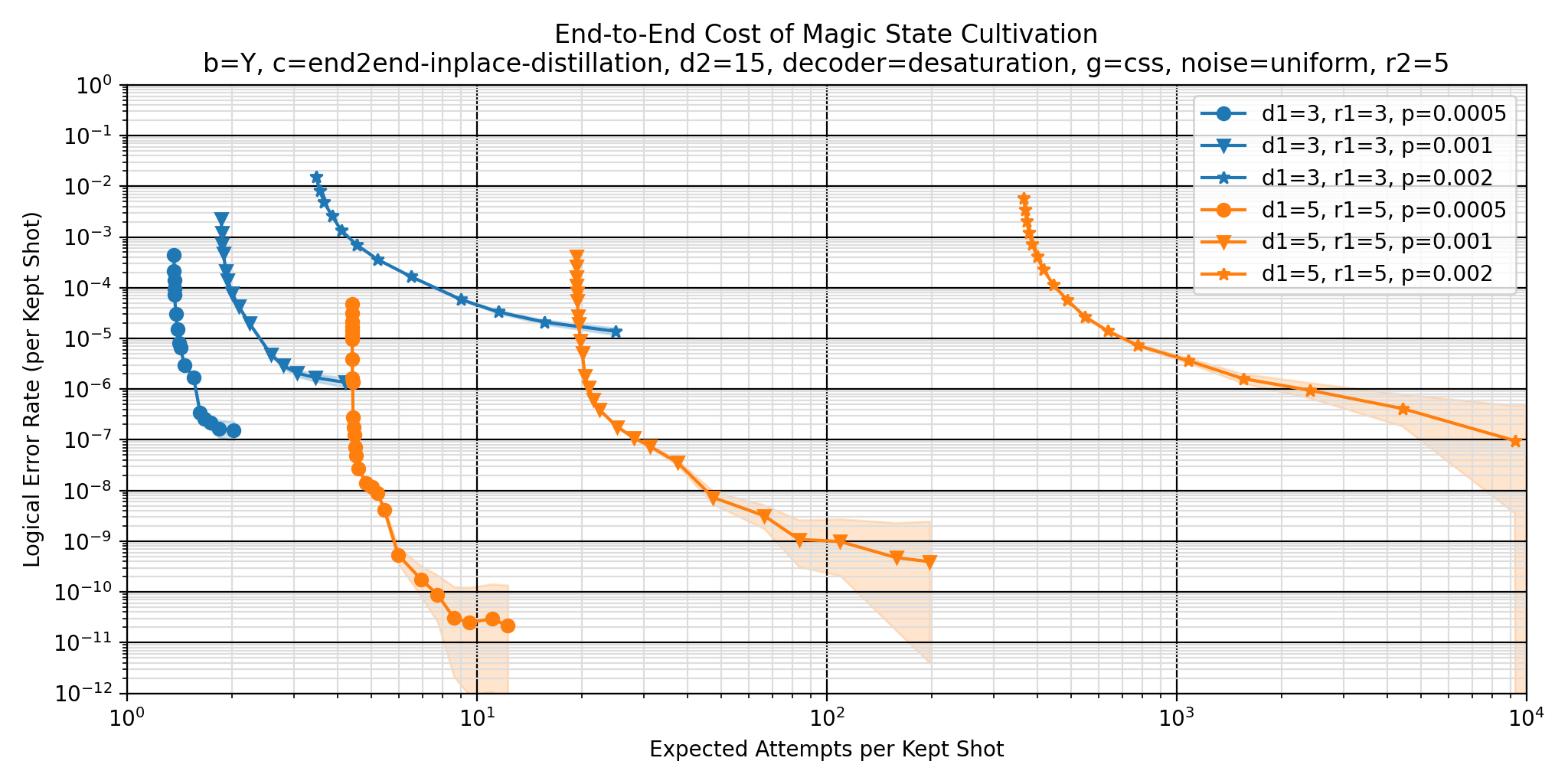}
        }
    \end{adjustwidth}
    \caption{
        \textbf{End to end simulation of magic state cultivation.}
        Curves show the effect of varying the complementary gap cutoff for rejecting shots, decreasing logical error rate at the cost of increasing expected attempts.
        \href{
            https://algassert.com/crumble\#circuit=Q(0.5,2.5)0;Q(0.5,4.5)1;Q(1,1)2;Q(1,2)3;Q(1,3)4;Q(1,4)5;Q(1,5)6;Q(1.5,0.5)7;Q(1.5,1.5)8;Q(1.5,2.5)9;Q(1.5,3.5)10;Q(1.5,4.5)11;Q(1.5,5.5)12;Q(2,0)13;Q(2,1)14;Q(2,2)15;Q(2,3)16;Q(2,4)17;Q(2,5)18;Q(2,6)19;Q(2.5,0.5)20;Q(2.5,1.5)21;Q(2.5,2.5)22;Q(2.5,3.5)23;Q(2.5,4.5)24;Q(2.5,5.5)25;Q(3,1)26;Q(3,2)27;Q(3,3)28;Q(3,4)29;Q(3,5)30;Q(3.5,0.5)31;Q(3.5,1.5)32;Q(3.5,2.5)33;Q(3.5,3.5)34;Q(3.5,4.5)35;Q(3.5,5.5)36;Q(4,0)37;Q(4,1)38;Q(4,2)39;Q(4,3)40;Q(4,4)41;Q(4,5)42;Q(4,6)43;Q(4.5,0.5)44;Q(4.5,1.5)45;Q(4.5,2.5)46;Q(4.5,3.5)47;Q(4.5,4.5)48;Q(4.5,5.5)49;Q(5,1)50;Q(5,2)51;Q(5,3)52;Q(5,4)53;Q(5,5)54;Q(5.5,0.5)55;Q(5.5,1.5)56;Q(5.5,2.5)57;Q(5.5,3.5)58;Q(5.5,4.5)59;Q(5.5,5.5)60;Q(6,0)61;Q(6,1)62;Q(6,2)63;Q(6,3)64;Q(6,4)65;Q(6,5)66;Q(6,6)67;Q(6.5,0.5)68;Q(6.5,1.5)69;Q(6.5,2.5)70;Q(6.5,3.5)71;Q(6.5,4.5)72;Q(6.5,5.5)73;Q(7,2)74;Q(7,4)75;POLYGON(0,0,1,0.25)69_70;POLYGON(0,0,1,0.25)71_72;POLYGON(0,0,1,0.25)11_24_25_12;POLYGON(0,0,1,0.25)5_11;POLYGON(0,0,1,0.25)15_27_33_16;POLYGON(0,0,1,0.25)8_15_16_4;POLYGON(0,0,1,0.25)4_16_23_5;POLYGON(0,0,1,0.25)7_20_32_27_15_8;POLYGON(0,0,1,0.25)31_44_45_32;POLYGON(0,0,1,0.25)59_72_73_60;POLYGON(0,0,1,0.25)35_48_49_36;POLYGON(0,0,1,0.25)47_58_59_48;POLYGON(0,0,1,0.25)57_70_71_58;POLYGON(0,0,1,0.25)45_56_57_46;POLYGON(0,0,1,0.25)55_68_69_56;POLYGON(0,0,1,0.25)23_34_35_24;POLYGON(0,0,1,0.25)33_46_47_34;POLYGON(1,0,0,0.25)20_7;POLYGON(1,0,0,0.25)44_31;POLYGON(1,0,0,0.25)68_55;POLYGON(1,0,0,0.25)73_60;POLYGON(1,0,0,0.25)49_36;POLYGON(1,0,0,0.25)25_12;POLYGON(1,0,0,0.25)5_23_24_11;POLYGON(1,0,0,0.25)15_27_33_34_23_16;POLYGON(1,0,0,0.25)8_15_16_4;POLYGON(1,0,0,0.25)7_27_15_8;POLYGON(1,0,0,0.25)20_31_32;POLYGON(1,0,0,0.25)32_45_46_33_27;POLYGON(1,0,0,0.25)48_59_60_49;POLYGON(1,0,0,0.25)24_35_36_25;POLYGON(1,0,0,0.25)34_47_48_35;POLYGON(1,0,0,0.25)58_71_72_59;POLYGON(1,0,0,0.25)46_57_58_47;POLYGON(1,0,0,0.25)56_69_70_57;POLYGON(1,0,0,0.25)44_55_56_45;TICK;R_22_9_21;RX_15_27_16_8_14_3_28_7_0_33;TICK;CX_15_22_27_21_16_9;TICK;CX_15_9_27_22;TICK;CX_15_21_16_22;TICK;CX_14_21_3_9_28_22;TICK;CX_21_14_9_3_22_28;TICK;CX_8_3_33_28;TICK;CX_8_14_28_33;TICK;CX_3_8;TICK;CX_14_8;TICK;S_DAG_8;TICK;CX_14_8;TICK;CX_3_8;TICK;CX_7_14_0_3;TICK;CX_14_7_3_0;TICK;R_21_28_9_4;RX_14_22_3;TICK;CX_14_21_3_9_0_4_22_28;TICK;CX_14_7_21_27_9_16_4_0_22_15;TICK;CX_14_8_21_15_9_4_22_16;TICK;CX_3_8_9_15_22_27_28_33;TICK;CX_8_3_15_9_27_22_33_28;TICK;CX_7_14_15_22_27_21_16_9;TICK;CX_8_14_15_21_4_9_16_22;TICK;CX_14_21_3_9_22_28;TICK;M_21_28_9_0;MX_14_22_3;DT(2.5,1.5,0)rec[-7];DT(3,3,0)rec[-6];DT(1.5,2.5,0)rec[-5];DT(0.5,2.5,0)rec[-4];DT(2,1,0)rec[-3];DT(2.5,2.5,0)rec[-2];DT(1,2,0)rec[-1];TICK;RX_28_22_21_14_9_2;TICK;S_DAG_7_8_15_4_27_16_33;TICK;CX_2_7_14_8_21_15_9_4_22_27_28_33;TICK;CX_8_2_15_9_22_28;TICK;CX_15_8_22_16;TICK;CX_15_22;TICK;MX_15;DT(1,2,1)rec[-1]_rec[-2]_rec[-3];TICK;RX_15;TICK;CX_15_22;TICK;CX_15_8_22_16;TICK;CX_8_2_15_9_22_28;TICK;CX_2_7_14_8_21_15_9_4_22_27_28_33;TICK;MX_28_22_21_14_9_2;DT(3,3,2)rec[-6];DT(2.5,2.5,2)rec[-5];DT(2,2,2)rec[-4]_rec[-7];DT(2,1,2)rec[-3];DT(1.5,2.5,2)rec[-2];DT(1,1,2)rec[-1];TICK;S_7_8_15_4_27_16_33;TICK;R_20_31_44_32_55_45_68_56_46_69_57_70_58_71_72_21_9_38_28_17_6_62_51_40_29_18_74_64_53_42_75_66;RX_5_23_11_34_24_12_47_35_25_48_36_59_49_60_73_13_14_3_37_26_22_10_1_61_50_39_63_52_41_30_19_65_54_43_67;TICK;CX_13_7_14_21_3_9_37_31_26_32_22_28_10_17_1_6_61_55_39_46_19_25_70_74_43_49_72_75_67_73;TICK;CX_13_20_14_7_37_44_21_27_9_16_22_15_61_68_28_34_17_24_69_74_19_12_71_75_43_36_67_60;TICK;CX_14_8_26_31_21_15_9_4_32_27_22_16_10_5_39_33_28_23_17_11;TICK;CX_3_8_26_20_9_15_44_38_22_27_50_55_39_45_28_33_17_23_68_62_56_51_46_40_34_29_24_18_63_69_52_57_41_47_30_35_70_64_58_53_48_42_65_71_54_59_72_66;TICK;CX_7_14_15_22_4_10_27_21_16_9_50_44_45_38_39_32_69_62_63_56_57_51_52_46_47_40_41_34_35_29_30_24_25_18_71_64_65_58_59_53_54_48_49_42_73_66;TICK;CX_8_14_31_38_15_21_4_9_32_27_16_22_5_10_55_62_50_56_45_51_33_40_23_29_11_18_63_70_57_64_52_58_47_53_41_48_35_42_30_36_65_72_59_66_54_60;TICK;CX_20_14_8_3_15_9_27_22_16_10_5_1_33_28_23_17_11_6;TICK;CX_14_21_3_9_32_38_22_28_10_17_1_6_50_45_56_62_46_51_34_40_24_29_12_18_63_57_52_47_41_35_30_25_58_64_48_53_36_42_65_59_54_49_60_66;TICK;M_21_9_38_28_17_6_62_51_40_29_18_74_64_53_42_75_66;MX_13_14_3_37_26_22_10_1_61_50_39_63_52_41_30_19_65_54_43_67;DT(2.5,1.5,3)rec[-37];DT(1.5,2.5,3)rec[-36];DT(4,1,3)rec[-35];DT(3,3,3)rec[-34];DT(6,1,3)rec[-31];DT(5,2,3)rec[-30];DT(7,2,3)rec[-26];DT(6,3,3)rec[-25];DT(7,4,3)rec[-22];DT(2,2,3)rec[-19]_rec[-44];DT(1,2,3)rec[-18]_rec[-44];DT(2.5,2.5,3)rec[-15]_rec[-44];DT(1.5,3.5,3)rec[-14];DT(0.5,4.5,3)rec[-13];DT(4,4,3)rec[-7];DT(3,5,3)rec[-6];DT(2,6,3)rec[-5];DT(5,5,3)rec[-3];DT(4,6,3)rec[-2];DT(6,6,3)rec[-1];TICK;R_21_9_38_28_17_6_62_51_40_29_18_74_64_53_42_75_66;RX_13_14_3_37_26_22_10_1_61_50_39_63_52_41_30_19_65_54_43_67;TICK;CX_13_7_14_21_3_9_37_31_26_32_22_28_10_17_1_6_61_55_39_46_19_25_70_74_43_49_72_75_67_73;TICK;CX_13_20_14_7_37_44_21_27_9_16_22_15_61_68_28_34_17_24_69_74_19_12_71_75_43_36_67_60;TICK;CX_14_8_26_31_21_15_9_4_32_27_22_16_10_5_39_33_28_23_17_11;TICK;CX_3_8_26_20_9_15_44_38_22_27_50_55_39_45_28_33_17_23_68_62_56_51_46_40_34_29_24_18_63_69_52_57_41_47_30_35_70_64_58_53_48_42_65_71_54_59_72_66;TICK;CX_7_14_15_22_4_10_27_21_16_9_50_44_45_38_39_32_69_62_63_56_57_51_52_46_47_40_41_34_35_29_30_24_25_18_71_64_65_58_59_53_54_48_49_42_73_66;TICK;CX_8_14_31_38_15_21_4_9_32_27_16_22_5_10_55_62_50_56_45_51_33_40_23_29_11_18_63_70_57_64_52_58_47_53_41_48_35_42_30_36_65_72_59_66_54_60;TICK;CX_20_14_8_3_15_9_27_22_16_10_5_1_33_28_23_17_11_6;TICK;CX_14_21_3_9_32_38_22_28_10_17_1_6_50_45_56_62_46_51_34_40_24_29_12_18_63_57_52_47_41_35_30_25_58_64_48_53_36_42_65_59_54_49_60_66;TICK;M_21_9_38_28_17_6_62_51_40_29_18_74_64_53_42_75_66;MX_13_14_3_37_26_22_10_1_61_50_39_63_52_41_30_19_65_54_43_67;DT(2.5,1.5,4)rec[-37]_rec[-74];DT(1.5,2.5,4)rec[-36]_rec[-73];DT(4,1,4)rec[-35]_rec[-72];DT(3,3,4)rec[-34]_rec[-71];DT(2,4,4)rec[-33]_rec[-70];DT(1,5,4)rec[-32]_rec[-69];DT(6,1,4)rec[-31]_rec[-68];DT(5,2,4)rec[-30]_rec[-67];DT(4,3,4)rec[-29]_rec[-66];DT(3,4,4)rec[-28]_rec[-65];DT(2,5,4)rec[-27]_rec[-64];DT(7,2,4)rec[-26]_rec[-63];DT(6,3,4)rec[-25]_rec[-62];DT(5,4,4)rec[-24]_rec[-61];DT(4,5,4)rec[-23]_rec[-60];DT(7,4,4)rec[-22]_rec[-59];DT(6,5,4)rec[-21]_rec[-58];DT(2,0,4)rec[-20]_rec[-57];DT(1,2,4)rec[-19]_rec[-55]_rec[-56];DT(2,1,4)rec[-18]_rec[-56];DT(4,0,4)rec[-17]_rec[-54];DT(3,1,4)rec[-16]_rec[-53]_rec[-56];DT(1.5,3.5,4)rec[-15]_rec[-51];DT(1.5,3.5,5)rec[-14];DT(0.5,4.5,4)rec[-13];DT(6,0,4)rec[-12]_rec[-49];DT(5,1,4)rec[-11]_rec[-48];DT(4,2,4)rec[-10]_rec[-47]_rec[-52];DT(6,2,4)rec[-9]_rec[-46];DT(5,3,4)rec[-8]_rec[-45];DT(4,4,4)rec[-7]_rec[-44];DT(3,5,4)rec[-6]_rec[-43];DT(2,6,4)rec[-5]_rec[-42];DT(6,4,4)rec[-4]_rec[-41];DT(5,5,4)rec[-3]_rec[-40];DT(4,6,4)rec[-2]_rec[-39];DT(6,6,4)rec[-1]_rec[-38];TICK;R_21_9_38_28_17_6_62_51_40_29_18_74_64_53_42_75_66;RX_13_14_3_37_26_22_10_1_61_50_39_63_52_41_30_19_65_54_43_67;TICK;CX_13_7_14_21_3_9_37_31_26_32_22_28_10_17_1_6_61_55_39_46_19_25_70_74_43_49_72_75_67_73;TICK;CX_13_20_14_7_37_44_21_27_9_16_22_15_61_68_28_34_17_24_69_74_19_12_71_75_43_36_67_60;TICK;CX_14_8_26_31_21_15_9_4_32_27_22_16_10_5_39_33_28_23_17_11;TICK;CX_3_8_26_20_9_15_44_38_22_27_50_55_39_45_28_33_17_23_68_62_56_51_46_40_34_29_24_18_63_69_52_57_41_47_30_35_70_64_58_53_48_42_65_71_54_59_72_66;TICK;CX_7_14_15_22_4_10_27_21_16_9_50_44_45_38_39_32_69_62_63_56_57_51_52_46_47_40_41_34_35_29_30_24_25_18_71_64_65_58_59_53_54_48_49_42_73_66;TICK;CX_8_14_31_38_15_21_4_9_32_27_16_22_5_10_55_62_50_56_45_51_33_40_23_29_11_18_63_70_57_64_52_58_47_53_41_48_35_42_30_36_65_72_59_66_54_60;TICK;CX_20_14_8_3_15_9_27_22_16_10_5_1_33_28_23_17_11_6;TICK;CX_14_21_3_9_32_38_22_28_10_17_1_6_50_45_56_62_46_51_34_40_24_29_12_18_63_57_52_47_41_35_30_25_58_64_48_53_36_42_65_59_54_49_60_66;TICK;M_21_9_38_28_17_6_62_51_40_29_18_74_64_53_42_75_66;MX_13_14_3_37_26_22_10_1_61_50_39_63_52_41_30_19_65_54_43_67;DT(2.5,1.5,6)rec[-37]_rec[-74];DT(1.5,2.5,6)rec[-36]_rec[-73];DT(4,1,6)rec[-35]_rec[-72];DT(3,3,6)rec[-34]_rec[-71];DT(2,4,6)rec[-33]_rec[-70];DT(1,5,6)rec[-32]_rec[-69];DT(6,1,6)rec[-31]_rec[-68];DT(5,2,6)rec[-30]_rec[-67];DT(4,3,6)rec[-29]_rec[-66];DT(3,4,6)rec[-28]_rec[-65];DT(2,5,6)rec[-27]_rec[-64];DT(7,2,6)rec[-26]_rec[-63];DT(6,3,6)rec[-25]_rec[-62];DT(5,4,6)rec[-24]_rec[-61];DT(4,5,6)rec[-23]_rec[-60];DT(7,4,6)rec[-22]_rec[-59];DT(6,5,6)rec[-21]_rec[-58];DT(2,0,6)rec[-20]_rec[-57];DT(1,2,6)rec[-19]_rec[-55]_rec[-56];DT(2,1,6)rec[-18]_rec[-56];DT(4,0,6)rec[-17]_rec[-54];DT(3,1,6)rec[-16]_rec[-53]_rec[-56];DT(1.5,3.5,6)rec[-15]_rec[-51];DT(1.5,3.5,7)rec[-14];DT(0.5,4.5,6)rec[-13];DT(6,0,6)rec[-12]_rec[-49];DT(5,1,6)rec[-11]_rec[-48];DT(4,2,6)rec[-10]_rec[-47]_rec[-52];DT(6,2,6)rec[-9]_rec[-46];DT(5,3,6)rec[-8]_rec[-45];DT(4,4,6)rec[-7]_rec[-44];DT(3,5,6)rec[-6]_rec[-43];DT(2,6,6)rec[-5]_rec[-42];DT(6,4,6)rec[-4]_rec[-41];DT(5,5,6)rec[-3]_rec[-40];DT(4,6,6)rec[-2]_rec[-39];DT(6,6,6)rec[-1]_rec[-38];TICK;R_0_21_9_38_28_17_6_62_51_40_29_18_74_64_53_42_75_66;RX_13_14_3_37_26_22_10_1_61_50_39_63_52_41_30_19_65_54_43_67;TICK;CX_13_7_14_21_3_9_37_31_26_32_22_28_10_17_1_6_61_55_39_46_19_25_70_74_43_49_72_75_67_73;TICK;CX_13_20_37_44_9_16_22_15_61_68_28_34_17_24_69_74_19_12_71_75_43_36_67_60;TICK;CX_3_0_26_31_9_4_32_27_22_16_10_5_39_33_28_23_17_11;TICK;CX_3_8_26_20_9_15_4_0_44_38_22_27_50_55_39_45_28_33_17_23_68_62_56_51_46_40_34_29_24_18_63_69_52_57_41_47_30_35_70_64_58_53_48_42_65_71_54_59_72_66;TICK;CX_7_14_8_3_4_10_27_21_16_9_50_44_45_38_39_32_69_62_63_56_57_51_52_46_47_40_41_34_35_29_30_24_25_18_71_64_65_58_59_53_54_48_49_42_73_66;TICK;CX_8_14_3_0_31_38_15_21_32_27_5_10_55_62_50_56_45_51_33_40_23_29_11_18_63_70_57_64_52_58_47_53_41_48_35_42_30_36_65_72_59_66_54_60;TICK;CX_20_14_8_3_15_9_27_22_16_10_5_1_33_28_23_17_11_6;TICK;CX_14_21_3_9_32_38_22_28_10_17_1_6_50_45_56_62_46_51_34_40_24_29_12_18_63_57_52_47_41_35_30_25_58_64_48_53_36_42_65_59_54_49_60_66;TICK;M_0_21_9_38_28_17_6_62_51_40_29_18_74_64_53_42_75_66;MX_13_14_3_37_26_22_10_1_61_50_39_63_52_41_30_19_65_54_43_67;DT(2.5,1.5,8)rec[-37]_rec[-75];DT(1.5,2.5,8)rec[-36]_rec[-38]_rec[-74];DT(4,1,8)rec[-35]_rec[-73];DT(3,3,8)rec[-34]_rec[-36]_rec[-72];DT(2,4,8)rec[-33]_rec[-71];DT(1,5,8)rec[-32]_rec[-70];DT(6,1,8)rec[-31]_rec[-69];DT(5,2,8)rec[-30]_rec[-68];DT(4,3,8)rec[-29]_rec[-67];DT(3,4,8)rec[-28]_rec[-66];DT(2,5,8)rec[-27]_rec[-65];DT(7,2,8)rec[-26]_rec[-64];DT(6,3,8)rec[-25]_rec[-63];DT(5,4,8)rec[-24]_rec[-62];DT(4,5,8)rec[-23]_rec[-61];DT(7,4,8)rec[-22]_rec[-60];DT(6,5,8)rec[-21]_rec[-59];DT(2,0,8)rec[-20]_rec[-58];DT(2,1,8)rec[-19];DT(1,2,8)rec[-18]_rec[-57];DT(4,0,8)rec[-17]_rec[-55];DT(3,1,8)rec[-16]_rec[-54]_rec[-57];DT(1.5,3.5,8)rec[-15]_rec[-52];DT(1.5,3.5,9)rec[-14];DT(0.5,4.5,8)rec[-13];DT(6,0,8)rec[-12]_rec[-50];DT(5,1,8)rec[-11]_rec[-49];DT(4,2,8)rec[-10]_rec[-48]_rec[-53];DT(6,2,8)rec[-9]_rec[-47];DT(5,3,8)rec[-8]_rec[-46];DT(4,4,8)rec[-7]_rec[-45];DT(3,5,8)rec[-6]_rec[-44];DT(2,6,8)rec[-5]_rec[-43];DT(6,4,8)rec[-4]_rec[-42];DT(5,5,8)rec[-3]_rec[-41];DT(4,6,8)rec[-2]_rec[-40];DT(6,6,8)rec[-1]_rec[-39];TICK;R_0_21_9_38_28_17_6_62_51_40_29_18_74_64_53_42_75_66;RX_13_14_3_37_26_22_10_1_61_50_39_63_52_41_30_19_65_54_43_67;TICK;CX_13_7_14_21_3_9_37_31_26_32_22_28_10_17_1_6_61_55_39_46_19_25_70_74_43_49_72_75_67_73;TICK;CX_13_20_37_44_9_16_22_15_61_68_28_34_17_24_69_74_19_12_71_75_43_36_67_60;TICK;CX_3_0_26_31_9_4_32_27_22_16_10_5_39_33_28_23_17_11;TICK;CX_3_8_26_20_9_15_4_0_44_38_22_27_50_55_39_45_28_33_17_23_68_62_56_51_46_40_34_29_24_18_63_69_52_57_41_47_30_35_70_64_58_53_48_42_65_71_54_59_72_66;TICK;CX_7_14_8_3_4_10_27_21_16_9_50_44_45_38_39_32_69_62_63_56_57_51_52_46_47_40_41_34_35_29_30_24_25_18_71_64_65_58_59_53_54_48_49_42_73_66;TICK;CX_8_14_3_0_31_38_15_21_32_27_5_10_55_62_50_56_45_51_33_40_23_29_11_18_63_70_57_64_52_58_47_53_41_48_35_42_30_36_65_72_59_66_54_60;TICK;CX_20_14_8_3_15_9_27_22_16_10_5_1_33_28_23_17_11_6;TICK;CX_14_21_3_9_32_38_22_28_10_17_1_6_50_45_56_62_46_51_34_40_24_29_12_18_63_57_52_47_41_35_30_25_58_64_48_53_36_42_65_59_54_49_60_66;TICK;M_0_21_9_38_28_17_6_62_51_40_29_18_74_64_53_42_75_66;MX_13_14_3_37_26_22_10_1_61_50_39_63_52_41_30_19_65_54_43_67;DT(0.5,2.5,10)rec[-38]_rec[-76];DT(2.5,1.5,10)rec[-37]_rec[-75];DT(1.5,2.5,10)rec[-36]_rec[-74];DT(4,1,10)rec[-35]_rec[-73];DT(3,3,10)rec[-34]_rec[-72];DT(2,4,10)rec[-33]_rec[-71];DT(1,5,10)rec[-32]_rec[-70];DT(6,1,10)rec[-31]_rec[-69];DT(5,2,10)rec[-30]_rec[-68];DT(4,3,10)rec[-29]_rec[-67];DT(3,4,10)rec[-28]_rec[-66];DT(2,5,10)rec[-27]_rec[-65];DT(7,2,10)rec[-26]_rec[-64];DT(6,3,10)rec[-25]_rec[-63];DT(5,4,10)rec[-24]_rec[-62];DT(4,5,10)rec[-23]_rec[-61];DT(7,4,10)rec[-22]_rec[-60];DT(6,5,10)rec[-21]_rec[-59];DT(2,0,10)rec[-20]_rec[-58];DT(2,1,10)rec[-19];DT(1,2,10)rec[-18]_rec[-56];DT(4,0,10)rec[-17]_rec[-55];DT(3,1,10)rec[-16]_rec[-54];DT(1.5,3.5,10)rec[-15]_rec[-52];DT(1.5,3.5,11)rec[-14];DT(0.5,4.5,10)rec[-13];DT(6,0,10)rec[-12]_rec[-50];DT(5,1,10)rec[-11]_rec[-49];DT(4,2,10)rec[-10]_rec[-48]_rec[-53];DT(6,2,10)rec[-9]_rec[-47];DT(5,3,10)rec[-8]_rec[-46];DT(4,4,10)rec[-7]_rec[-45];DT(3,5,10)rec[-6]_rec[-44];DT(2,6,10)rec[-5]_rec[-43];DT(6,4,10)rec[-4]_rec[-42];DT(5,5,10)rec[-3]_rec[-41];DT(4,6,10)rec[-2]_rec[-40];DT(6,6,10)rec[-1]_rec[-39];TICK;MPP_X68*Z12*X69*Z25*X70*Z36*X71*Z49*X72*Z60*Y73_X7*X20_X8*X15*X4*X16_X31*X44_X32*X27*X45*X33*X46_X5*X23*X11*X24_X34*X47*X35*X48;DT(2,0,12)rec[-6]_rec[-27];DT(1,2,12)rec[-5]_rec[-25];DT(4,0,12)rec[-4]_rec[-24];DT(4,2,12)rec[-3]_rec[-17]_rec[-22];DT(1,4,12)rec[-2];DT(4,4,12)rec[-1]_rec[-14];OI(0)rec[-7]_rec[-161]_rec[-165]_rec[-166]_rec[-167]_rec[-168]_rec[-169]_rec[-173]_rec[-174]_rec[-177]_rec[-178]_rec[-180]_rec[-181]_rec[-184]_rec[-185]_rec[-186]_rec[-189]_rec[-190]_rec[-195]_rec[-196]_rec[-199]_rec[-200]_rec[-201];TICK;MPP_Z7*Z20*Z8*Z15*Z32*Z27_Z4*Z16*Z5*Z23_X44*X55*X45*X56_Z33*Z46*Z34*Z47_Z11*Z24*Z12*Z25_Z69*Z70_Z35*Z48*Z36*Z49_X58*X71*X59*X72_X60*X73;DT(2.5,1.5,13)rec[-9]_rec[-53];DT(2,4,13)rec[-8]_rec[-49];DT(5,1,13)rec[-7]_rec[-27];DT(4,3,13)rec[-6]_rec[-45];DT(2,5,13)rec[-5]_rec[-43];DT(7,2,13)rec[-4]_rec[-42];DT(4,5,13)rec[-3]_rec[-39];DT(6,4,13)rec[-2]_rec[-20];DT(6,6,13)rec[-1]_rec[-17];TICK;MPP_X20*X31*X32_Z8*Z4_X15*X27*X16*X33*X23*X34_Z5*Z11_X55*X68_Z45*Z56*Z46*Z57_X24*X35*X25*X36_Z47*Z58*Z48*Z59_Z71*Z72;DT(3,1,14)rec[-9]_rec[-41];DT(0.5,2.5,14)rec[-8]_rec[-63];DT(1.5,3.5,14)rec[-7]_rec[-39];DT(1,5,14)rec[-6]_rec[-57];DT(6,0,14)rec[-5]_rec[-37];DT(5,2,14)rec[-4]_rec[-55];DT(3,5,14)rec[-3]_rec[-31];DT(5,4,14)rec[-2]_rec[-49];DT(7,4,14)rec[-1]_rec[-47];TICK;MPP_Z31*Z44*Z32*Z45_Z15*Z16_Z27*Z33_Z55*Z68*Z56*Z69_Z23*Z34*Z24*Z35_X46*X57*X47*X58_X12*X25_X48*X59*X49*X60;DT(4,1,15)rec[-8]_rec[-68];DT(1.5,2.5,15)rec[-7]_rec[-69];DT(3,3,15)rec[-6]_rec[-67];DT(6,1,15)rec[-5]_rec[-64];DT(3,4,15)rec[-4]_rec[-61];DT(5,3,15)rec[-3]_rec[-41];DT(2,6,15)rec[-2]_rec[-38];DT(5,5,15)rec[-1]_rec[-36];TICK;MPP_X56*X69*X57*X70_X36*X49_Z59*Z72*Z60*Z73;DT(6,2,16)rec[-3]_rec[-45];DT(4,6,16)rec[-2]_rec[-38];DT(6,5,16)rec[-1]_rec[-57];TICK;MPP_Z57*Z70*Z58*Z71;DT(6,3,17)rec[-1]_rec[-62]
        }{Click here to open an example end-to-end circuit in Crumble.}
        Shaded regions indicate error rate hypotheses with a likelihood within a factor of 1000 of the max likelihood hypothesis, given the sampled data.
    }
    \label{fig:end-to-end-error}
\end{figure}

In \fig{end-to-end-error} we show the results of end-to-end stabilizer simulations of $S|+\rangle$ cultivation under uniform depolarizing noise (see \app{noise-model}).
These simulations start with injection and end in a distance 15 matchable code.
They don't include consuming the state to perform gate teleportation.

From the results of the end-to-end simulations, it's clear that magic state cultivation is extremely sensitive to the physical noise strength $p$ and that this impacts the optimal choice of $d_1$.
For example, at $p=2 \cdot 10^{-3}$ it would take 1000 attempts per kept shot for $d_1=5$ cultivation to break even with $d_1=3$ cultivation using 20 attempts per kept shot.
At $p=10^{-3}$ the $d_1=5$ cultivation beats $d_1=3$ cultivation by three orders of magnitude using 100 attempts per kept shot (versus 4 attempts per kept shot for $d_1=3$).
At $p=10^{-3}/2$, the $d_1=5$ cultivation is four or five orders of magnitude better than the $d_1=3$ cultivation, while needing only 10 attempts per kept shot (versus 2).
It's plausible at $p=10^{-3}/2$ that even $d_1=7$ cultivation is becoming viable, but we didn't simulate this case.

\begin{figure}
    \centering
    \resizebox{\linewidth}{!}{
        \includegraphics{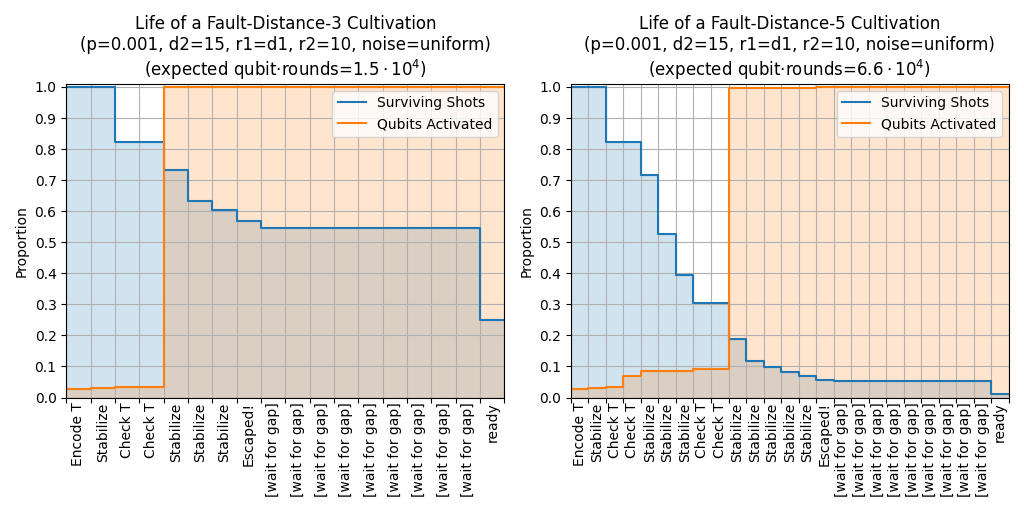}
    }
    \caption{
        \textbf{Growth and survival during magic state cultivation}.
        Each column is a labelled circuit cycle, which performs roughly ten layers of one/two qubit gates and ends with one or two layers of dissipative gates.
        The proportion of activated qubits starts small during the injection stage, gradually increases during the cultivation stage, and then jumps for the escape stage.
        Survival rates drop continuously due to postselected detectors failing, then hold steady for 10 cycles while waiting for the complementary gap to be computed.
        Expected qubit·rounds is computed by integrating the product of the survival rate times the qubit proportion, then multiplying by the total number of qubits and dividing by the final survival rate.
        This is slightly optimistic, due to ignoring factors like packing efficiency.
    }
    \label{fig:cultivation-lifetime}
\end{figure}

To quantify the spacetime volume of the construction, we performed simulations that eagerly discarded shots as soon as a postselected detector triggered.
We also tracked how many qubits had been activated as the simulation progressed.
The result is \fig{cultivation-lifetime}, which shows the gradual loss of shots alongside the gradual (then sudden) gain of qubits.
The integral of the product of the surviving shots and activated qubits over time gives a simple spacetime cost for the construction.
We make slightly more pessimistic assumptions for the volumes shown in \fig{historical_progression}, to account for packing constraints.

\begin{figure}
    \centering
    \resizebox{\linewidth}{!}{
        \includegraphics{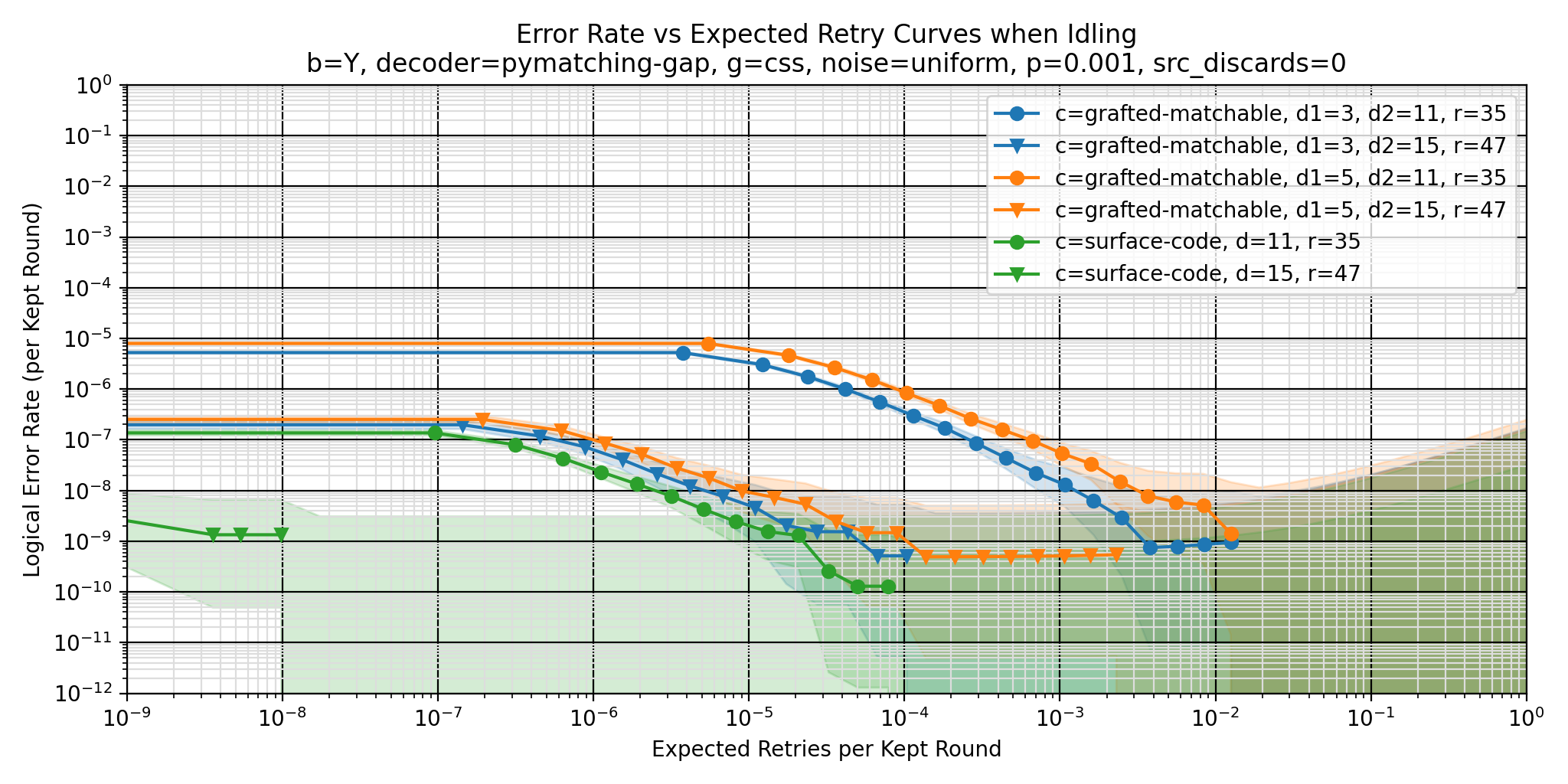}
    }
    \caption{
        \textbf{Cost of idling} after the escape stage ends.
        Shows that the $d=15$ grafted matchable code we currently end in is roughly equivalent to a $d=11$ surface code.
        Suggests our construction would be improved by ending in a normal surface code.
        Indicates that maintaining a $10^{-9}$ fault rate after escaping would require incurring around $10^{-4}$ additional retries per additional round of storing the output state (at a noise strength of $10^{-3}$).
        Shaded regions indicate error rate hypotheses with a likelihood within a factor of 1000 of the max likelihood hypothesis, given the sampled data.
    }
    \label{fig:idling-gap}
\end{figure}

Finally, in \fig{idling-gap}, we quantify the reliability of the grafted matchable code that our end-to-end simulations end in.
They show that the $d_2=15$ grafted matchable code is comparable to a $d=11$ surface code.
They further show that at $d_2=15$, performing light postselection incurring around a $10^{-4}$ chance of retry per round is sufficient to maintain an error rate of $10^{-9}$ per round.
This would be acceptable in contexts where the surrounding computation can be easily retried (such as inputs into a following magic state factory), but suggests we should grow to a larger code distance in contexts where the state will be used unconditionally.
As a result, we expect future work to adjust the escape stage (e.g. having it grow to a larger code in two medium sized steps instead of to $d_2=15$ in one big step).
We expect the cost of cultivation to remain roughly similar, despite these adjustments, because targeting a larger final distance loosens the packing constraints on the earlier stages.

\section{Conclusion}
\label{sec:conclusion}

In this work, we presented a construction for preparing magic states that we call magic state cultivation.
Cultivation achieves target error rates between $10^{-4}$ and $10^{-9}$ with spacetime costs an order of magnitude lower than in prior work.
Cultivation is also extremely responsive to improvements in noise strength.
For example, a 2x noise strength improvement (from $10^{-3}$ to $5 \cdot 10^{-4}$) becomes a 50x logical error rate improvement (from $2 \cdot 10^{-9}$ to $4 \cdot 10^{-11}$) and a 10x cost reduction.
This suggests that even modest ongoing improvements to physical qubits could result in cultivation outpacing the needs of quantum computers in practice, obsoleting magic state distillation.

Our construction, though more refined than prior work, is still rough.
The qubit utilization in \fig{cat-check-d5} is very low; maybe it can be increased.
Our escape stage is horribly complicated and inelegant; maybe it can be streamlined.
Maybe a sufficiently good color code decoder could be written, so that the escape stage could simply grow the color code.
Maybe the color code stabilizer measurement cycle can be optimized further.
Maybe non-local connectivity could be used to accelerate cultivation on non-planar architectures.
Maybe a different code with transversal Clifford gates could be used.
Maybe--- suffice it to say there's many opportunities for others to improve on our work.

In this work, we only simulated the production of T states.
T states are produced in order to perform T gates.
A major way to improve on our results would be to perform end-to-end simulations of a logical T \emph{gate}, rather than just the T state.
Ideally these simulations would include the parallel attempts at producing a T state, real time retries when postselection fails, the lattice surgery operations needed to attach the T state to a target logical qubit, and the real time Clifford correction needed to complete the gate teleportation.
\fig{historical-comparison} suggests that cultivating a T state is roughly as expensive as performing a CNOT gate of a similar reliability.
We expect the full T gate construction to be around twice this cost but can't say for sure without a full simulation.

Another major limitation of our simulations is our reliance on \ass{tbys} to extrapolate the behavior of T gate circuits from the behavior of S gate circuits.
It's clear from the changing ratio in \fig{t-vs-s} that this assumption has a limited regime of applicability.
Ideally, we would directly simulate the T gate circuits.
Hopefully, new simulation techniques can be found to make this tractable.
Additionally, we should note that our simulations assume a digitized noise model.
This may not be reasonable for low fault distance processes, like injection, that are present in our circuits.
Ultimately, experiments on real quantum computers are needed to decide how well the construction works.
The injection and cultivation stages are small enough to fit on existing quantum computers, so experiments are a surprisingly tractable avenue for benchmarking much of the magic state cultivation process.

Cultivation marks a major reduction in the expected cost of fault tolerant T gates, but it's difficult to guess how this will impact the overall cost of quantum algorithms.
For example, consider an algorithm implementation that used ripple carry adders to minimize T count (such as \cite{gidney2021factor}).
The first order effect of cheap T states is the adders get a bit cheaper.
The second order effect is the adders can be replaced by faster T-hungrier carry-lookahead adders (the Jevons paradox), completely changing the available spacetime trade-offs.
The first order effect is easy to estimate, but irrelevant compared to the second order effect.
We look forward to seeing how large the second order effects are, as the details of quantum algorithms change in response to the existence of magic state cultivation.

\section{Contributions}

Craig Gidney built the construction, simulated its performance, and wrote the paper.
Cody Jones made many suggestions for improving the construction, and supervised the project.
Noah Shutty saved the project by showing that initial discouraging simulations of the escape stage (which suggested it would be a fatal bottleneck) were limitations of the decoder rather than inherent limitations of the stage.

\section{Acknowledgements}

We thank the Google Quantum AI team, for providing an environment where this work was possible.
We thank Matt McEwen, Alexis Morvan, and Oscar Higgott for comments that improved the paper.
We thank Michael Newman for pushing us to include the CNOT gate comparison.

\printbibliography

\appendix

\clearpage
\section{Noise Model}
\label{app:noise-model}

All circuits in this paper were simulated with a uniform depolarizing noise model, defined in \fig{noise_model}, so they could more easily be compared to prior work.

\begin{figure}[H]
    \centering
    \begin{tabular}{|r|l|}
    \hline
    Noise channel & Probability distribution of effects
    \\
    \hline
    $\text{MERR}(p)$ & $\begin{aligned}
        1-p &\rightarrow \text{(report previous measurement correctly)}
        \\
        p &\rightarrow \text{(report previous measurement incorrectly; flip its result)}
    \end{aligned}$
    \\
    \hline
    $\text{XERR}(p)$ & $\begin{aligned}
        1-p &\rightarrow I
        \\
        p &\rightarrow X
    \end{aligned}$
    \\
    \hline
    $\text{ZERR}(p)$ & $\begin{aligned}
        1-p &\rightarrow I
        \\
        p &\rightarrow Z
    \end{aligned}$
    \\
    \hline
    $\text{DEP1}(p)$ & $\begin{aligned}
        1-p &\rightarrow I
        \\
        p/3 &\rightarrow X
        \\
        p/3 &\rightarrow Y
        \\
        p/3 &\rightarrow Z
    \end{aligned}$
    \\
    \hline
    $\text{DEP2}(p)$ & $\begin{aligned}
        1-p &\rightarrow I \otimes I
        &\;\;
        p/15 &\rightarrow I \otimes X
        &\;\;
        p/15 &\rightarrow I \otimes Y
        &\;\;
        p/15 &\rightarrow I \otimes Z
        \\
        p/15 &\rightarrow X \otimes I
        &\;\;
        p/15 &\rightarrow X \otimes X
        &\;\;
        p/15 &\rightarrow X \otimes Y
        &\;\;
        p/15 &\rightarrow X \otimes Z
        \\
        p/15 &\rightarrow Y \otimes I
        &\;\;
        p/15 &\rightarrow Y \otimes X
        &\;\;
        p/15 &\rightarrow Y \otimes Y
        &\;\;
        p/15 &\rightarrow Y \otimes Z
        \\
        p/15 &\rightarrow Z \otimes I
        &\;\;
        p/15 &\rightarrow Z \otimes X
        &\;\;
        p/15 &\rightarrow Z \otimes Y
        &\;\;
        p/15 &\rightarrow Z \otimes Z
    \end{aligned}$
    \\
    \hline
    \end{tabular}
    \caption{
        \textbf{Definitions of various noise channels}.
        Used by \fig{noise_model}.
    }
    \label{fig:noise_channels}
\end{figure}

\begin{figure}[H]
    \centering
    \begin{tabular}{|r|l|}
    \hline
    Ideal gate & Noisy gate
    \\
    \hline
    (single qubit unitary, including idle) $U_1$ & $\text{DEP1}(p) \cdot U_1$
    \\
    $CX$ & $\text{DEP2}(p) \cdot CX$
    \\
    \hline
    (reset) $R_X$ & $\text{ZERR}(p) \cdot R_X$
    \\
    $R_Z$ & $\text{XERR}(p) \cdot R_Z$
    \\
    $M_X$ & $\text{DEP1}(p) \cdot \text{MERR}(p) \cdot M_X$
    \\
    $M_Z$ & $\text{DEP1}(p) \cdot \text{MERR}(p) \cdot M_Z$
    \\
    \hline
    \end{tabular}
    \caption{
        \textbf{The uniform depolarizing circuit noise model}, used by simulations in this paper.
        Note $B \cdot A$ means $B$ is applied after $A$.
        Noise channels are defined in \fig{noise_channels}.
    }
    \label{fig:noise_model}
\end{figure}

\clearpage
\section{Chunked Circuit Construction}
\label{app:chunk}

To transform our construction into a circuit, and to iteratively improve upon the pieces of that circuit, we used a process similar to what's described in \cite{gidney2024ybasis}.
Every cycle of our construction was created, iterated, and optimized in ``chunks''.

A chunk is a quantum stabilizer circuit annotated with a list of ``stabilizer flows''~\cite{mcewenmidoutsurfaces2023}.
A stabilizer flow $A \xrightarrow{M} B$ has an input stabilizer $A$, an output stabilizer $B$, and a set of measurements $M$.
A circuit has the stabilizer flow $A \xrightarrow{M} B$ if it transforms $A$ into $B$ with a sign determined by the parity of $M$.
The flows of a chunk are similar to the signature of a function; they're a simplified and verifiable summary of what the chunk does and how it can be used.
For example, if a chunk is annotated with the flow $A \xrightarrow{m_1,m_2} 1$, that is an assertion that its circuit will measure the stabilizer $A$ and that the measurement result is recovered by computing $m_1 \oplus m_2$.
Stim~\cite{gidney2021stim} has a class \href{https://github.com/quantumlib/Stim/blob/main/doc/python_api_reference_vDev.md#stim.Flow}{stim.Flow} for representing flows, and a method \href{https://github.com/quantumlib/Stim/blob/main/doc/python_api_reference_vDev.md#stim.Circuit.has_flow}{stim.Circuit.has\_flow} for verifying that a given circuit implements a flow.

When two chunks are concatenated, we verify that the output stabilizers of flows declared by the first chunk match up exactly with the input stabilizers of flows declared by the second chunk.
We then attach the matching flows to each other, forming longer flows over the concatenated circuit.
Often this will produce flows entirely internal to the concatenated circuit, with an empty input and an empty output but a non-empty set of measurements.
This guarantees the set of measurements must have deterministic parity under noiseless execution, meaning those measurements form a detector.
In this way, flow annotations can be automatically converted into detector annotations while concatenating chunks.

Compared to other approaches for making circuits that we've tried, the chunk based approach has four main benefits: decoupling, contracts, artistic freedom, and micro-management.

By ``decoupling'' we mean that the details of a chunk can stay fixed when varying surrounding chunks.
For example, during development we tried both superdense~\cite{gidney2023colorcode} and Bell flagged~\cite{baireuther2019bellflaggedcolor} color code cycles.
A superdense cycle measures all stabilizers of the color code, while a Bell flagged cycle only measures the stabilizers in one basis.
When two superdense cycles are adjacent, all stabilizer measurements of the second cycle are compared to measurements in the first cycle in order to form detectors.
But if the first cycle is changed to a Bell flagged X basis cycle, then the Z stabilizer measurements of the superdense cycle will need to compare to something earlier.
When annotating the measurements to compare (as in a Stim circuit), this means the second cycle changes when the first cycle changes.
In a chunk based approached, varying the first chunk doesn't change the second chunk; the flows remain the same.
The differing comparisons only appear at a later stage, when automatically matching up flows between concatenated chunks.
Decoupling makes it easier to mix and match different approaches in each part of a construction.

By ``contracts'' we mean the ability to write a specification for what a chunk should do and then verify that the chunk does it.
The flows are of course a specification of what the circuit should do, but we can also write specifications for which flows are expected to be present.
For example, if a chunk is supposed to measure all the stabilizers of a code, we can write a unit test that iterates over the stabilizers of the code and searches for a corresponding empty-output flow in the chunk.
Having the ability to specify and verify these kinds of contracts makes debugging faster, because when two chunks don't match we can narrow down the problem by comparing them to the contract instead of to each other.

By ``artistic freedom'' we mean the ability to create chunks in a variety of ways.
For example, some of the chunks in our construction are small and irregular.
Writing code to generate these chunks is tedious and error-prone; it's more convenient to create them in a visual editor like \href{https://algassert.com/crumble}{Crumble}.
By contrast, huge repetitive circuits are easier to generate with code.
Another detail we vary is the amount of automation during chunk creation.
For example, the measurement sets of flows can be solved for via \href{https://github.com/quantumlib/Stim/blob/main/doc/python_api_reference_vDev.md#stim.Circuit.flow_generators}{Gaussian elimination}.
Solving the measurements in this way is convenient, and works well on chunks corresponding to a single cycle, but is quite slow and makes bad choices on multi-cycle circuits.
A chunk doesn't have to be made in one specific way; it has no memory of the strategy that was used to make it.
All that's required is that the flows agree with the circuit.
So we're free to choose the best creation strategy for each chunk.

By ``micro-management'' we mean that we have guaranteed control over the detector basis.
For example, if we have three measurements $A$, $B$, $C$ and we know $A=B=C$ then there are several different ways to express this as detectors.
We can define a detector $A \oplus B = 0$ and a detector $B \oplus C = 0$.
Alternatively we can define a detector $A \oplus B = 0$ and a detector $A \oplus C = 0$.
Or we could redundantly define three detectors: $A \oplus B = 0$, $A \oplus C = 0$, and $B \oplus C = 0$.
These choices are technically all equivalent, in that they all imply $A=B=C$, but often decoders require specific comparison structures to function correctly.
For example, a matching based decoder won't function if all measurements are compared to measurements from the third round, instead of to measurements from the previous round.
By having flows be specified as an explicit list, instead of inferred automatically from the circuit, we can guarantee these kinds of decoding requirements are met.

Another benefit of micro-management is that it avoids ambiguity.
For example, consider a toric code memory experiment circuit.
The toric code has two logical qubits.
Is the circuit intended to benchmark one of the logical qubits? The other? Both?
This can't be determined just by looking at the quantum operations performed by the circuit.
Similarly, the toric code has global spacelike invariants: without noise the product of all X measurements in a round would equal +1 (similarly for Z).
Is the circuit also benchmarking this invariant~\cite{gidney2022stability}?
Ignoring it?
Using it as a coarse global flag check?
It's impossible to tell.
As a result of this ambiguity, systems that attempt to fully automate the discovery of detectors/observables from a circuit, with no ability for a manual override, are inherently inflexible.
For chunks, the explicit list of flows is the manual override.

\clearpage
\section{Bonus Figures}
\label{app:additional-figures}

\begin{figure}[H]
    \centering
    \begin{adjustwidth}{-2.5cm}{-2.5cm}
        \resizebox{\linewidth}{!}{
            \includegraphics{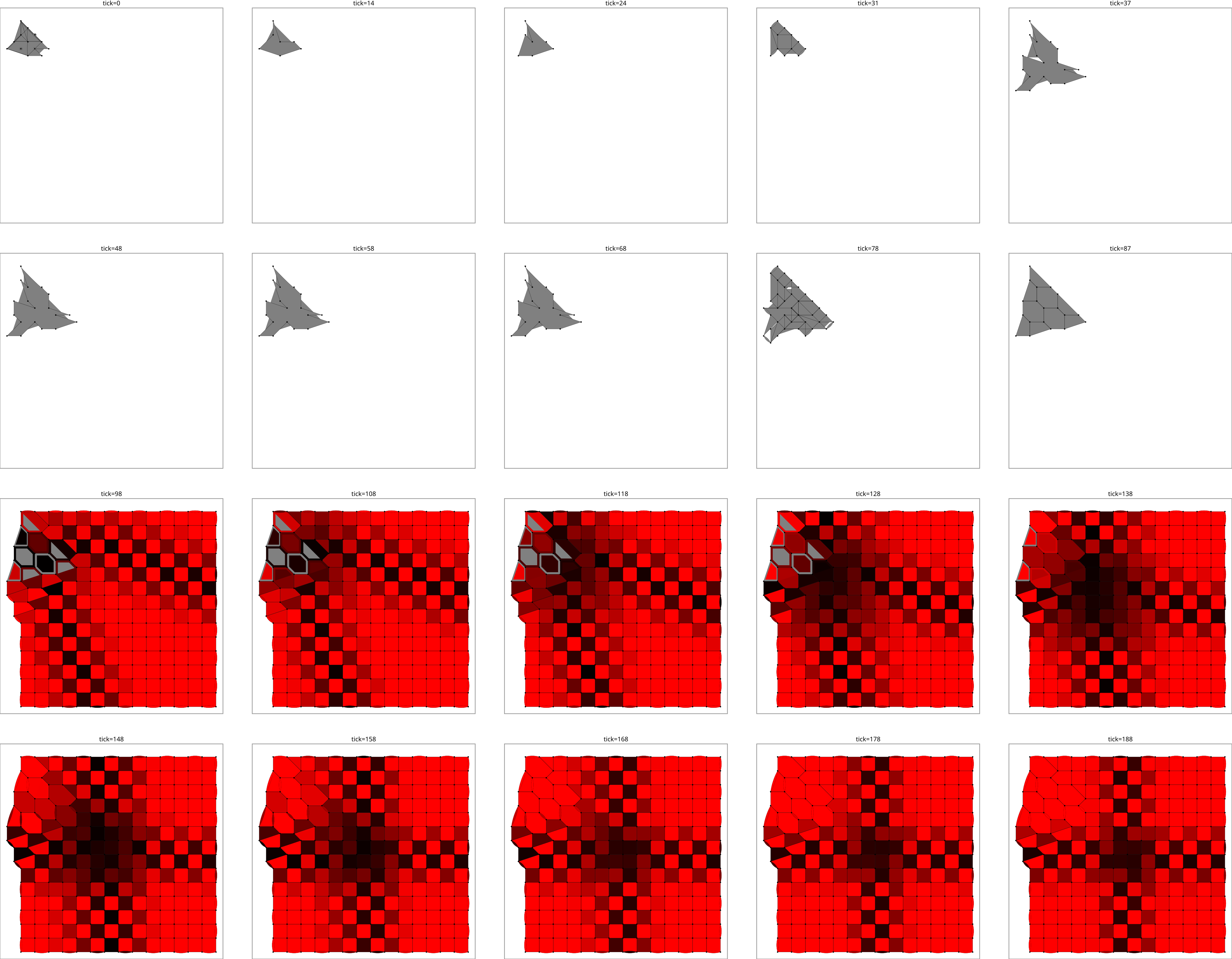}
        }
    \end{adjustwidth}
    \caption{
        \textbf{Decoder confidence given individual detection events}.
        Each outlined box is a detector slice from a circuit cycle of end-to-end cultivation.
        Polygons correspond to the support of instantaneous stabilizers checked by a detector.
        Color indicates decoder confidence (red = very confident, black = not confident) if that detector is excited, without any other detector being excited.
        Gray means the detector is postselected, so decoding never reaches a point of computing the confidence.
        Note how the dark region (the region of maximum uncertainty) shifts towards the middle of the patch over time.
        This is an indication of the logical state transitioning from being stored in the color code to being stored in the full code.
    }
    \label{fig:desaturation-gap}
\end{figure}

\begin{figure}
    \centering
    \begin{adjustwidth}{-2.0cm}{-2.0cm}
        \centering
        \resizebox{\linewidth}{!}{
            \includegraphics{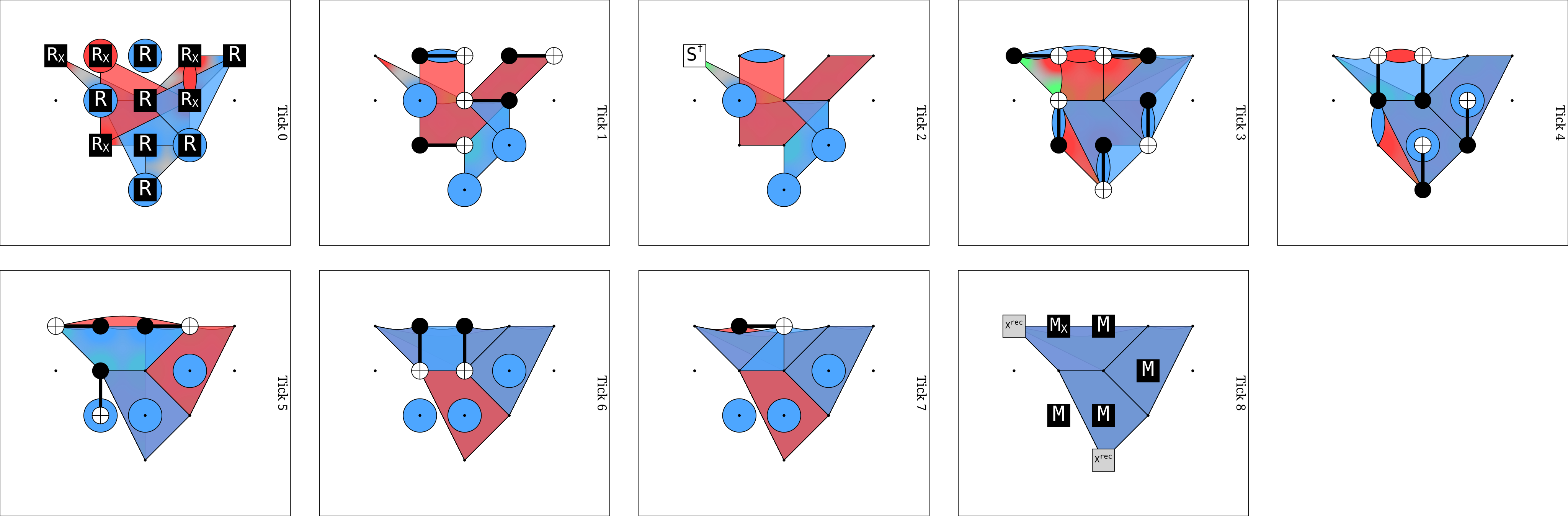}
        }
    \end{adjustwidth}
    \caption{
        \textbf{Detector slice diagram of Bell injection}, using XYZ=RGB color convention.
        Replace the S gate with a T gate to prepare $|T\rangle$ instead of $|i\rangle$.
        Prepares a single data qubit in the state $T^k|+\rangle$.
        This data qubit is then treated as a distance 1 color code, and the code is grown to distance 3 by initializing six new data qubits into Bell pairs that imply the values of the target color code's four stabilizers that don't touching the initial data qubit.
        Afterward, all six stabilizers of the distance 3 color code are measured to stabilize the code.
        Fast feedback is needed, to ensure the Z stabilizers are in the +1 eigenbasis during cultivation.
        \href{
            https://algassert.com/crumble\#circuit=Q(0,0)0;Q(1,0)1;Q(1,1)2;Q(1,2)3;Q(2,0)4;Q(2,1)5;Q(2,2)6;Q(2,3)7;Q(3,0)8;Q(3,1)9;Q(3,2)10;Q(4,0)11;POLYGON(0,0,1,0.25)8_11_10_5;POLYGON(0,1,0,0.25)7_10_5_2;POLYGON(1,0,0,0.25)0_8_5_2;POLYGON(1,0,1,0.25)2_7;POLYGON(1,0,1,0.25)5_10;POLYGON(1,0,1,0.25)11_8;TICK;R_6_2_7_5_10_11_4;RX_0_3_9_8_1;TICK;CX_1_4_3_6_8_11_9_5;TICK;S_DAG_0;TICK;CX_0_1_3_2_6_7_8_4_9_10;TICK;CX_2_1_5_4_7_6_10_9;TICK;CX_1_0_2_3_4_8;TICK;CX_1_2_4_5;TICK;CX_1_4;TICK;M_3_6_9_4;MX_1;DT(1,2,0)rec[-5];DT(2,2,0)rec[-4];DT(3,1,0)rec[-3];TICK;TICK;MPP_X0*X2*X5*X8;DT(1,0,1)rec[-1]_rec[-2];TICK;MPP_Z0*Z2*Z5*Z8;DT(2,0,2)rec[-1]_rec[-4];TICK;MPP_X2*X5*X7*X10;DT(1,1,3)rec[-1];TICK;MPP_Z2*Z5*Z7*Z10;DT(2,0,4)rec[-1]_rec[-6];TICK;MPP_X5*X8*X10*X11;DT(2,1,5)rec[-1];TICK;MPP_Z5*Z8*Z10*Z11;DT(2,1,6)rec[-1];TICK;MPP_Y0*Y2*Y5*Y7*Y8*Y10*Y11;OI(0)rec[-1]
        }{Click here to open this circuit in Crumble.}
    }
    \label{fig:d3-init-bell}
\end{figure}

\begin{figure}
    \begin{adjustwidth}{-2.0cm}{-2.0cm}
        \centering
        \resizebox{\linewidth}{!}{
            \includegraphics{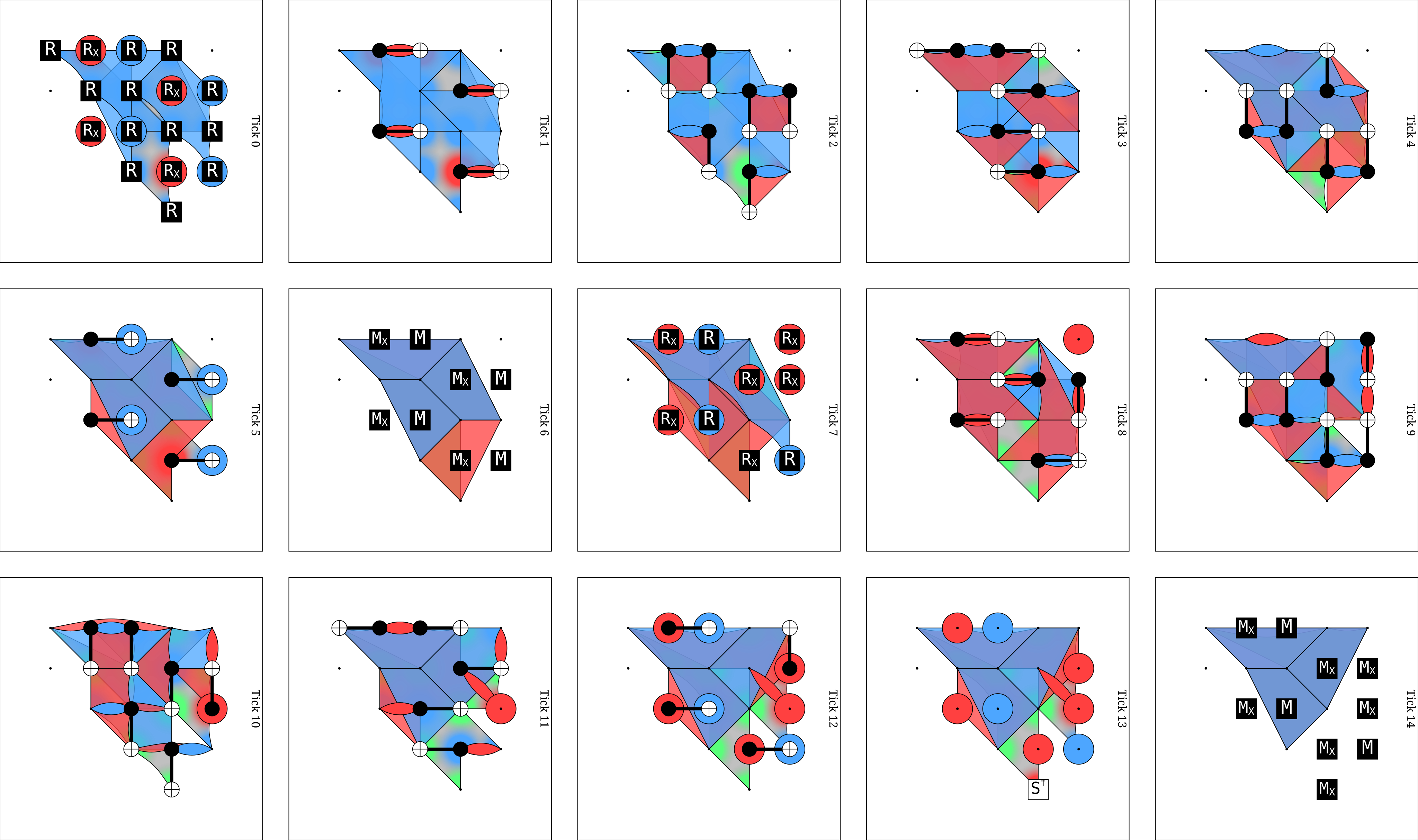}
        }
    \end{adjustwidth}
    \caption{
        \textbf{Detector slice diagram of teleport injection}, using XYZ=RGB color convention.
        Replace the S gate with a T gate to prepare $|T\rangle$ instead of $|i\rangle$.
        Creates a four-tile $[[8, 0, \infty]]$ color code by initializing its data qubits in the Z basis, and then measuring its X stabilizers.
        Then applies $T^k$ and $M_X$ to one of the qubits touched by only one tile, leaving behind a $[[7,1,3]]$ color code storing a $T^k|+\rangle$ state (up to deferrable Pauli Z feedback).
        \href{
            https://algassert.com/crumble\#circuit=Q(0,0)0;Q(1,0)1;Q(1,1)2;Q(1,2)3;Q(2,0)4;Q(2,1)5;Q(2,2)6;Q(2,3)7;Q(3,0)8;Q(3,1)9;Q(3,2)10;Q(3,3)11;Q(3,4)12;Q(4,0)13;Q(4,1)14;Q(4,2)15;Q(4,3)16;POLYGON(0,0,1,0.25)10_15_8_5;POLYGON(0,1,0,0.25)7_10_5_2;POLYGON(1,0,0,0.25)7_12_15_10;POLYGON(1,0,0,0.25)2_5_8_0;TICK;R_0_8_15_10_5_2_12_7_6_4_16_14;RX_3_1_9_11;TICK;CX_1_4_3_6_9_14_11_16;TICK;CX_1_2_4_5_6_7_9_10_11_12_14_15;TICK;CX_1_0_4_8_6_10_9_5_11_7;TICK;CX_3_2_6_5_9_8_11_10_16_15;TICK;CX_1_4_3_6_9_14_11_16;TICK;M_6_4_16_14;MX_3_1_9_11;DT(2,2,0)rec[-8];DT(2,0,0)rec[-7];DT(4,3,0)rec[-6];DT(4,1,0)rec[-5];TICK;R_6_4_16;RX_3_1_9_11_14_13;TICK;CX_1_4_3_6_9_5_11_16_14_15;TICK;CX_3_2_6_5_9_8_11_10_13_14_16_15;TICK;CX_1_2_4_5_6_7_9_10_11_12_15_14;TICK;CX_1_0_4_8_6_10_9_14_11_7;TICK;CX_1_4_3_6_11_16_14_13;TICK;S_DAG_12;TICK;M_16_4_6;MX_14_9_15_12_1_3_11;DT(4,3,1)rec[-10];DT(2,0,1)rec[-9];DT(2,2,1)rec[-8];DT(4,1,1)rec[-7];DT(4,2,1)rec[-5];DT(3,1,1)rec[-5]_rec[-6]_rec[-12];DT(1,0,1)rec[-3]_rec[-13];DT(1,2,1)rec[-2]_rec[-14];DT(3,3,1)rec[-1]_rec[-11];TICK;POLYGON(0,0,1,0.25)13_8_5_10;POLYGON(0,1,0,0.25)7_10_5_2;POLYGON(1,0,0,0.25)2_5_8_0;TICK;MPP_X0*X2*X5*X8;DT(1,0,2)rec[-1]_rec[-4];TICK;MPP_Z0*Z2*Z5*Z8;DT(0,0,3)rec[-1];TICK;MPP_X2*X5*X7*X10;DT(1,2,4)rec[-1]_rec[-5];TICK;MPP_Z2*Z5*Z7*Z10;DT(1,1,5)rec[-1];TICK;MPP_X5*X8*X10*X13;DT(3,1,6)rec[-1]_rec[-11]_rec[-12];TICK;MPP_Z5*Z8*Z10*Z13;DT(2,1,7)rec[-1];TICK;MPP_Y0*Y2*Y5*Y7*Y8*Y10*Y13;OI(0)rec[-1]_rec[-11]_rec[-12]_rec[-14]_rec[-15]_rec[-17]_rec[-18]_rec[-20]_rec[-22]_rec[-23]
        }{Click here to open this circuit in Crumble.}
    }
    \label{fig:d3-init-teleport}
\end{figure}

\begin{figure}
    \centering
    \resizebox{\linewidth}{!}{
        \includegraphics{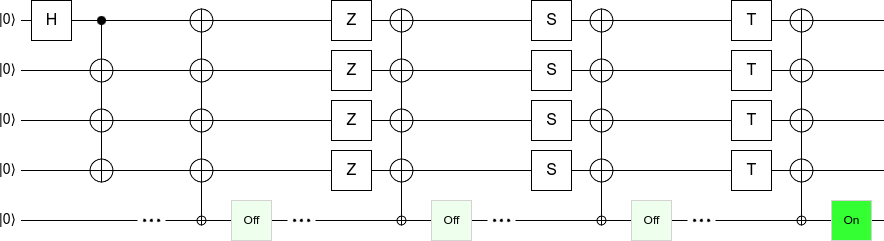}
    }
    \caption{
        \textbf{Danger of using S as a proxy for T}.
        A simple example of why \ass{tbys} doesn't hold in general.
        Replacing Z or S gates with T gates can cause a circuit to behave very differently.
        \href{
            https://algassert.com/quirk\#circuit=\%7B\%22cols\%22\%3A\%5B\%5B\%22H\%22\%5D\%2C\%5B\%22\%E2\%80\%A2\%22\%2C\%22X\%22\%2C\%22X\%22\%2C\%22X\%22\%5D\%2C\%5B1\%2C1\%2C1\%2C1\%2C\%22\%E2\%80\%A6\%22\%5D\%2C\%5B\%22X\%22\%2C\%22X\%22\%2C\%22X\%22\%2C\%22X\%22\%2C\%22\%E2\%8A\%96\%22\%5D\%2C\%5B1\%2C1\%2C1\%2C1\%2C\%22Chance\%22\%5D\%2C\%5B1\%2C1\%2C1\%2C1\%2C\%22\%E2\%80\%A6\%22\%5D\%2C\%5B\%22Z\%22\%2C\%22Z\%22\%2C\%22Z\%22\%2C\%22Z\%22\%5D\%2C\%5B\%22X\%22\%2C\%22X\%22\%2C\%22X\%22\%2C\%22X\%22\%2C\%22\%E2\%8A\%96\%22\%5D\%2C\%5B1\%2C1\%2C1\%2C1\%2C\%22Chance\%22\%5D\%2C\%5B1\%2C1\%2C1\%2C1\%2C\%22\%E2\%80\%A6\%22\%5D\%2C\%5B\%22Z\%5E\%C2\%BD\%22\%2C\%22Z\%5E\%C2\%BD\%22\%2C\%22Z\%5E\%C2\%BD\%22\%2C\%22Z\%5E\%C2\%BD\%22\%5D\%2C\%5B\%22X\%22\%2C\%22X\%22\%2C\%22X\%22\%2C\%22X\%22\%2C\%22\%E2\%8A\%96\%22\%5D\%2C\%5B1\%2C1\%2C1\%2C1\%2C\%22Chance\%22\%5D\%2C\%5B1\%2C1\%2C1\%2C1\%2C\%22\%E2\%80\%A6\%22\%5D\%2C\%5B\%22Z\%5E\%C2\%BC\%22\%2C\%22Z\%5E\%C2\%BC\%22\%2C\%22Z\%5E\%C2\%BC\%22\%2C\%22Z\%5E\%C2\%BC\%22\%5D\%2C\%5B\%22X\%22\%2C\%22X\%22\%2C\%22X\%22\%2C\%22X\%22\%2C\%22\%E2\%8A\%96\%22\%5D\%2C\%5B1\%2C1\%2C1\%2C1\%2C\%22Chance\%22\%5D\%5D\%7D
        }{Click here to open this circuit in Quirk.}
    }
    \label{fig:t-differs-from-s-and-z}
\end{figure}

\begin{figure}
    \centering
    \resizebox{\linewidth}{!}{
        \includegraphics{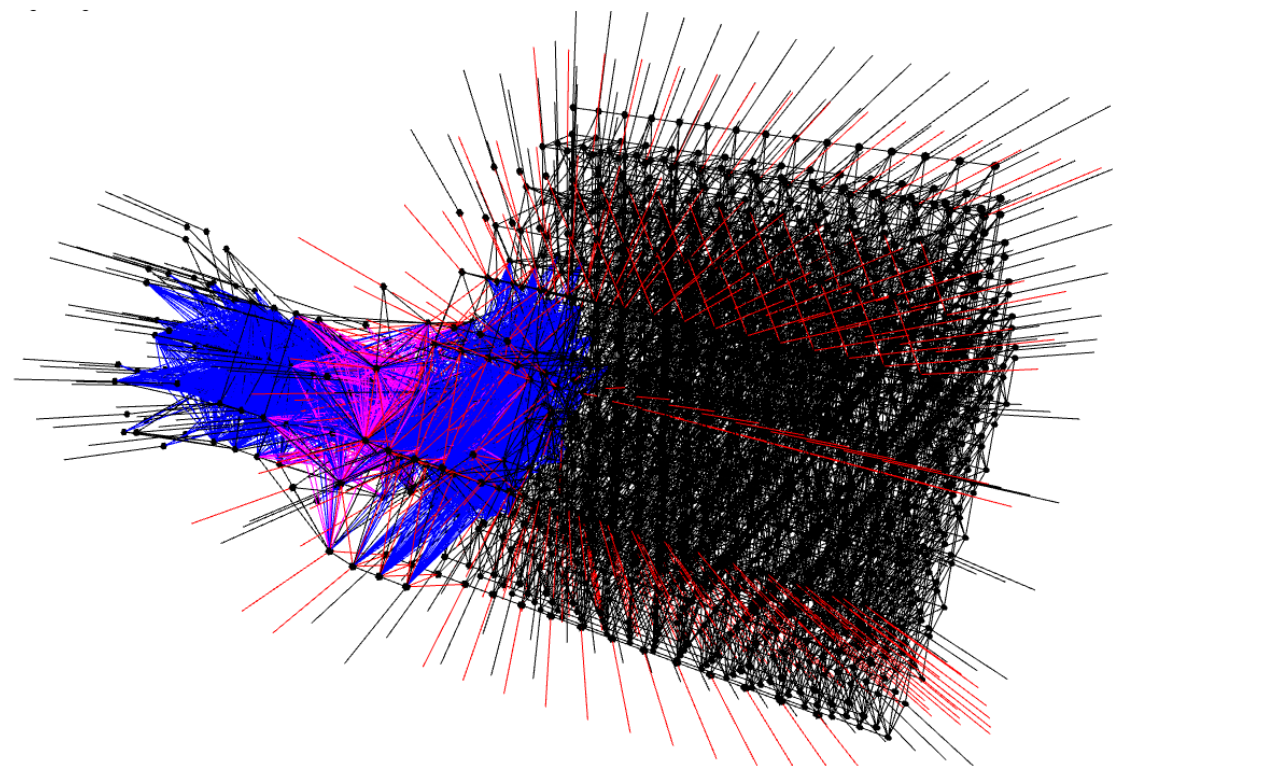}
    }
    \caption{
        \textbf{Detector error model of magic state cultivation}.
        Black circles are detectors.
        Black lines are the X and Z parts of matchable errors (errors that cause at most two X basis detection events and at most two Z basis detection events).
        Red lines are matchable errors that flip the logical X or Z observable.
        Blue lines meeting at a common point are hyper errors (errors that aren't decodable by matching).
        Magenta lines are hyper errors that flip the logical X or Z observable.
        The blue blob is a dense nest of hyper errors, corresponding to regions of the cultivation process where a color code is present and non-negligible postselection costs are being incurred.
    }
    \label{fig:dem-diagram}
\end{figure}

\begin{figure}
    \begin{adjustwidth}{-2.5cm}{-2.5cm}
        \centering
        \resizebox{0.48\linewidth}{!}{
            \includegraphics{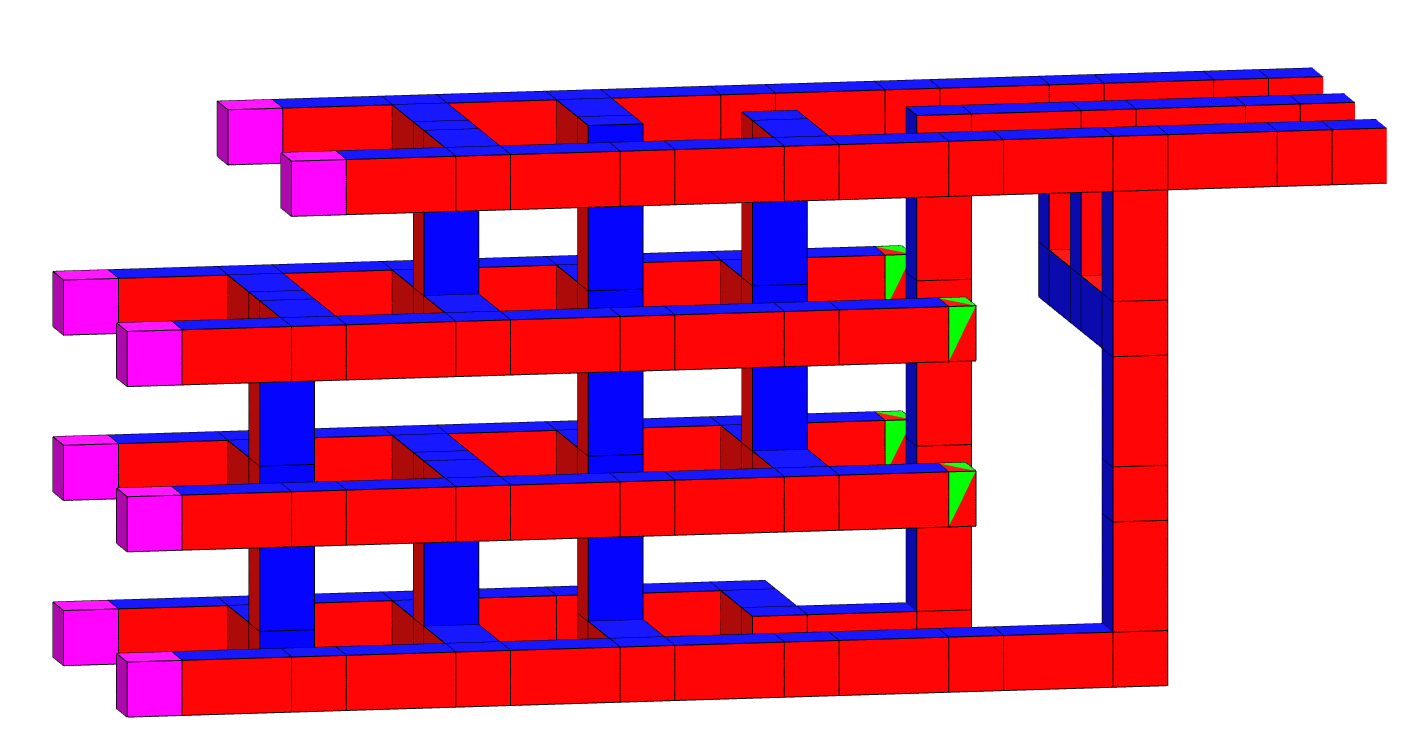}
        }
        \resizebox{0.48\linewidth}{!}{
            \includegraphics{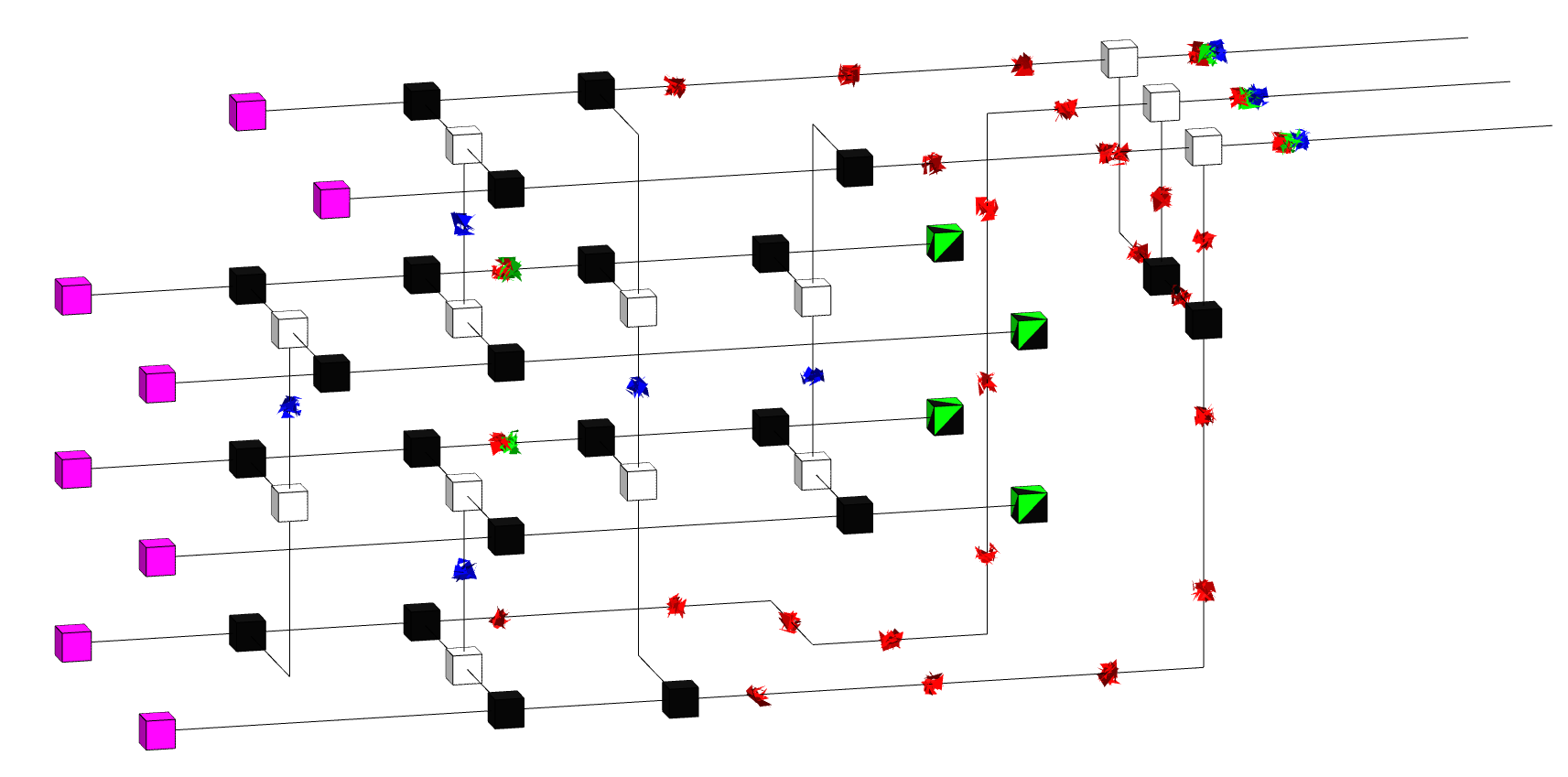}
        }
    \end{adjustwidth}
    \caption{
        \textbf{Layout of an 8T-to-1CCZ distillation factory}.
        A more optimized version of the factory from \cite{gidney2019catalyzeddistillation}.
        Left: spacetime defect diagram of the factory.
        Pink boxes are noisy T state inputs, presumably created by magic state cultivation.
        Red surfaces are X type boundaries.
        Blue surfaces are Z type boundaries.
        Green-and-red boxes are transversal X basis measurement or inplace Y basis measurement~\cite{gidney2024ybasis}, depending on feedback from earlier measurements.
        Right: ZX diagram of the factory, with annotated error sensitivities.
        White boxes are X spiders, black boxes are Z spiders, magenta boxes are $\pi/4$ X spiders, and green+black boxes are Z spiders with an angle of 0 or $\pi/2$ depending on feedback.
        Red (X), green (Y), and blue (Z) bursts on edges correspond to locations where a topological Pauli error of that type can cause the factory to silently fail.
        When a location only has a red burst, or only has a blue burst, it's partially protected by the distillation process.
        The pipe at that location can be squished to save qubit·rounds, as in \cite{litinski2019notascostly}.
        Crossbars only have blue bursts due to the usage of Temporally Encoded Lattice Surgery~\cite{chamberland2022tels}.
    }
    \label{fig:ccz-factory}
\end{figure}

\begin{figure}
    \centering
    \resizebox{\linewidth}{!}{
        \includegraphics{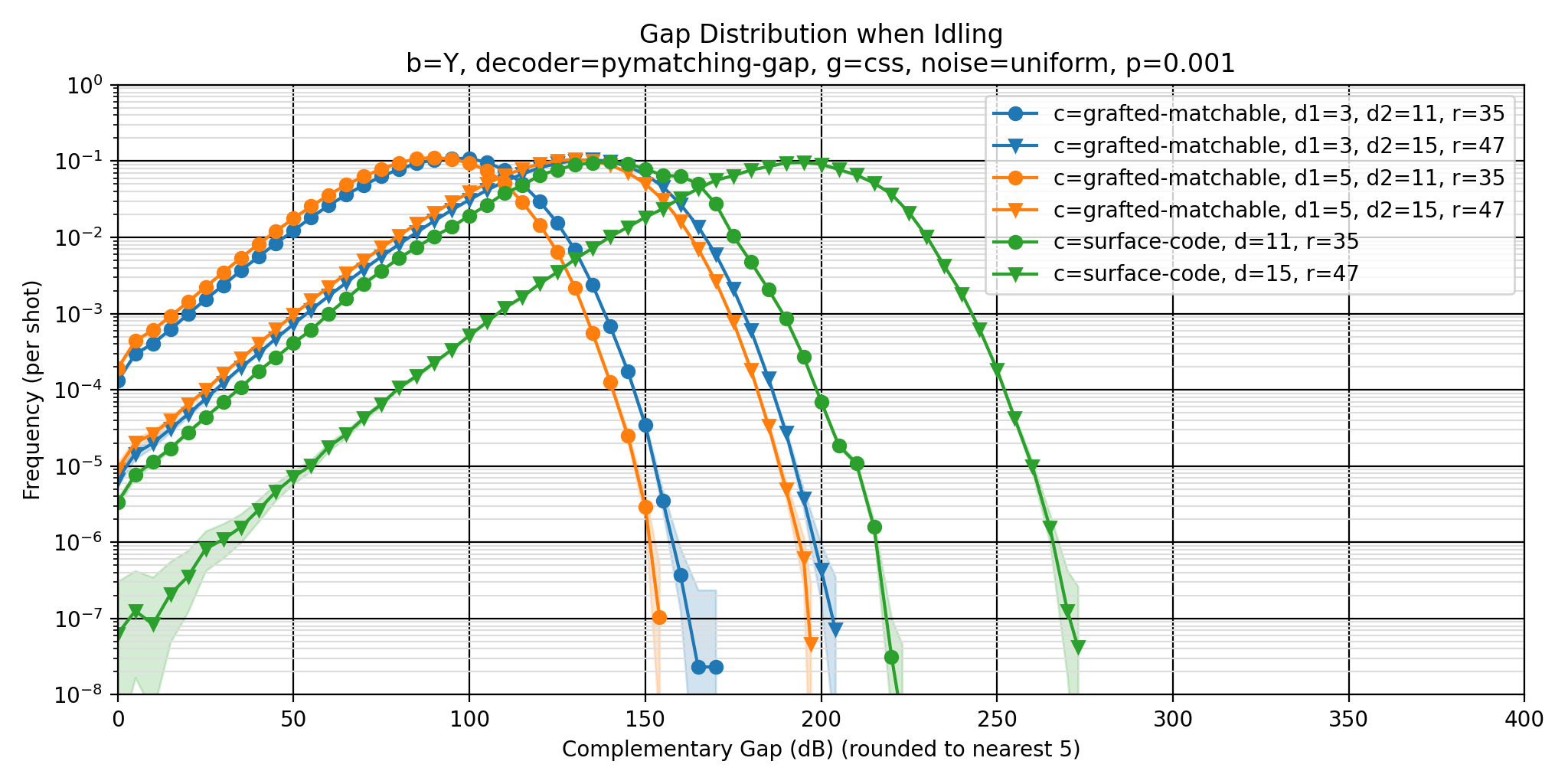}
    }
    \caption{
        \textbf{Distribution of complementary gaps observed when idling} in the surface code and the grafted matchable code that our magic state cultivation construction current ends in.
        Note the similar shapes, with the main distinction being the offsets of the peaks.
        This suggests the grafted matchable code is well behaved, despite its unusual stabilizers.
        Shaded regions indicate frequency hypotheses with a likelihood within a factor of 1000 of the max likelihood hypothesis, given the sampled data.
    }
    \label{fig:gap-distribution}
\end{figure}

\begin{figure}
    \centering
    \begin{adjustwidth}{-2.0cm}{-2.0cm}
        \resizebox{\linewidth}{!}{
            \includegraphics{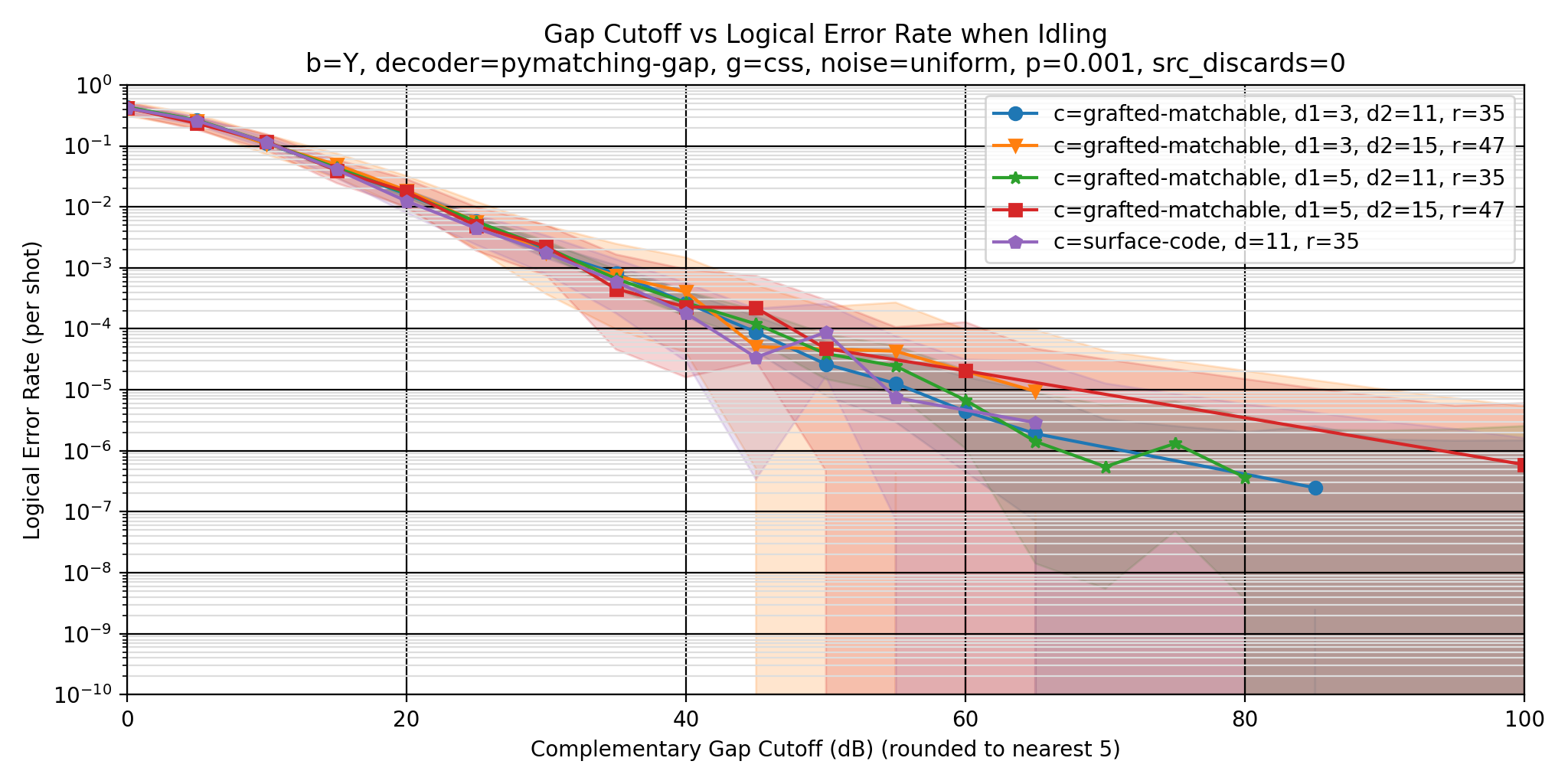}
        }
    \end{adjustwidth}
    \caption{
        \textbf{Calibration of complementary gaps when idling.}
        Sampled logical error rates, conditioned on sampled complementary gaps, when idling in the surface code and the grafted matchable code that our magic state cultivation construction current ends in.
        Shows that the complementary gap is strongly predictive of the logical error rate, both in the surface code and in the grafted matchable code.
        Shaded regions indicate error rate hypotheses with a likelihood within a factor of 1000 of the max likelihood hypothesis, given the sampled data.
    }
    \label{fig:gap-calibration}
\end{figure}

\begin{figure}
    \begin{adjustwidth}{-2.5cm}{-2.5cm}
        \centering
        \resizebox{\linewidth}{!}{
            \includegraphics{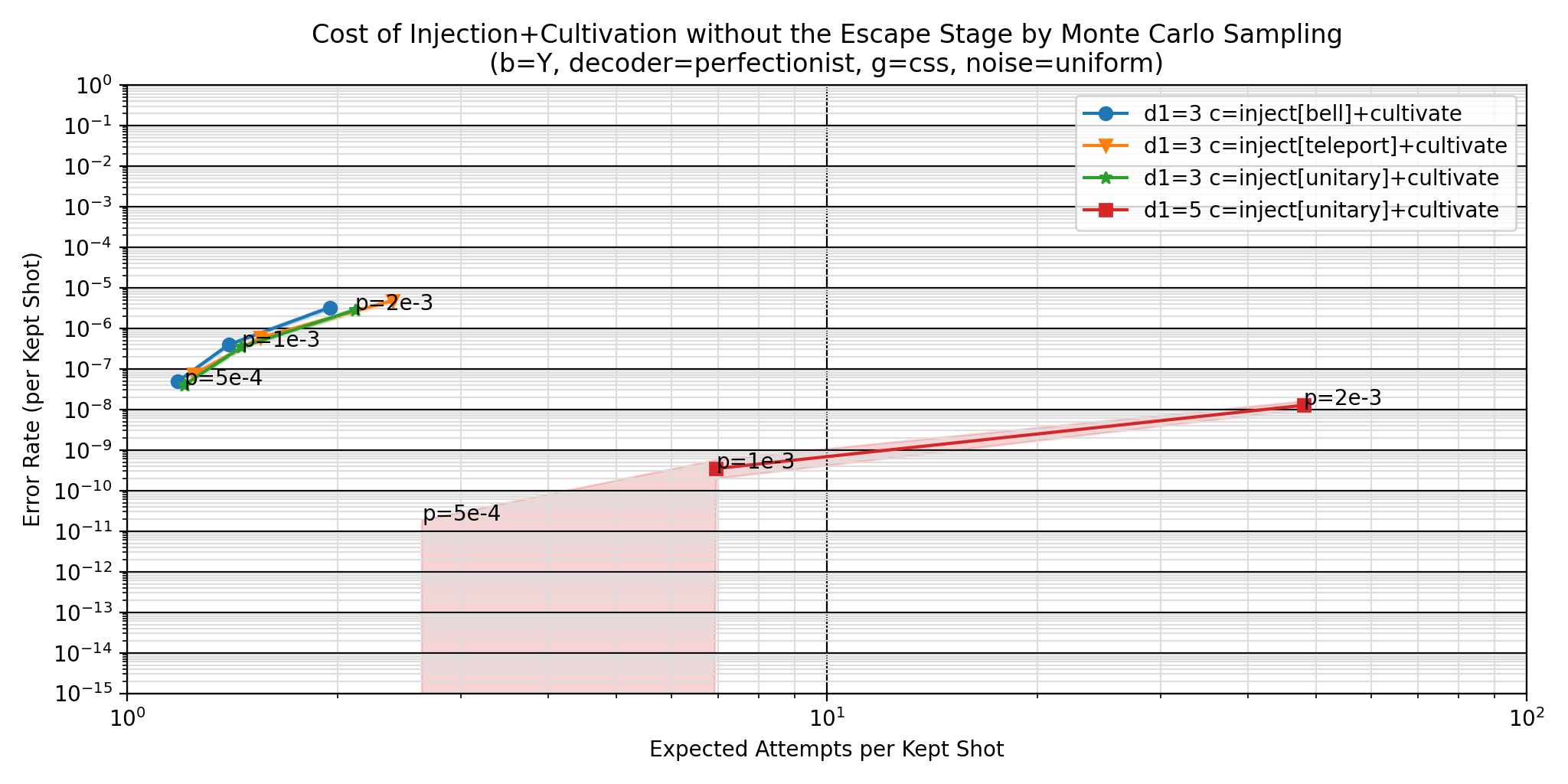}
        }
    \end{adjustwidth}
    \caption{
        \textbf{Monte-Carlo sampling of cultivation without an escape stage}.
        This data was collected to validate \fig{inject-only-enumerated}.
        Shaded regions indicate error rate hypotheses with a likelihood within a factor of 1000 of the max likelihood hypothesis, given the sampled data.
    }
    \label{fig:inject-only-sampled}
\end{figure}

\begin{figure}
    \begin{adjustwidth}{-2.5cm}{-2.5cm}
        \centering
        \resizebox{\linewidth}{!}{
            \includegraphics{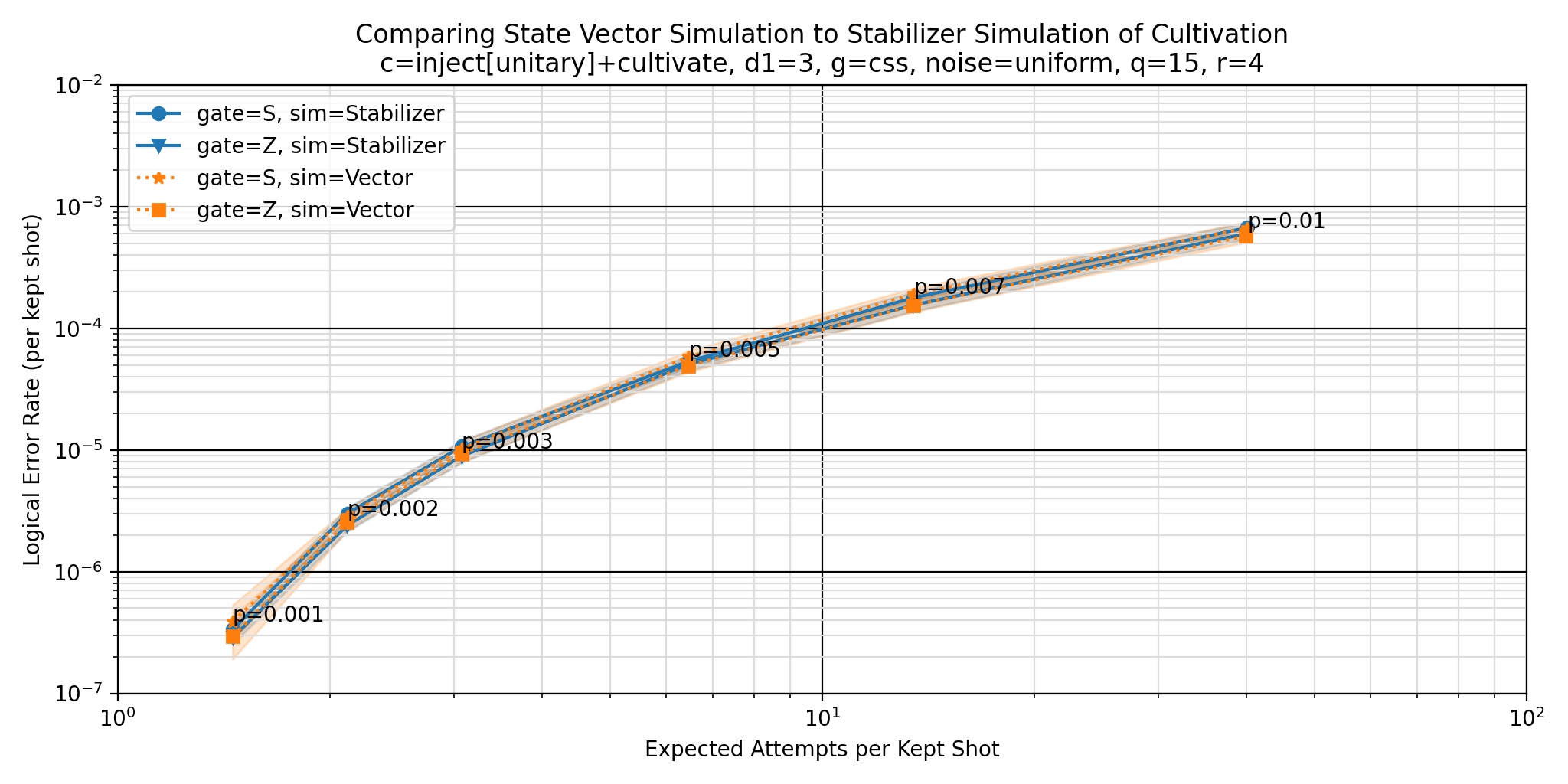}
        }
    \end{adjustwidth}
    \caption{
        \textbf{Comparing stabilizer simulations to state vector simulations}.
        Shows that the state vector simulator is functioning correctly.
        Shaded regions indicate error rate hypotheses with a likelihood within a factor of 1000 of the max likelihood hypothesis, given the sampled data.
    }
    \label{fig:vec-vs-stabilizer}
\end{figure}

\begin{figure}
    \begin{adjustwidth}{-2.5cm}{-2.5cm}
        \resizebox{\linewidth}{!}{
            \includegraphics{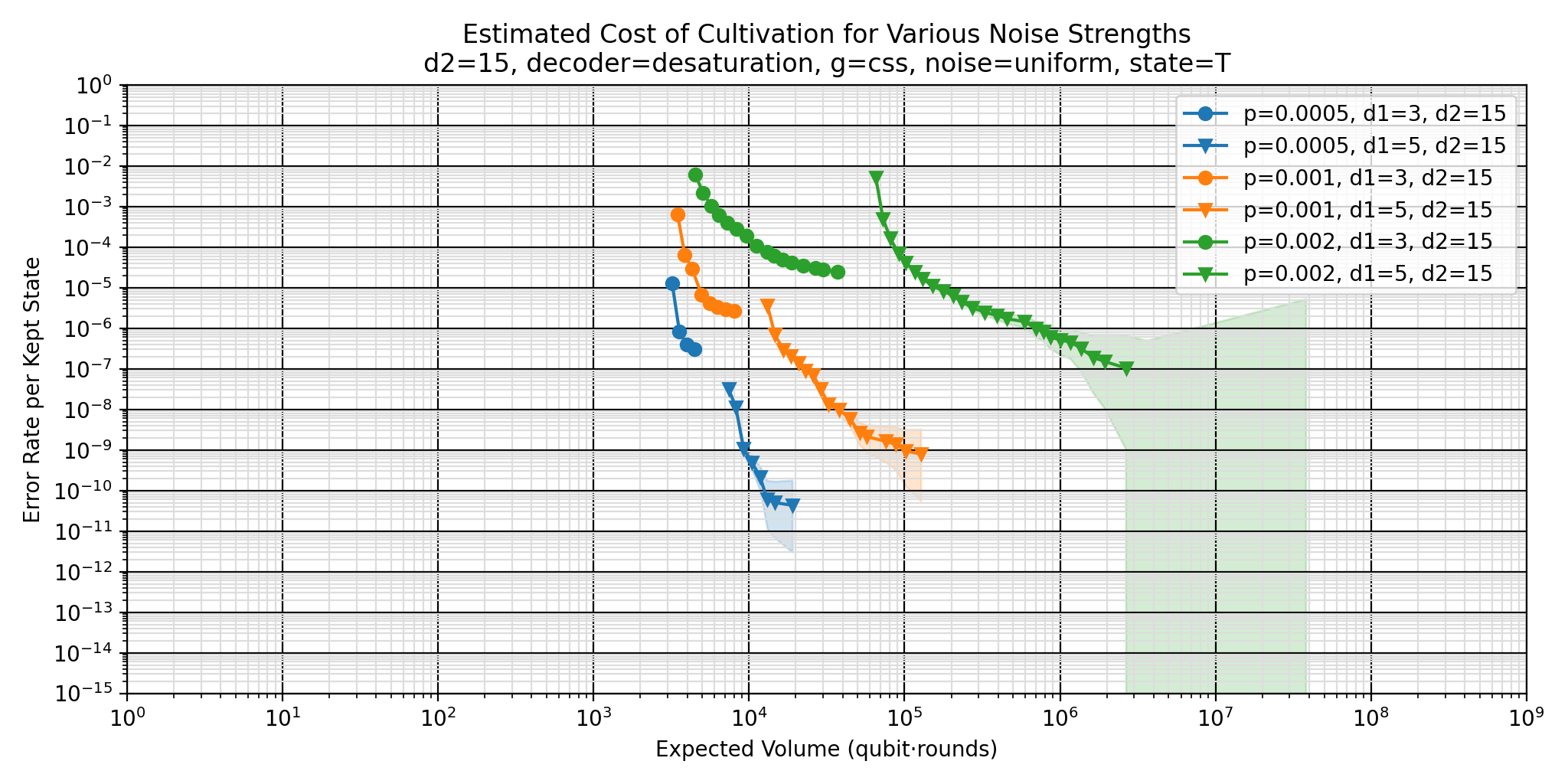}
        }
    \end{adjustwidth}
    \caption{
        \textbf{Cost trade-offs of cultivation under different noise strengths}.
        Same as \fig{historical_progression}, but with different noise strengths instead of different constructions.
        Derived from \fig{end-to-end-error}.
        Shaded regions indicate error rate hypotheses with a likelihood within a factor of 1000 of the max likelihood hypothesis, given the sampled data.
    }
    \label{fig:future-comparison}
\end{figure}

\begin{figure}
    \begin{adjustwidth}{-2.5cm}{-2.5cm}
        \centering
        \resizebox{\linewidth}{!}{
            \includegraphics{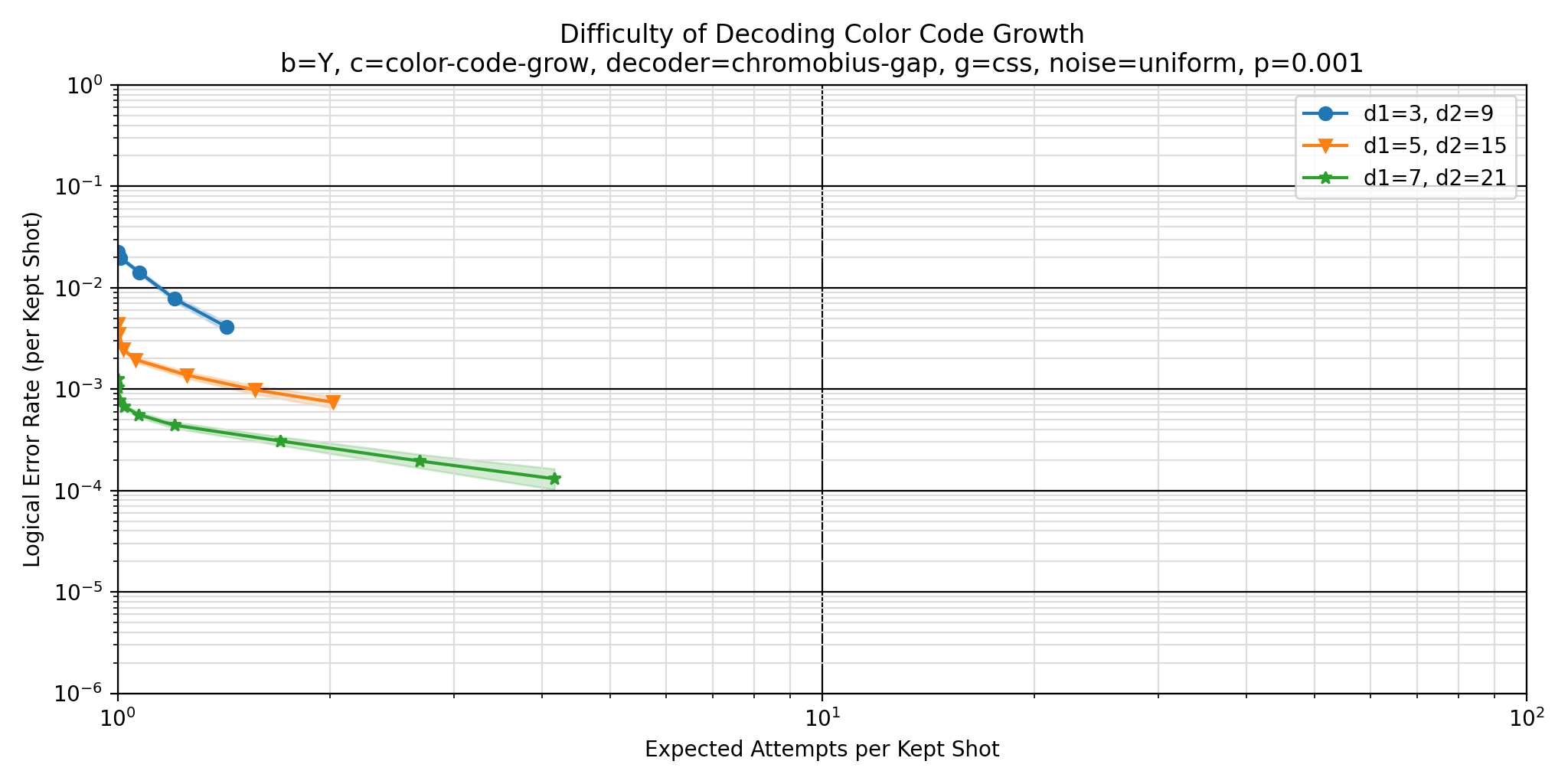}
        }
    \end{adjustwidth}
    \caption{
        \textbf{Decoding color code growth using Chromobius}~\cite{gidney2023colorcode}.
        Curves show a variety of confidence cutoffs, based on complementary gaps.
        Gaps are computed similar to how they are done for matching: by adding a detector equivalent to an observable to the boundary of the patch then decoding once with it forced on and once with it forced off in order to take a weight difference.
        The added detector runs along a boundary, and is colored to complete the boundary (e.g. for the red-green boundary the detector would be colored blue).
        This figure shows that Chromobius isn't good enough to be used for cultivation (e.g. limited to $5 \cdot 10^{-3}$ instead of $2 \cdot 10^{-6}$ when growing from distance 3).
        Shaded regions indicate error rate hypotheses with a likelihood within a factor of 1000 of the max likelihood hypothesis, given the sampled data.
    }
    \label{fig:color-code-growth-difficulty}
\end{figure}

\begin{figure}
    %\begin{adjustwidth}{-2.5cm}{-2.5cm}
        \centering
        \resizebox{\linewidth}{!}{
            \includegraphics{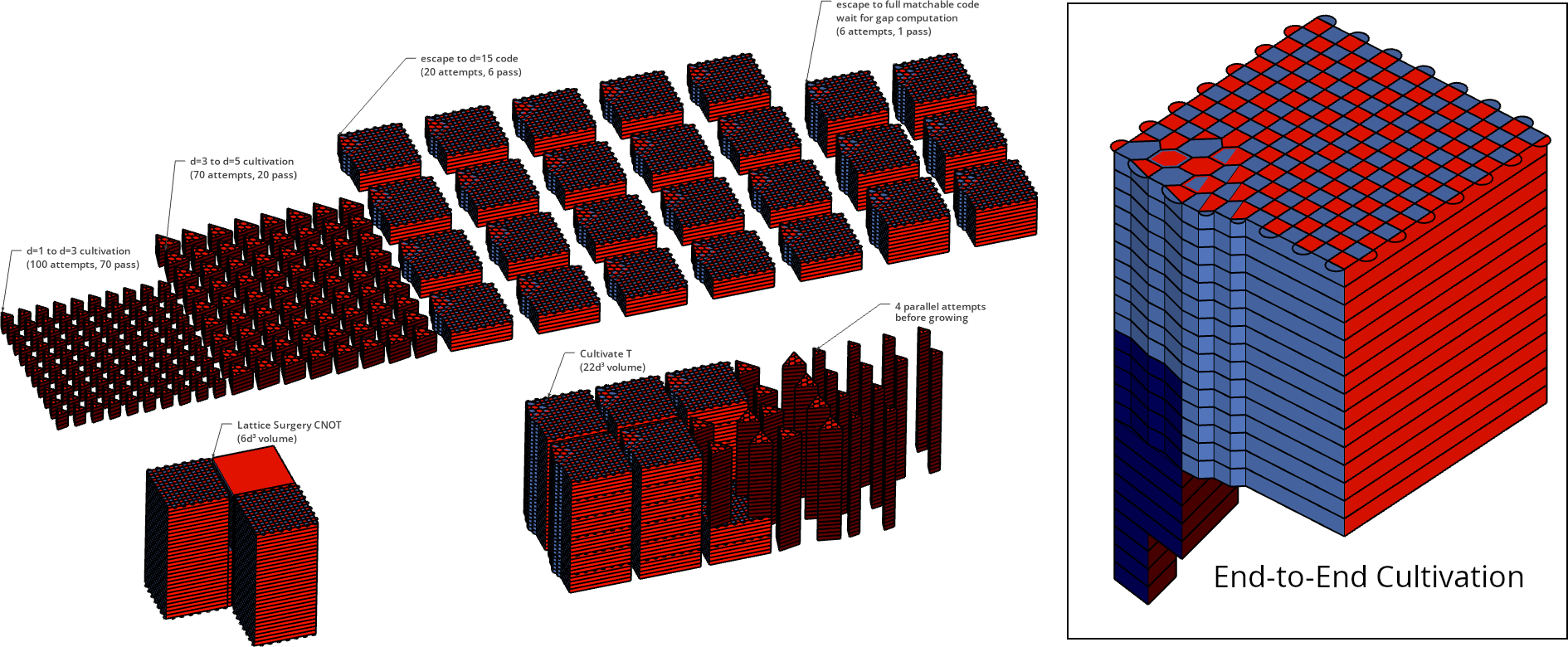}
        }
    %\end{adjustwidth}
    \caption{
        \textbf{Visualizing the size of cultivation}, based on quantities shown in \fig{cultivation-lifetime}.
        Right: a spacetime defect diagram of the full cultivation process, with dark red/blue/green for color code boundaries and light red/blue for X/Z surface code boundaries.
        Top left: defect diagrams of each piece of the cultivation, multiplied by how many are expected per successful cultivation.
        Bottom left: defect diagram of a distance 15 surface code lattice surgery CNOT~\cite{horsman2012latticesurgery}.
        Bottom middle: same as top left but rearranged into stacks for easy comparison to the CNOT.
        The cultivation's spacetime volume is nearly four times larger, when the CNOT uses the same code distance.
        (\fig{historical_progression} shows the distance of the CNOT needs to be larger to have the same reliability.)
    }
    \label{fig:cultivation-layout}
\end{figure}

\begin{figure}
    \centering
    \begin{adjustwidth}{-2.5cm}{-2.5cm}
        \centering
        \resizebox{\linewidth}{!}{
            \includegraphics{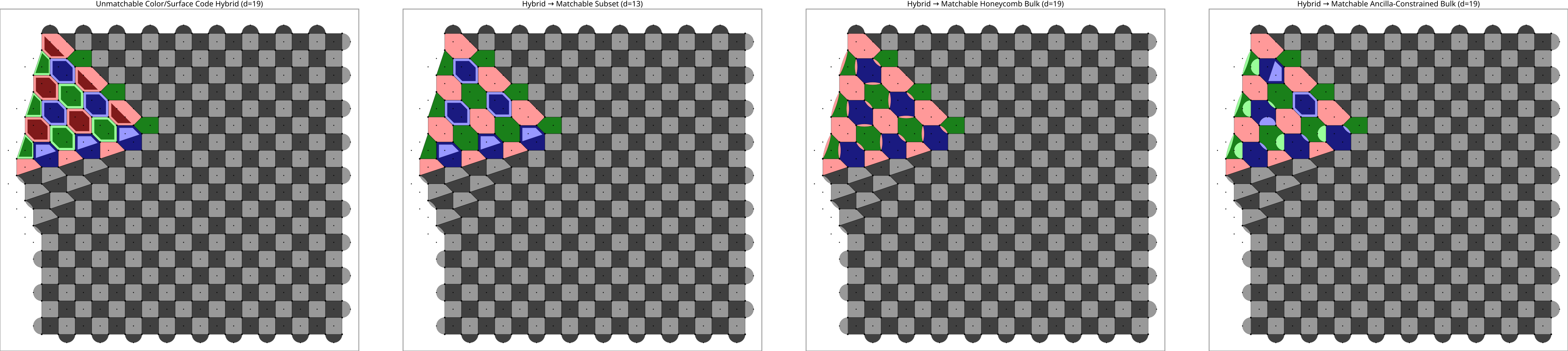}
        }
    \end{adjustwidth}
    \caption{
        \textbf{Various matchable codes the escape stage could end in}.
        Dark/light indicates the basis of the stabilizer (dark=X, light=Z).
        Red/green/blue shows the 3-coloring of the color code.
        Left: a grafted color/surface code.
        Left middle: a matchable code can created purely subtractively, by dropping stabilizers from the grafted code.
        Produces a region equal to a detector slice of the honeycomb code~\cite{hastings2021dynamically,gidney2022planarhoneycomb}.
        Has a code distance of $d_\text{surface} - d_\text{color}$ instead of $d_\text{surface}$.
        Right middle: a matchable code created by replacing the bulk of the color code region with the instantaneous stabilizers of a honeycomb code.
        Preserves the code distance, but has more stabilizers than available measurement qubits.
        Right: the code we actually transition into.
        Adds a subset of the instantaneous honeycomb code stabilizers, without oversubscribing the measurement qubits.
        Has full code distance, but the fault distance of transitioning into this code is slightly lower than $d_\text{surface}$.
    }
    \label{fig:code-transition-alternates}
\end{figure}

\begin{figure}
    \centering
    \begin{adjustwidth}{-2.5cm}{-2.5cm}
        \centering
        \resizebox{\linewidth}{!}{
            \includegraphics{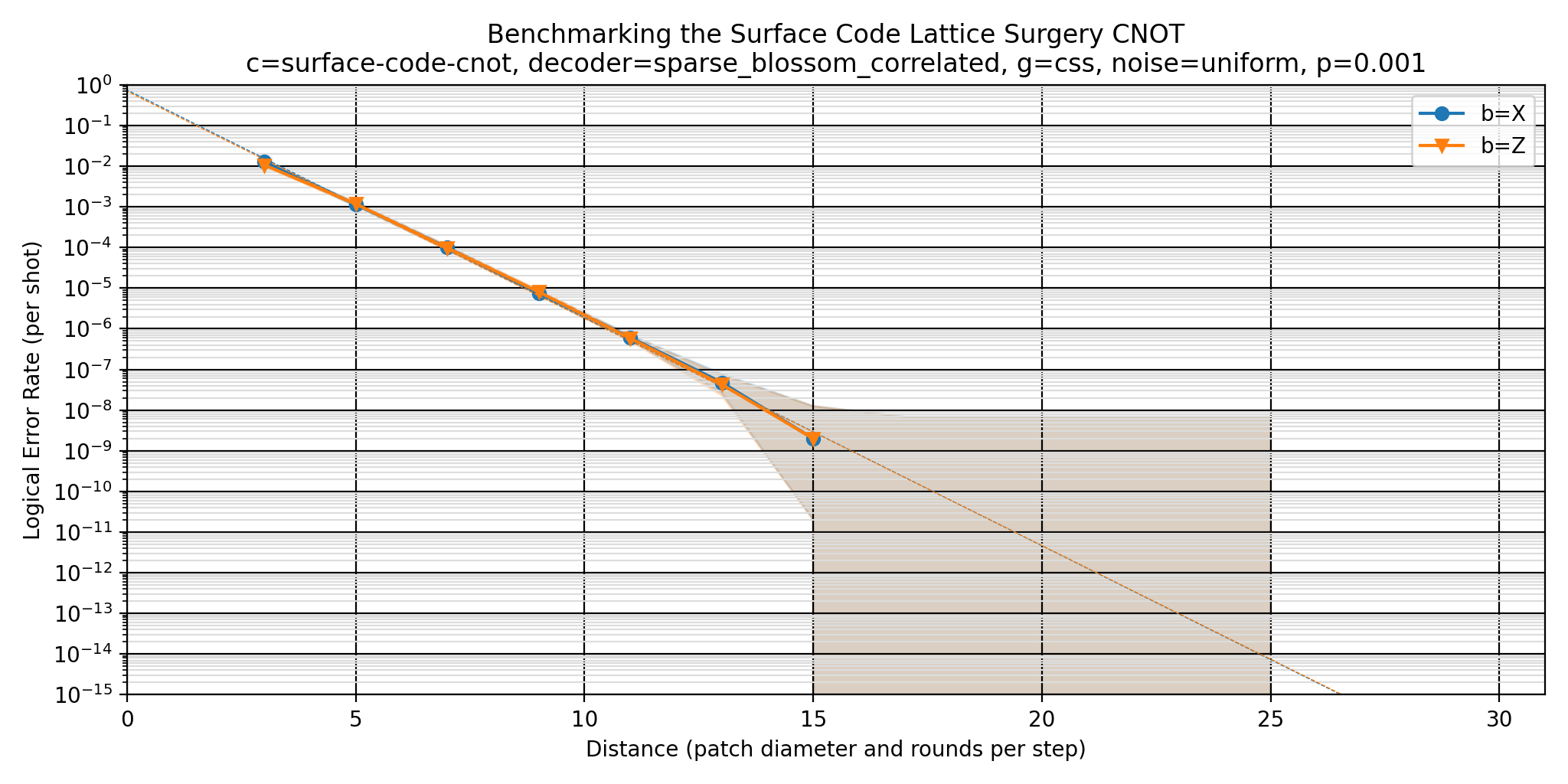}
        }
    \end{adjustwidth}
    \caption{
        \textbf{Performance of a surface code lattice surgery CNOT}.
        Used to extrapolate the curve shown in \fig{historical_progression}.
        The control and target patches are diagonally adjacent, so they touch different boundaries of a common ancilla patch, as in \cite{horsman2012latticesurgery}.
    }
    \label{fig:benchmark-cnot}
\end{figure}

\end{document}